\documentclass[cernpreprint,UKenglish,texlive=2011,txfonts,paper=a4]{latex/atlasdoc}
\pdfoutput=1

\usepackage[block=space]{latex/atlaspackage}
\usepackage{latex/atlasbiblatex}

\usepackage{latex/atlascontribute}

\usepackage{latex/atlasphysics}

\graphicspath{{logos/}{figures/}}

\AtlasTitle{The performance of the jet trigger for the ATLAS detector during 2011 data taking}

\author{The ATLAS Collaboration}

\newcommand{\hvbf}{\fontfamily{phv}\fontseries{b}\selectfont}

\AtlasRefCode{ATLAS-TRIG-2012-01}

\PreprintIdNumber{CERN-EP-2016-031}

\AtlasJournal{EPJC}
\AtlasJournalRef{Eur. Phys. J. C76 (2016) 526}
\AtlasDOI{10.1140/epjc/s10052-016-4325-0}

\newcommand{\jetabstract}{
The performance of  the jet trigger for the ATLAS  detector at the LHC
during the 2011  data taking period is described.  During 2011 the LHC
provided proton--proton collisions with a  
centre-of-mass  energy  of   7\,\TeV\  and 
heavy ion collisions with a 2.76\,\TeV\ per nucleon--nucleon  collision energy.
The ATLAS trigger is  a three level system designed to
reduce the rate of events from the 40\,MHz nominal maximum bunch crossing rate
to the approximate 400\,Hz which can be written to offline storage. 
The ATLAS jet trigger is the primary means for the online selection of
events containing jets. 
Events are accepted by the trigger if they contain 
one or more jets above some transverse energy threshold.  
During 2011  data taking  the jet trigger was  fully efficient
for jets with transverse energy above 25\,\GeV\ for triggers seeded
randomly at Level 1.  For triggers which require a jet to be
identified at each of the three trigger levels, full efficiency is
reached for offline jets with transverse energy above 60\,\GeV.  
Jets reconstructed in the final trigger level and corresponding to 
offline jets with transverse energy greater than 
60\,\GeV, are reconstructed 
with a resolution in transverse energy with respect to offline jets, 
of better than 4\% in the central region and better than 2.5\% in 
the forward direction.
}

\AtlasAbstract{
\jetabstract
}

\AtlasCoverSupportingNote{ATL-COM-DAQ-2013-021}{https://cds.cern.ch/record/1544006/}
\AtlasCoverAnalysisTeam{Nuno Anjos, Michael Begel, {\hvbf Mario Campanelli}, Robert Cantrill, 
{\hvbf Susan Cheatham}, Rebecca Chislett, Patricia Conde, Ingrid Deigaard, {\hvbf Ricardo Gon\c{c}alo}, 
Robert Keyes, Alexandre Lopez, Joana Machado Miguens, David Miller, Nicolas Ortiz, Steven Robertson, 
Francesco Rubbo, Martin Rybar, Helena Santos, Martin Spousta, Michael Stoebe, {\hvbf Mark Sutton}, Louren\c{c}o Vaz}

\AtlasCoverEdBoardMember{Eric Feng}
\AtlasCoverEdBoardMember{Peter Krieger (chair)}
\AtlasCoverEdBoardMember{Martin zur Nedden}
\AtlasCoverEdBoardMember{Brigitte Vachon}

\AtlasCoverEgroupEditors{atlas-atlas-trig-2012-01-editors@cern.ch}

\AtlasCoverEgroupEdBoard{atlas-trig-2012-01-editorial-board@cern.ch}

\hypersetup{pdftitle={ATLAS draft},pdfauthor={The ATLAS Collaboration}}

\newcommand {\pythia}    {\normalfont{\scshape Pythia}}
\newcommand {\herwig}    {\normalfont{\scshape Herwig}}
\newcommand {\hijing}    {\normalfont{\scshape Hijing}}
\newcommand {\fastjet}   {\normalfont{\scshape FastJet}}
\newcommand{\antikt}{\mbox{anti-$k_{t}$}}

\renewcommand{\ET}{\mbox{$E_{\mathrm T}$}}
\renewcommand{\et}{\mbox{$E_{\mathrm T}$}}

\newcommand{\etoffline}{\mbox{$E_{\mathrm T}^{\,\mathrm{Offline}}$}}

\newcommand{\ptoffline}{\mbox{$p_{\mathrm T}^{\mathrm{Offline}}$}}

\newcommand{\deta}{\mbox{$\Delta\eta$}}
\newcommand{\dphi}{\mbox{$\Delta\phi$}}
\newcommand{\mus}{\mbox{$\mu$s}}
\newcommand{\largeR}{\mbox{large-$R$}}
\newcommand{\LargeR}{\mbox{Large-$R$}}
\newcommand{\offline}{\mbox{$^{\mathrm{Offline}}$}}

\renewcommand{\epsilon}{\varepsilon}
\newcommand{\deltaR}{\mbox{$\Delta R$}}
\newcommand{\delR}{\deltaR}
\newcommand{\pseudorapidity}{\mbox{pseudorapidity}}

\newcommand{\multijet}{\mbox{multi-jet}}
\newcommand{\Multijet}{\mbox{Multi-jet}}

\newcommand{\pileup}{\mbox{pile-up}}
\newcommand{\Pileup}{\mbox{Pile-up}}
\newcommand{\spacer}{}

\renewcommand{\HT}{\mbox{$H_{\mathrm T}$}}

\begin{document}

\maketitle

\tableofcontents

\section{Introduction}
\label{section:introduction}

ATLAS~\cite{atlas} is one of two general purpose detectors at the
Large Hadron Collider (LHC)~\cite{lhc}. 
During the 2011
running period the LHC operated with a collision energy of $\sqrt{s} =
7$\,\TeV, allowing ATLAS to collect an integrated luminosity of
5.25\,fb$^{-1}$ during proton--proton ({$pp$}) collisions, and
158\,$\mu$b$^{-1}$ during lead--lead (Pb+Pb) collisions with 
centre-of-mass energy of 2.76\,\TeV\ for each pair of 
colliding nucleons
in the interaction.

The large event rate at the LHC makes the online selection of
interesting physics events essential for achieving the physics goals
of the LHC programme.  During the 2011 data taking period, the LHC ran
with a bunch spacing of 50\,ns providing a nominal rate of 20\,MHz, and with a mean of more than 20
separate $pp$ interactions per bunch crossing (known as \pileup)
towards the end of data taking. To reduce the rate of events to be read out
from the detector to a rate of around 400\,Hz which can be written to offline
storage, a rejection factor greater than $10^5$ is required. This is achieved 
by the ATLAS trigger~\cite{atlastrigger} which is divided into the Level 1
(L1) trigger and the High Level Trigger (HLT). In 2011 the HLT itself 
consisted of two levels: Level 2 (L2) followed by the Event Filter (EF).

The jet trigger system of the ATLAS detector is the primary means to
select events containing jets with high transverse energy (\ET).  It
selects collision events to be used in jet physics
analyses~\cite{atlasjets1,atlasjets2,atlasjets3,atlasjets4,atlasjets5,atlasjets6,atlasjets7,atlasjets8},
as well as in many other analyses where one or more jets may be
required, perhaps in conjunction with additional physics signatures such
as an isolated lepton candidate. 
In this paper, the design and performance of the ATLAS jet trigger during
the 2011 data taking is described.
 
The outline of this paper is as follows:
Section~\ref{section:triggerDesign} describes the design of the ATLAS
jet trigger.  Section~\ref{section:menu} provides an overview of the
jet based event selection defined in the ATLAS trigger menu and
explains the nomenclature used for trigger names.  The timing, or CPU budget, of each
trigger level is outlined in Section~\ref{section:timing}.
Various aspects of
jet trigger performance are described in 
Section~\ref{section:performance},
which outlines the measures used for the evaluation of
the trigger performance, and includes the selection efficiencies 
for inclusive single jet and \multijet\ triggers. 
Descriptions of specialised
triggers designed for specific physics selections are provided in
Section~\ref{section:physicsTriggers}. These include selections for
triggering on large summed scalar \ET\ or boosted objects that can
decay into multiple narrow jets.  Event selection in the Pb+Pb
programme is described in Section~\ref{section:HI}.

\subsection{The ATLAS detector and trigger system}
\label{section:detector}

\newcommand{\atlastrigger}{
\begin{figure}[t!]
  \begin{center}
  \includegraphics[width=0.5\textwidth]{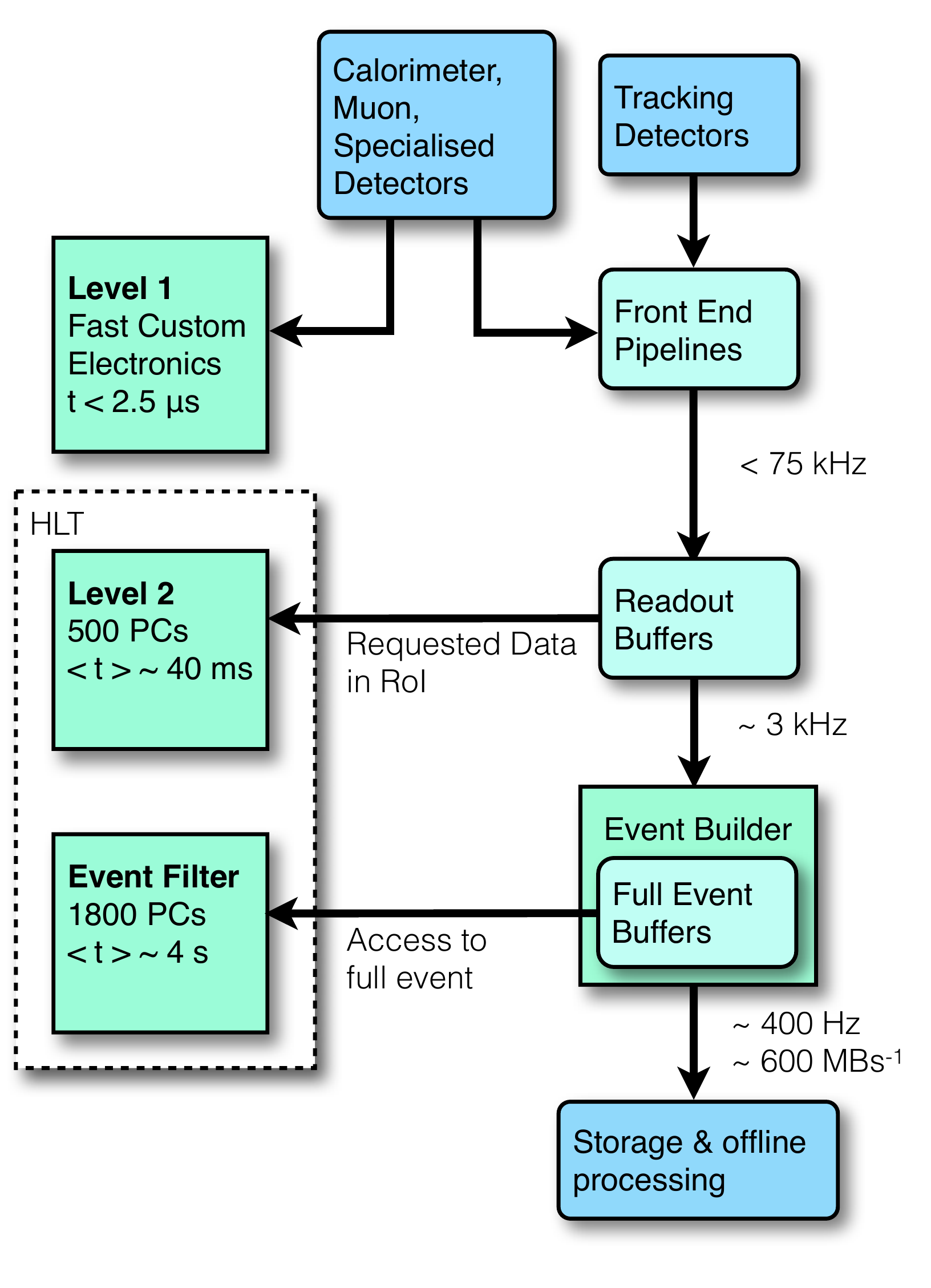}
  \end{center}
  \caption{The ATLAS trigger system.}
  \label{fig:atlastrigger}
\end{figure}
}

The ATLAS detector is a multi-purpose particle detector with a
forward-backward symmetric cylindrical geometry and a near $4\pi$
coverage in solid angle\footnote{ATLAS uses a
  right-handed coordinate system with its origin at the nominal
  interaction point (IP) in the centre of the detector and the
  $z$-axis along the beam pipe.  The $x$-axis points from the IP to
  the centre of the LHC ring, and the $y$-axis points upwards.
  Cylindrical coordinates $(r,\phi)$ are used in the transverse plane,
  $\phi$ being the azimuthal angle around the $z$-axis.  The
  \pseudorapidity\ is defined in terms of the polar angle $\theta$ as
  $\eta = -\ln \tan(\theta/2)$.}.  
Owing to the cylindrical geometry, subdetector
components are described 
as being in the {\em central region}, if they are part of the barrel, 
at small absolute \pseudorapidity,
or described as {\em forward}, if part of 
the endcaps at large absolute \pseudorapidity.
Outwards from the beam pipe, ATLAS consists of an
inner tracking detector surrounded by a thin superconducting solenoid
providing a 2\,T axial magnetic field, electromagnetic and hadronic
calorimeters, and a muon spectrometer.  The inner tracking detector
covers the pseudorapidity range $|\eta| < 2.5$ and consists of silicon
pixel, silicon microstrip, and transition radiation tracking
detectors.

The calorimeters cover the region $|\eta| < 4.9$ and consist of
electromagnetic (EM), 
and hadronic 
subsystems. The EM, the hadronic endcap (HEC), and the forward
calorimeters (FCal) use liquid argon (LAr) as the active medium, and 
either a lead, copper or tungsten absorber technology. The EM calorimeter is divided into a barrel part,
$|\eta|<1.475$, and two endcap components with $1.375<|\eta|<3.2$.
The central hadronic calorimeter, referred to as the tile calorimeter,
uses steel absorber layers interleaved with plastic scintillator
covering the pseudorapidity range $|\eta| < 1.7$.  A presampler is
installed in front of the EM calorimeter for $|\eta| < 1.8$.  For the
calorimeter subsystems, there are two separate readout paths: the
first, a very fast readout of combined towers of calorimeter cells,
is used at Level 1, while the second is the slower readout of the full
calorimeter cell information for use in the HLT and offline.

The muon spectrometer surrounds the calorimeters and is based on three
large air-core toroid superconducting magnets with eight coils each.
The toroid bending power ranges from 2.0 to 7.5\,Tm.  The
muon spectrometer 
includes a system of precision tracking chambers and fast detectors
for triggering.

The ATLAS trigger~\cite{TDAQTDR,Aad:1125884,atlastrigger} for 2011
consisted of three processing levels, each allowing increasingly
detailed reconstruction and selection algorithms. This approach
enables the successive identification of potentially interesting
features and the early rejection of less interesting events. The L1
trigger runs hardware algorithms over data with reduced spatial granularity 
from the calorimeter and muon subsystems to identify geometrical 
{\em regions of interest} (RoI) in the detector, containing candidate
physical objects which should be examined more closely in subsequent
trigger levels. The L1 trigger has a fixed maximum latency of 2.5\,\mus, and
a rate for accepting events up to 75 kHz.  For standard triggers, events with at 
least one RoI passing the L1 selection are passed to the
L2 system, which runs software algorithms on a farm of commodity CPUs.
The L2 algorithms have access to the data at the full detector
granularity but only from those detector elements that lie within an
RoI.  The number of processors in the L2 farm and the time taken to
process each event provides a limit on the rate at which events 
can be accepted by the L1 system.

Following the L2 processing, all events with RoIs that satisfy a set
of predefined selection criteria are passed to the event builder which
reads out the detector at full granularity. These fully built events
are then processed by the EF, which also consists of a farm of
commodity CPUs. The EF farm runs modified versions of the offline
reconstruction algorithms, simplified to improve the speed of execution.
Although the full event data are available at the EF, 
for many trigger signatures the EF trigger reconstruction takes
place within RoIs for reasons of speed. This is not the case for the 
jet trigger, for which
the whole detector is read out.  The rate of L2 accepted events passed
to the EF during 2011 was approximately 3\,kHz, and the rate at which
events were read out for offline storage was approximately 400\,Hz.
The ATLAS trigger is illustrated in Figure~\ref{fig:atlastrigger}.   

\atlastrigger

In 2011 the full jet trigger was operated in rejection mode for the
first time, allowing events to be discarded at each of the three
trigger levels.  Prior to 2011, the ATLAS trigger selection for events
containing jets was based purely on the algorithms running at L1 and
L2, with the EF algorithms executed in commissioning mode only. In
this mode, events were processed by the EF but not rejected should
they have failed the EF requirements. The resulting trigger decision was
stored in the event stream for commissioning purposes.

\section{Jet trigger design overview}
\label{section:triggerDesign}

The jet trigger is an integral part of the ATLAS trigger system,
processing events based on successively more detailed detector
information at the L1, L2 and EF stages. Hadronic and
electromagnetic energy deposits in the calorimeter subsystems are used 
to reconstruct jets; fast, custom jet algorithms are used at L1 and L2; 
and for the EF, the \antikt~\cite{antikt} algorithm in the four-momentum 
recombination scheme, implemented in the \fastjet~\cite{fastjet} package 
is used.
In each of the three
stages, the bandwidth allocated to the jet trigger is about 10\% of
the total.  Jet trigger signatures, simply referred to as jet
triggers, are divided into either {\em central} or {\em forward}, with
the central jet triggers using detector data from the central and endcap
calorimeters ($|\eta|< 3.2$) and the forward jet triggers in the region
$3.2 <|\eta|< 4.9$ using data from the FCal. Different electronics 
are used for each to take account~\cite{cscnote} of the more coarse FCal
detector granularity in the forward direction.

The L1 calorimeter trigger system (L1Calo)~\cite{Achenbach:1080560}, 
is the first stage of the jet trigger. This
reconstructs jets from the combined
energy deposits in the LAr and tile calorimeters, using collections of
calorimeter cells projecting back to the nominal interaction point,
known as {\em trigger towers}.  A square sliding window of
$0.8\times0.8$ in $\deta\times\dphi$ is used to identify regions
where the summed transverse energy within the central $0.4\times0.4$ region of the 
window is large and corresponds to a local maximum~\cite{Lampl:2008zz, Mehdiyev:1999xfa}.

The jet candidate \ET\ values are then compared to a set of predefined
\ET\ thresholds to decide which candidates should form an RoI.  The
trigger thresholds are discussed in Section~\ref{section:menu}.
Information about the regions of the detector that contain jet
candidates -- specifically the multiplicity of candidates exceeding each threshold -- is sent to the central trigger
processor (CTP) and used in the generation of the global L1 decision. This is then
distributed to the detector front-end electronics, to initiate the data
readout, and the subsequent stages of the trigger. Information on which 
jet thresholds from L1 have been satisfied
can also be combined with information from other L1 trigger subsystems,
such as electron or muon triggers, to produce multi-object triggers.

The data from events which pass the L1 selection are processed by the
L2 trigger, which has access to the calorimeter cells within the RoIs
identified by L1. Limiting the data processed in this way allows the
detailed trigger reconstruction of any potentially interesting object,
whilst requiring typically only 1--2\% of the full detector data 
corresponding to the detector elements within the RoIs to be read out. 
The L2 jet trigger
runs a feature extraction algorithm consisting of a simple, iterative
cone algorithm (described in Section~\ref{sec:l2jets}) to build jets
using the full detector granularity.  The L1 RoI corresponding to the
jet is said to {\em seed} the L2 processing in the HLT.  The
characteristics of jets found using the iterative cone algorithm are
tested with a {\em hypothesis algorithm} to determine if they fulfil
the predetermined L2 trigger selection criteria. These criteria may
include minimum values for the jet transverse energy, and selection on
the jet \pseudorapidity\ and quality. Each event selected at L2 is 
then fully built from the various fragments temporarily stored in
memory in the data acquisition system.

The final stage of the trigger, the EF, must perform jet
reconstruction in the full event within approximately 4\,s before
making a decision on whether to write the event to offline
storage. Due to the larger available latency at the EF compared to L2,
more sophisticated reconstruction algorithms can be applied.  To the
maximum extent possible, the EF uses standard ATLAS event
reconstruction algorithms developed for offline analysis, as well as
final offline detector calibrations.  Since the EF runs after the full
event has been built by the event builder, it is able to access
information from the complete detector, rather than just that from
detector elements in an RoI. The EF jet trigger reconstructs
\antikt\ jets in the full calorimeter, in the same manner as the
standard offline jet reconstruction, rather than separately processing
data within individual RoIs.

The ability of the EF to operate on the full calorimeter data permits
seeding by triggers which select, at random, some fraction of events
from L1 at a predefined rate irrespective of whether any RoI is 
present. Using the random trigger in this way allows the EF to trigger 
on jets free from any bias that might be introduced by the jet
reconstruction at either the L1 or L2 stages. This is particularly
useful for lower \ET\ jet thresholds, where such biases can be large.

\subsection{Level 1}

The L1 trigger decision is based on analogue sums of signals from
calorimeter elements within 7,168 projective regions (trigger towers),
independent of the precision readout used in the HLT and offline.
Trigger towers have a size of approximately $\Delta\eta \times
\Delta\phi = 0.1 \times 0.1$ in the central part of the calorimeter
within $|\eta| < $2.5, and are larger and less regular in the more
forward regions. Electromagnetic and hadronic calorimeters have
separate trigger towers. The 7,168 analogue inputs to the L1
calorimeter trigger are first digitised and then assigned to a
particular LHC bunch crossing.

Two separate processor systems, working in parallel, run the trigger
algorithms. One system, the cluster processor, uses the full L1
trigger granularity information in the central region to look for
small localised calorimeter energy clusters typical of electrons, photons or the products
of tau lepton decays.  The other, used for jet and missing energy triggers,
uses coarser granularity {\em jet elements}, 
to identify jet candidates and form global
\ET\ sums: missing \ET, total \ET, and the scalar sum of all jet \ET.  
The jet elements consist of $2\times2$
arrays of trigger towers in the central region and fewer in the foward region where the trigger towers are larger. 
The \ET\ of
individual energy depositions and the \ET\ sums are compared to
preprogrammed trigger thresholds to form the trigger decision. Jet
RoIs are defined as $4\times4$ jet element windows for which the
summed electromagnetic and hadronic calorimeter \ET\ exceeds 
predefined thresholds
and which encompass a $2\times2$ jet element core where the 
hadronic calorimeter \ET\ is a local maximum. The location of the centre of 
this $2\times2$ array defines the coordinates of the jet RoI.

\subsection{Level 2}
\label{sec:level2}

In order to handle the large event rate from the detector, following a
decision to accept an event at L1, the L2 decision must arrive within
approximately 40\,ms. Even with the reduction in data volume
from reading out only those data corresponding to the RoIs identified
by L1, the data preparation at L2 still represents a large
contribution to the overall processing time.  In this section the data
preparation and jet finding stages of the L2 system are discussed.

\subsubsection{Level 2 data preparation}
\label{sec:dataPreparation}

The data preparation for the L2 jet trigger is a crucial part of the
L2 processing.  It provides the collection of data from detector
readout drivers (RODs)~\cite{TDAQTDR}, delivery to the L2 processing 
units, and the conversion from the raw data into the objects used by 
the HLT algorithms.  The RODs receive data from the calorimeter front-end
boards via optical fibres.  These boards are installed on the detector
and contain electronics for the amplification, shaping, sampling,
pipelining, and digitisation of the calorimeter
signals~\cite{LArReadiness,TileReadiness}. Due to the large number of
calorimeter readout channels, approximately 2$\times10^5$, and in
order to meet the L2 timing performance goal of 40\,ms per event, the
data volume read out should be kept to the minimum required to avoid
compromising algorithm performance. For each detector element
(calorimeter cell) within the RoI window, the direction from the
nominal interaction point to the element position is binned in a grid
in the $\eta$ -- $\phi$ plane, for use in the L2 jet reconstruction
algorithm.

\subsubsection{Level 2 jet reconstruction algorithm}
\label{sec:l2jets}

At L2, jets are defined as cone-shaped objects~\cite{cscnote} in the
$\eta$ -- $\phi$ plane with a given radius, $R$, such that they
contain energy deposits with a separation
$\delR\equiv\sqrt{(\Delta\eta)^2+(\Delta\phi)^2} < R$, where $\Delta\eta$
and $\Delta\phi$ are defined with respect to the jet axis.  The value of the 
radius parameter, $R$, is set during the trigger configuration.  The jet
energy and position are found through an iterative procedure using the grid in 
$(\eta, \phi)$ populated by the cell energies, with the
following steps:
\begin{itemize}
\item First, an initial reference jet, $j_0$, is defined by the L1 jet RoI
  position with the predefined cone radius $R$. Note that the
  possible directions of the reference jet are discrete due to the 0.2
  $\times$ 0.2 granularity at L1.
\item The $k$ elements from the binned distribution that fall within
  the ($\eta$, $\phi$) region encompassed by the reference jet, $j_0$,
  are used to recalculate the jet energy 
  and the energy weighted
  average position of the jet, to define a new, updated 
  reference jet $j_1$, according to
\begin{eqnarray}
  E_{j_1}    &=& \sum_{i=1}^{k} E_{i},\label{ene}\\ 
  \eta_{j_1} &=& \frac{\sum_{i=1}^{k}E_{i}\eta_{{i}}}{\sum_{{i=1}}^{k}E_{i}},\label{eneweightedpos1}\\ 
  \phi_{j_1} &=&
  \phi_{{j_0}}+\frac{\sum_{{i=1}}^{{k}}E_{{i}}\times(\phi_{{i}}-\phi_{
      {j_0}})}{\sum_{{i=1}}^{{k}}E_{{i}}}. \label{eneweightedpos2}
\end{eqnarray}
  where the sum runs over the $k$ grid elements whose centroids are
  contained within the cone of radius $R$ centred on the reference jet
  $j_0$.  The total energy, and coordinates $\eta_{j_1}$ and
  $\phi_{j_1}$, are computed from Equations~(\ref{ene}), ~(\ref{eneweightedpos1}) and
  (\ref{eneweightedpos2}).
\item The previous step is repeated with $j_0$ replaced by $j_1$ in
  Equations~(\ref{eneweightedpos1}) and (\ref{eneweightedpos2}), and
  so on to form jet $j_i$ from jet $j_{i-1}$, updating the jet energy $E_{j_i}$ and 
  the coordinates
  $\eta_{{j_i}}$ and $\phi_{{j_i}}$.  The iteration is repeated $N$
  times to create jet $j_{{N}}$.  A configurable number of iterations
  are executed. Typically, $N=3$ is used, having been found 
  sufficient to achieve the required performance~\cite{cscnote}.
\end{itemize}

The result of this algorithm is a jet defined by the reconstructed
($\eta$, $\phi$) direction, and the total jet energy. This energy 
is evaluated at the
electromagnetic calorimeter energy scale, by summing the
energy depositions in the electromagnetic and hadronic parts of the
calorimeter without applying any further calibration.

For the central jet trigger, $R=0.4$ is used. For the forward jet trigger, 
because of the coarse granularity of the FCal,
the radii used for the first and second iterations are 1.0 and 0.7 respectively, 
to ensure that the energy deposits are fully contained given the coarse 
position available for the L1 jet. For the final iteration, the radius $R=0.4$ is used.

\subsubsection{Level 2 full scan trigger}

Towards the end of data taking in 2011 a new 
{\em Level 2 full scan} trigger~\cite{Tamsett:proc,susy:l1.5}, using 
the lower granularity trigger tower data from Level 1, was introduced. Here, the trigger tower data 
for the full calorimeter for each event was read out by the Level 2 system and processed 
on the Level 2 CPU farm with  the \antikt\ algorithm.
This trigger was running in commissioning mode only during the heavy ion run
at the end of 2011 and was not deployed for production data taking in the proton--proton 
jet trigger until 2012.

\subsection{Event Filter}
\label{sec:ef}

\newcommand{\tcm}{{\em TrigCaloMaker}}
\newcommand{\ttm}{{\em TrigCaloTowerMaker}}
\newcommand{\tjr}{{\em TrigJetRec}}
\newcommand{\tjh}{{\em TrigEFJetHypo}}

The EF is the last stage in the trigger and is responsible for the
final decision of whether an event should be sent to offline storage
or discarded. The jet trigger at the EF is modular and makes use of
three general stages; data preparation (calorimeter unpacking and
pre-clustering), jet finding, and hypothesis
testing.

\newcommand{\designfigtwo}{
\begin{figure*}[thp]
  \includegraphics[width=\textwidth]{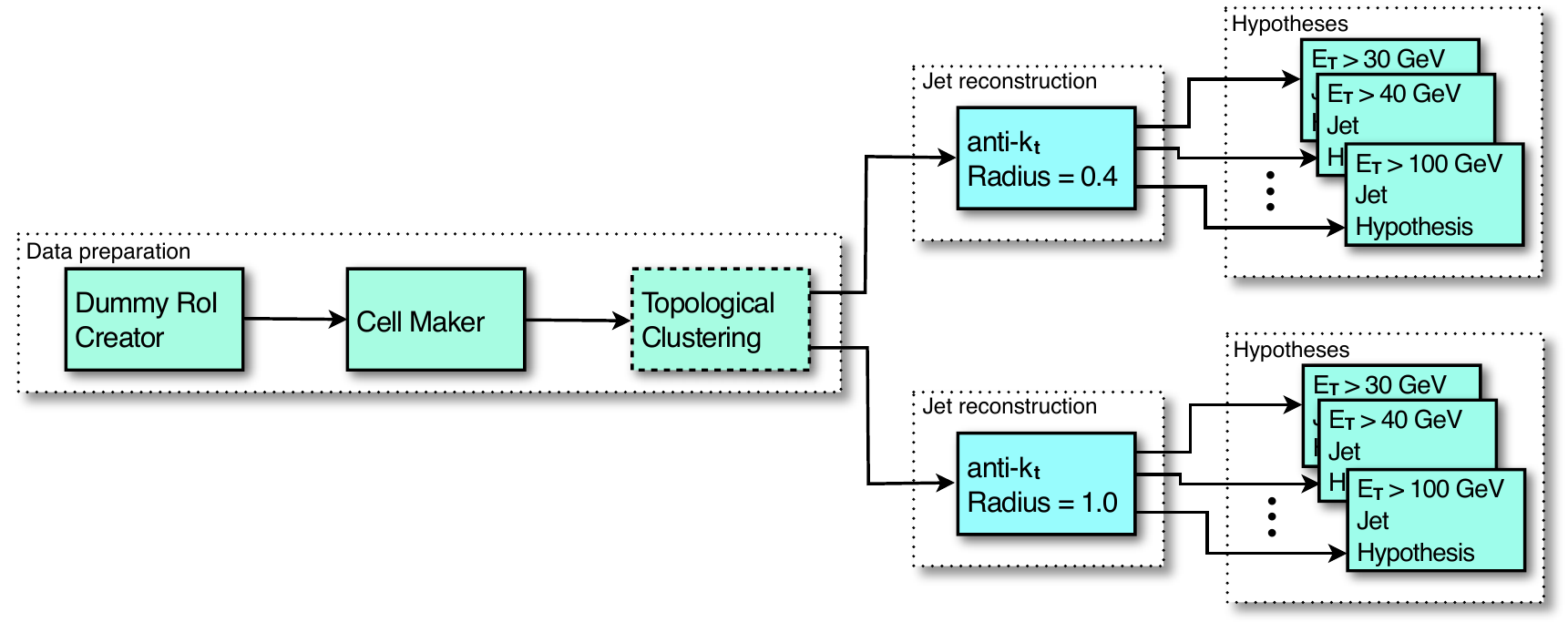}
  \caption{The stages of algorithm processing in the Event Filter for
    several inclusive single jet triggers with different
    \et\ thresholds. The case illustrated shows two sets of
    signatures, each set with a different jet radius parameter.}
  \label{fig:designfigtwo}
\end{figure*}
}

In contrast to the RoI-based approach used at L2, the EF runs the jet
finding algorithm once per event for each configured jet radius, using
data from the complete calorimeter. This is referred to as a {\em full
  scan}. The full scan approach has several advantages for jet
reconstruction with respect to the RoI based approach used at L2.  The
large RoIs required at L2 to ensure that any jet is completely
contained has the unfortunate disadvantage that RoIs may overlap in
events with high jet multiplicity, resulting in some parts of the detector
being processed multiple times. This can result in jets being fully,
or partially reconstructed in several RoIs, which may cause the double
counting of energy deposits and jets, which would affect the
\multijet\ signatures.  The full scan approach completely eliminates
the multiple processing of regions of the detector 
and, as a consequence, leads to faster processing in high
occupancy events, although it takes longer in low occupancy events,
where the processing time is in any case low.

Since the output from L2 is in the form of lists of RoIs passing each trigger 
threshold, a slightly different approach is required to seed the EF processing.
In this case, the first jet RoI to be processed by the EF initiates the
creation of a dummy, full scan RoI, encompassing the entire detector, 
required to ensure that the entire calorimeter 
is processed. 
The calorimeter cell data for this full detector RoI is then extracted by the 
{\em cell maker} and processed 
to provide the objects upon which the jet finding will then run.
Following the jet finding, hypothesis algorithms are executed. These
determine whether any specific jet selection signatures 
are satisfied, for example, typical selections are those based on 
specific jet \et\ thresholds.

\designfigtwo

The objects from both the data preparation and the jet finding stages are cached for 
this full scan RoI. When evaluating
any additional trigger signature requiring jets passing a different \ET\ threshold 
in the same event, 
the trigger 
can establish that this dummy RoI has already been created and 
will not start the sequence for the data preparation and jet finding again, instead simply 
retrieving the jets from the cache.
The hypothesis algorithm for this different \et\ threshold will then be executed.

Since the cell data are cached following the data preparation stage, the jet algorithms with 
different radius parameters run over the cached data and only the jet finding itself will be 
executed again for each different required radius.
In this way 
the data preparation is common to all jet finding, which is in turn executed only once for each
jet radius  required. 
The full sequence for multiple thresholds and multiple jet radius parameters is illustrated in 
Figure~\ref{fig:designfigtwo} and the individual stages are discussed in more detail below.

\subsubsection{Event Filter data preparation}
\label{sec:inputToJetFinder}

The jet finder stage can operate with a number of different types of
input objects produced by the data preparation from the raw cell data.
In early 2011 the primary objects used as input to jet finding were
projective calorimeter towers constructed from the raw calorimeter
cell information. From May 2011, so-called {\em
topological clusters}~\cite{Lampl:2008zz} were used. These are 
discussed later.
Since the
offline jet reconstruction also uses topological clusters, this improves the EF
jet resolutions with respect to offline reconstruction, although the
topological clustering algorithm does add additional processing time 
to the data preparation stage. 

The topological clustering algorithm creates clusters of topologically
related energy deposits.  The algorithm starts with a seed calorimeter
cell, with an energy deposit with absolute value greater than four
standard deviations above the expected noise. All cells directly
neighbouring these seed cells, in all three dimensions, are
collected into the cluster. Cells adjacent to the cluster are then
added, if they have an energy with an absolute value exceeding the 
noise by two standard deviations, iterating until all such adjacent 
cells have been used. Finally, a ring of guard cells is added to complete 
the cluster.  After the initial clusters have been formed, they are
analysed to identify local maxima, and split should more than one such
maxima be found in a cluster~\cite{Lampl:2008zz}.

\subsubsection{\Pileup\ noise suppression} 
\label{sec:pileupnoise}

Jet reconstruction in the trigger is affected by the presence of
\pileup\ interactions, which give rise to energy deposits in the
calorimeter that are unrelated to the primary interaction of interest.
The overlap of these energy deposits with those of the jets of
interest can distort the reconstructed direction and \ET\ of the
jet. Due to the long integration time of the calorimeter electronics
-- up to 600\,ns~\cite{atlas} -- the detector response is also
dependent on energy deposits arriving earlier or later than the
nominal beam crossing.  The size and likelihood of contributions due
to \pileup\ depend on the number of interactions per bunch crossing.
To account for this, the noise thresholds applied during the
topological clustering process were tuned at the start of the 2011
running period to reflect the expected mean number of
interactions per bunch crossing.

\subsubsection{Jet finding and hypothesis testing}
\label{sec:hypo}

Jet finding can be performed using any of the available offline jet
algorithms.  
Due to problems with the infrared and collinear safety of cone
algorithms~\cite{irsafety}, ATLAS has adopted
$k_\perp$-ordered sequential combination
algorithms~\cite{snowmass,kt}, and specifically the
\antikt~\cite{antikt} algorithm in the four-momentum recombination 
scheme as the jet algorithm of choice for
physics analyses~\cite{atlasjets5,atlasjets1,atlasjets2,atlasjets3}.
To match this offline choice, the \antikt\ algorithm was chosen for
use in the EF for 2011 data taking, to replace the ATLAS
cone~\cite{atlascone,cscnotejets} jet algorithm used in the trigger prior to 2011. Two
different values of the radius parameter, $R$,  were used in the EF trigger
reconstruction in 2011, $R=0.4$ and 1.0, the larger radius being
useful for the study of hadronic decays of boosted heavy particles.

Should any additional calibrations be required for the final jets
themselves, the jet reconstruction process can run a post-processing
stage to apply them to jets. As in the case of the offline processing, 
the EF jet algorithm runs on the full calorimeter information. 
Differences between the trigger and offline jets
generally only arise because the final offline
calibrations are not normally available at the time of data
taking. During the 2011 data taking the jet trigger was operated at
the electromagnetic scale, i.e. with no jet calibration applied.

For an inclusive single jet trigger, the hypothesis algorithm that executes following the 
jet finding, accepts events which
have at least one jet which satisfies the required criteria.  
Since the jets for each event are cached in memory, subsequent calls to 
hypothesis algorithms with different selection thresholds simply
use this cached jet collection. Identifying \multijet\ events is also
simply a case of iterating over the reconstructed jets to identify
combinations which pass the relevant selections for each signature.
Different \multijet\ signatures are possible, including those where
the \ET\ of each jet in the event is required to exceed a different
\ET\ threshold. 
The hypothesis algorithm takes as parameters the required jet
multiplicity, $n$, the \eta\ range within which the jets must lie,
$\eta_{\mathrm {min}} \le |\eta_{\mathrm {jet}}| < \eta_{\mathrm {max}}$, and the
required \ET\ thresholds for each of the required $n$ jets.

\section{The jet trigger menu}
\label{section:menu}

\newcommand{\qt}{\em}

The trigger system is configured via a menu which includes the 
specification of the list of event signatures to be accepted for events
written to offline storage. For the jet
trigger, this includes the number of jets, \ET\ thresholds, $\eta$
ranges, and other parameters such as jet-quality criteria, to be applied
at each of the three trigger levels. The aim of the menu design is to
deploy a complementary and robust set of selections for physics
channels of interest, compatible with the given bandwidth
limitations. The trigger menu determines the configuration of the
L1 firmware and the algorithms executed at the HLT. Corresponding triggers
in each of the three trigger levels constitute a trigger {\em chain}.

The names of the trigger selections used
in this document consist of the jet multiplicity followed
by the \et\ threshold separated with a {\qt j} for L2 and the EF, or
{\qt J} for L1.  This is preceded by the trigger level separated by an
underscore, so for instance {\qt EF\_j100} would be a 100\,\GeV\
single-jet trigger at the EF, and {\qt L1\_5J10 } would be a five jet trigger at
L1 with a 10\,\GeV\ transverse energy requirement on each jet. Additional items 
may be included in the name for specialised triggers, such as {\qt FJ} for 
forward jets which are required to have $|\eta|>3.2$.
Typically the item names also include
information regarding the specific jet algorithm.  For instance {\qt a4tc} or
{\qt a10tc} indicate that the \antikt\ algorithm was used, with radius
parameters 0.4 or 1.0 
 respectively, and running
on topological clusters ({\qt tc}).  Where this string is omitted, 
\antikt\ jets with radius parameter $R=0.4$ should be assumed. 
All the jet triggers used at the 
EF during 2011 were full scan triggers, and as such had names 
appended by {\em EFFS} to indicate the EF full scan; however, for the following discussion, 
the {\em EFFS} may be omitted from the trigger name for brevity.

Trigger selections at each level are designed to reduce the CPU usage at
later trigger levels by maximising event rejection at early
stages. Trigger thresholds in the higher levels are tightened to avoid the
distortion of the efficiency curve from the slower-rising efficiency
of previous levels.  Triggers can operate in {\em pass-through} mode, which
entails executing the trigger algorithms but accepting the event
irrespective of the algorithm decision. This allows the trigger
selections and algorithms to be validated, to ensure that they are
robust against the varying beam and detector conditions, which are
hard to predict before data taking.  Partial pass-through mode allows only 
a certain percentage of events to be passed through the trigger in this way, 
the rest being subject to the usual trigger selection. This operational 
mode was used during data taking for several triggers.  
Passing events through in
this way allows data to be collected by the higher threshold triggers
for performance evaluation and debugging, with as little bias as possible.

Further flexibility is provided by defining {\em bunch groups}, which
allow triggers to include specific requirements on the LHC bunches
colliding in ATLAS.  Not all bunch crossings contain protons; those
that do are called {\qt filled} bunches.  For the random trigger,
filled bunch crossings were required, indicated in the trigger name by
{\qt FILLED} at L1, and {\qt filled} at L2.  Non-collision triggers
require a coincidence with an {\qt empty} or {\qt unpaired} bunch
crossing, which correspond respectively to no protons in either LHC
beam or a filled bunch in only one beam.  For some of the lowest
threshold physics triggers, a corresponding non-collision trigger was
included in the chain for background studies.

As well as the trigger chains  selecting jets at both L1 and L2, there
were chains running  at the EF, which were seeded  by a random trigger
at L1,  and passed the events  through L2 without  running a selection
algorithm.  These allowed  triggering on very low \et\  jets at the EF
without the biases introduced by the L1 jet reconstruction at low \et.

In addition to the more common jet triggers such as inclusive single jet,
and \multijet\ triggers, some specialised jet triggers, dedicated to more specific
physics signatures, were used in 2011:
\begin{itemize}
\item Event Filter triggers that reconstruct \HT, the total scalar sum
  of \ET\ of all jets in an event. Such triggers are useful for
  physics analyses which study or search for events with a large
  summed \ET\ in the final state, as the requirement of large \HT\ can
  help to control the trigger rate without requiring e.g. a very
  energetic leading jet;
\item jet triggers where the jet algorithm is executed with a
  \largeR\ parameter, useful for searching for heavy particles
  decaying into boosted hadronic final states; the \antikt\ algorithm
  was used with $R=1.0$ (denoted {\qt a10});
\item heavy ion triggers, used for the Pb+Pb data taking period at the
  end of 2011, having a total transverse energy requirement in
  \GeV\ denoted by {\qt TE}, differing with respect to the
  \HT\ requirement used in proton runs in that TE is the sum of all
  transverse energy in the calorimeter, not only of that clustered in
  jets.
\end{itemize}

\begin{figure}[thp]
  \setlength{\unitlength}{1mm}
  
  \subfigure[]{
    \includegraphics[width=0.49\textwidth]{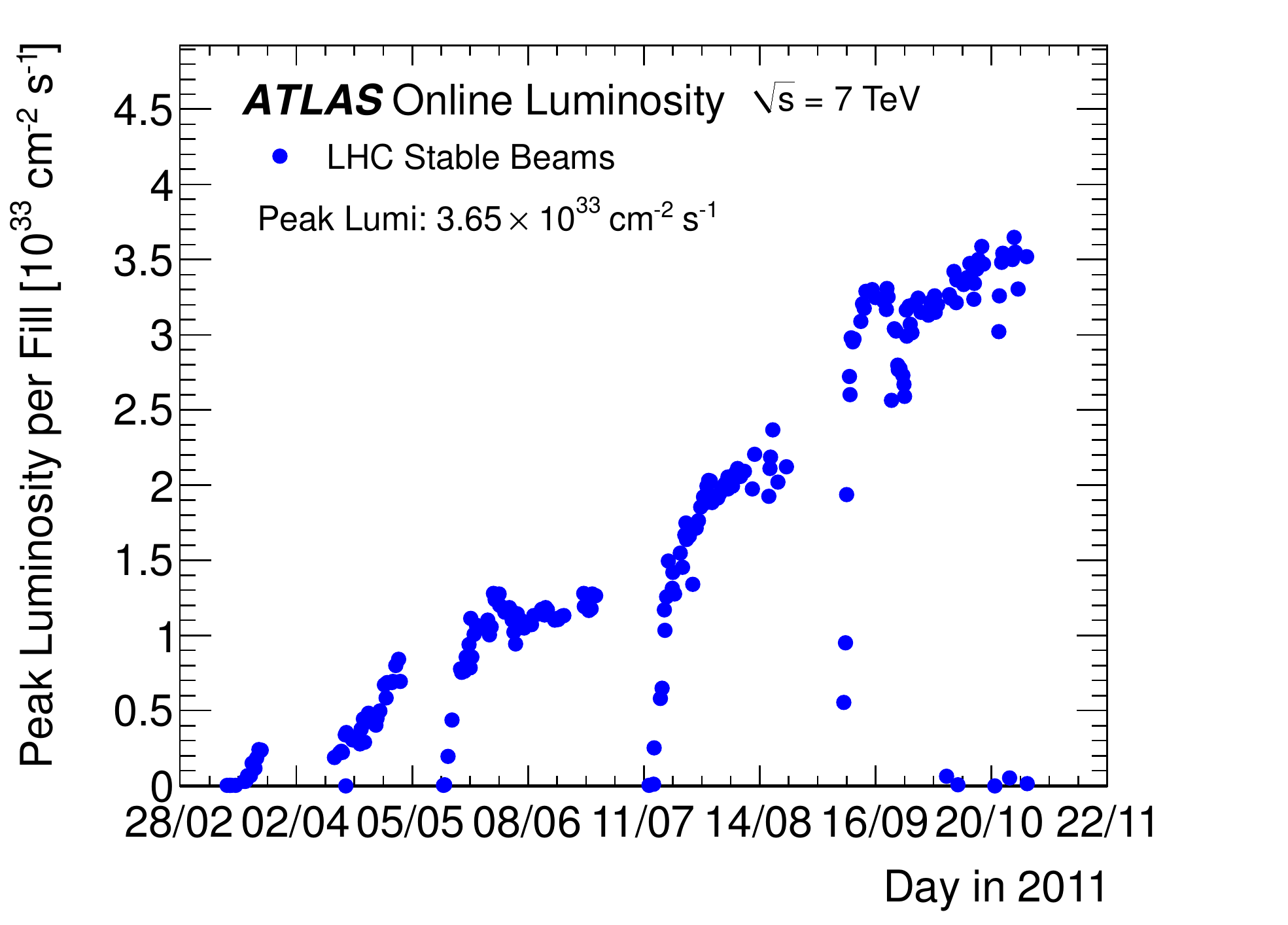}}
  \subfigure[]{
    \includegraphics[width=0.49\textwidth]{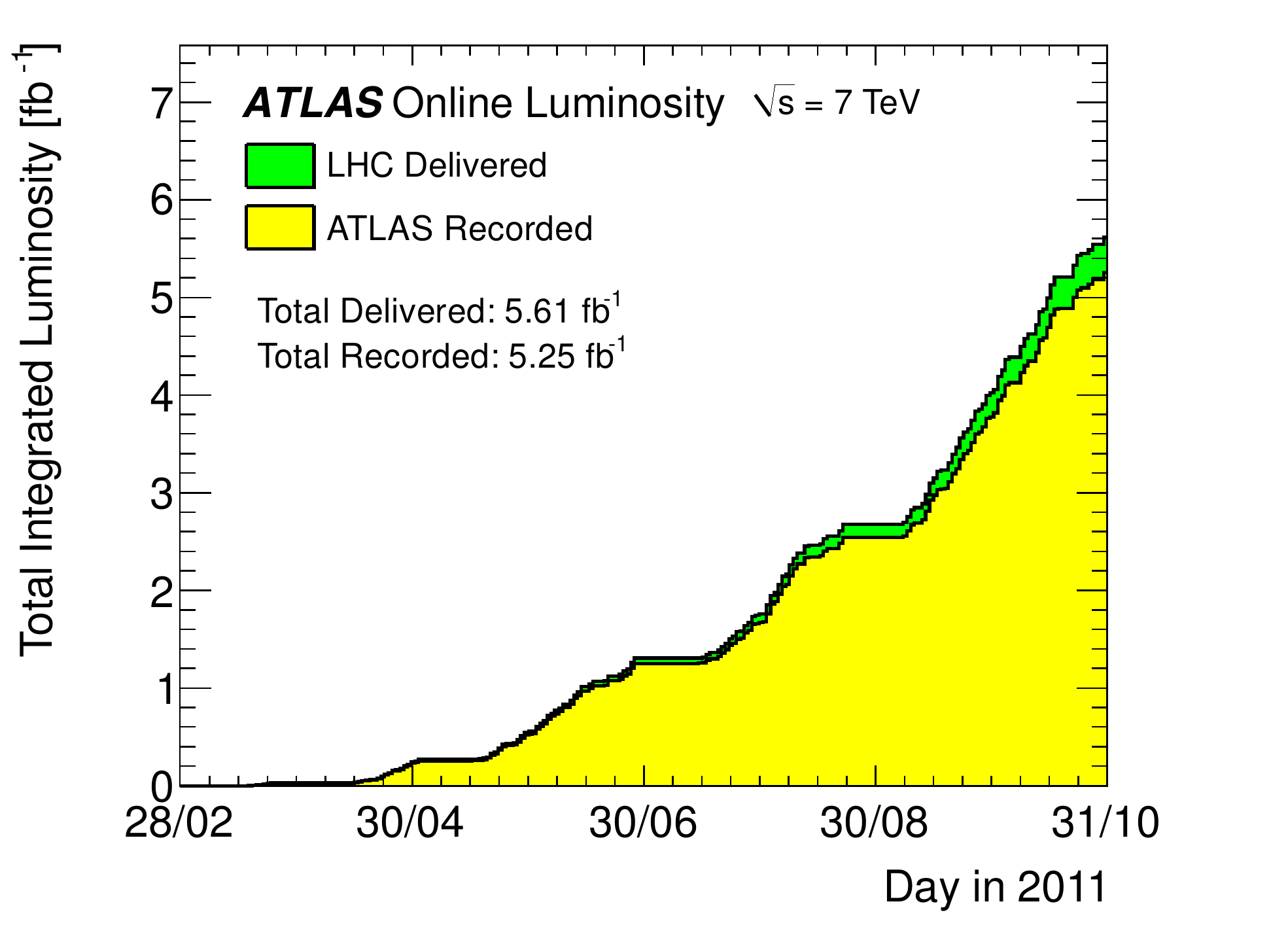}}
  \caption{The luminosity measurement at the ATLAS interaction region for
    2011 data taking~\cite{LuminosityPublicResults}: (a)~the maximum
    instantaneous luminosity versus day delivered to ATLAS during
    stable beam operation; (b)~the cumulative luminosity versus day
    delivered to (green), and recorded by (yellow) ATLAS during stable
    beam operation for $pp$ collisions at 7\,\TeV\ centre-of-mass energy.
  }
  \label{fig:LHC_delivered_lumi}
\end{figure}

The first time ATLAS used both the L2 and EF stages of the HLT in
event rejection mode was in 2011.  A number of key improvements were
introduced during that year, including the ability to use topological
clusters rather than calorimeter towers at the EF, as discussed in
Section~\ref{sec:inputToJetFinder}, which was found to increase the 
stability of the algorithm in the presence of \pileup. During the 2011 data taking
period the LHC peak instantaneous luminosity increased by more than an
order of magnitude, from $10^{32}\,\textrm{cm}^{-2}\textrm{s}^{-1}$ to
$3.6\times10^{33}\,\textrm{cm}^{-2}\textrm{s}^{-1}$.
Figure~\ref{fig:LHC_delivered_lumi} shows the maximum instantaneous
luminosity and the integrated luminosity delivered to ATLAS during
2011 as a function of time.  The highest values for the mean number of 
interactions per bunch crossing reached $\sim$20 towards the end of running in
2011.  The jet trigger menu evolved during this period to adapt to the
changing LHC conditions.

\begin{figure*}[pht]
  \subfigure[]{
    \includegraphics[width=0.49\textwidth]{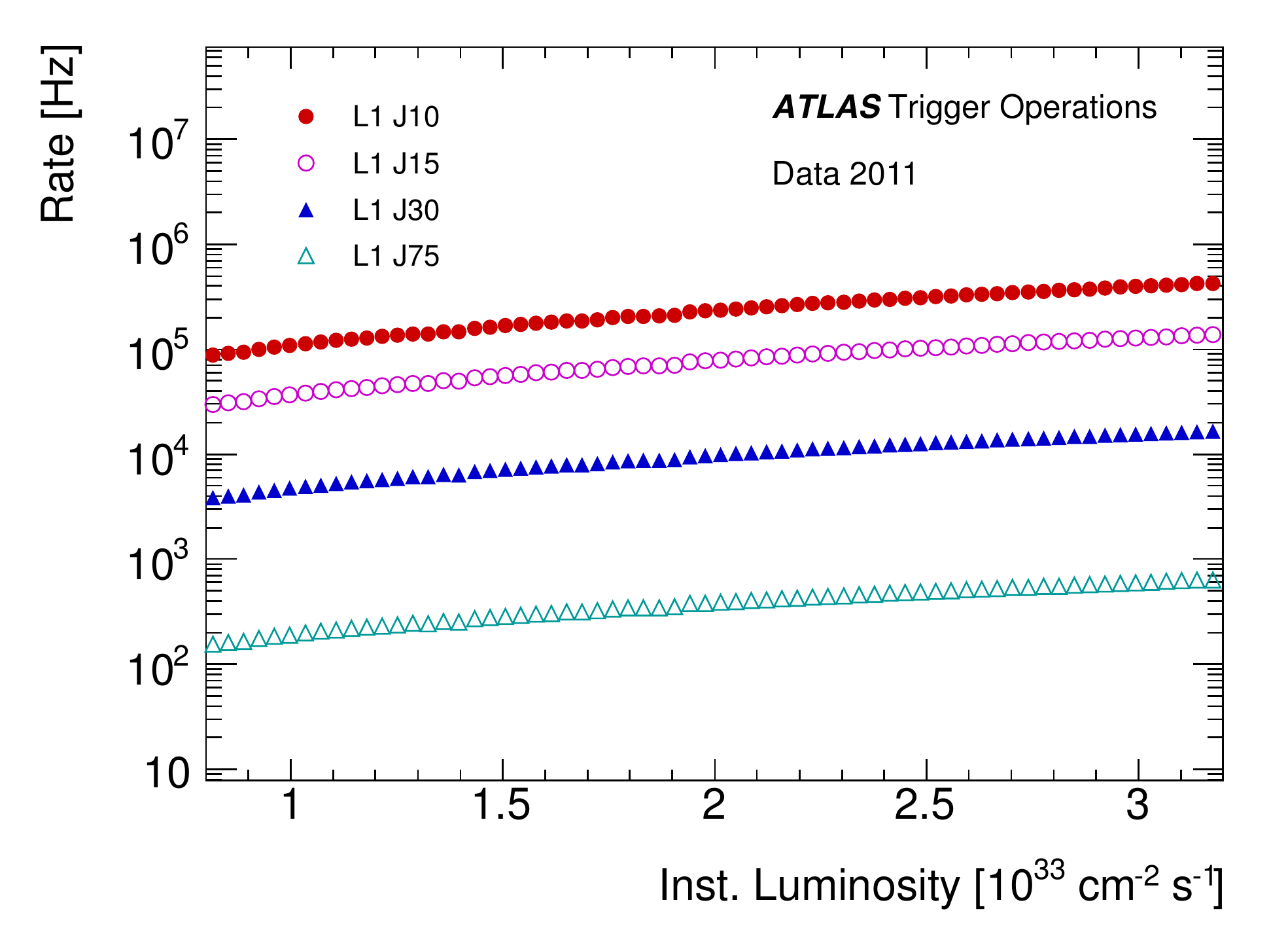}}
  \subfigure[]{
    \includegraphics[width=0.49\textwidth]{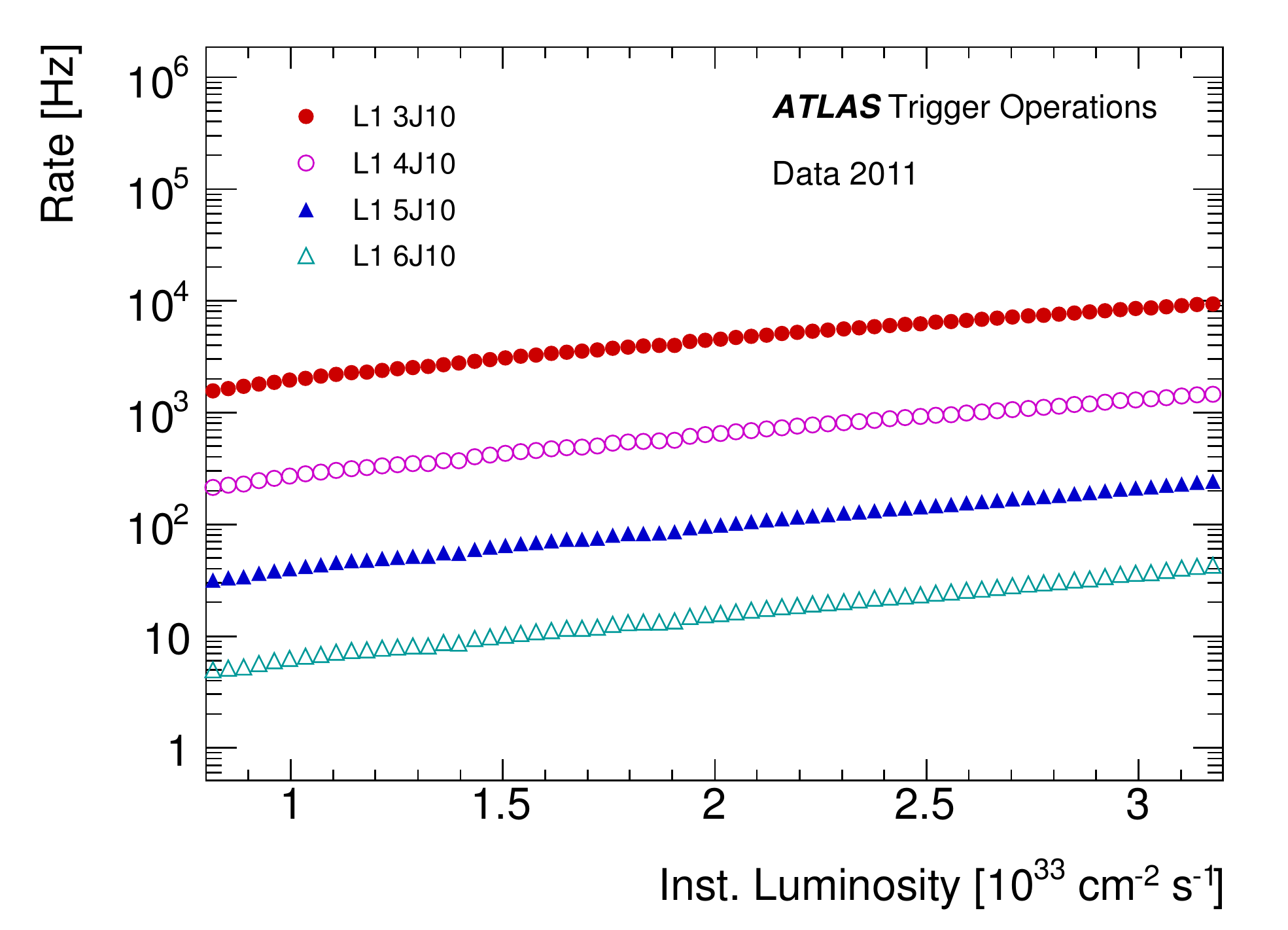}}
  \subfigure[]{
    \includegraphics[width=0.49\textwidth]{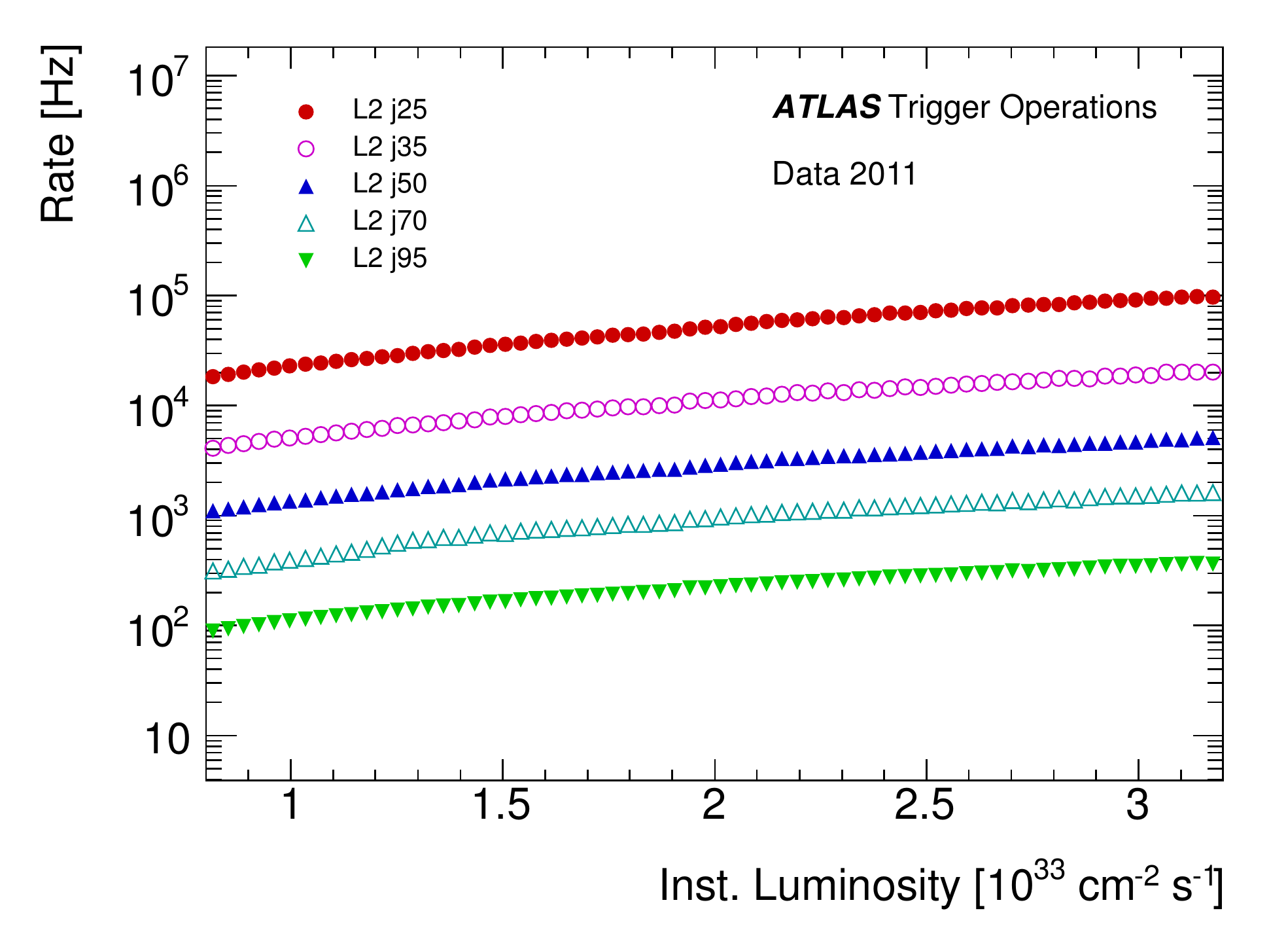}}
  \subfigure[]{
    \includegraphics[width=0.49\textwidth]{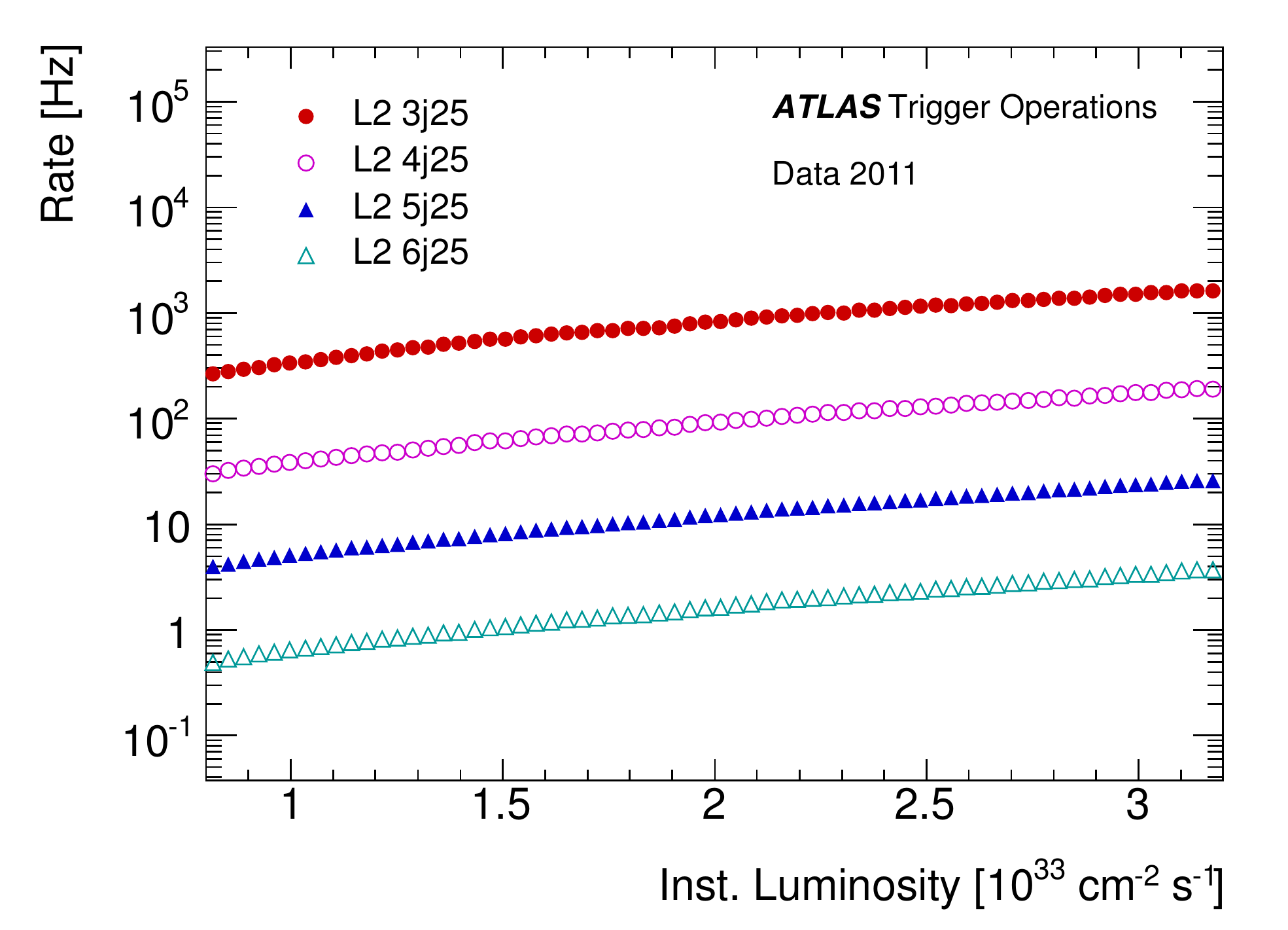}}
  \subfigure[]{
    \includegraphics[width=0.49\textwidth]{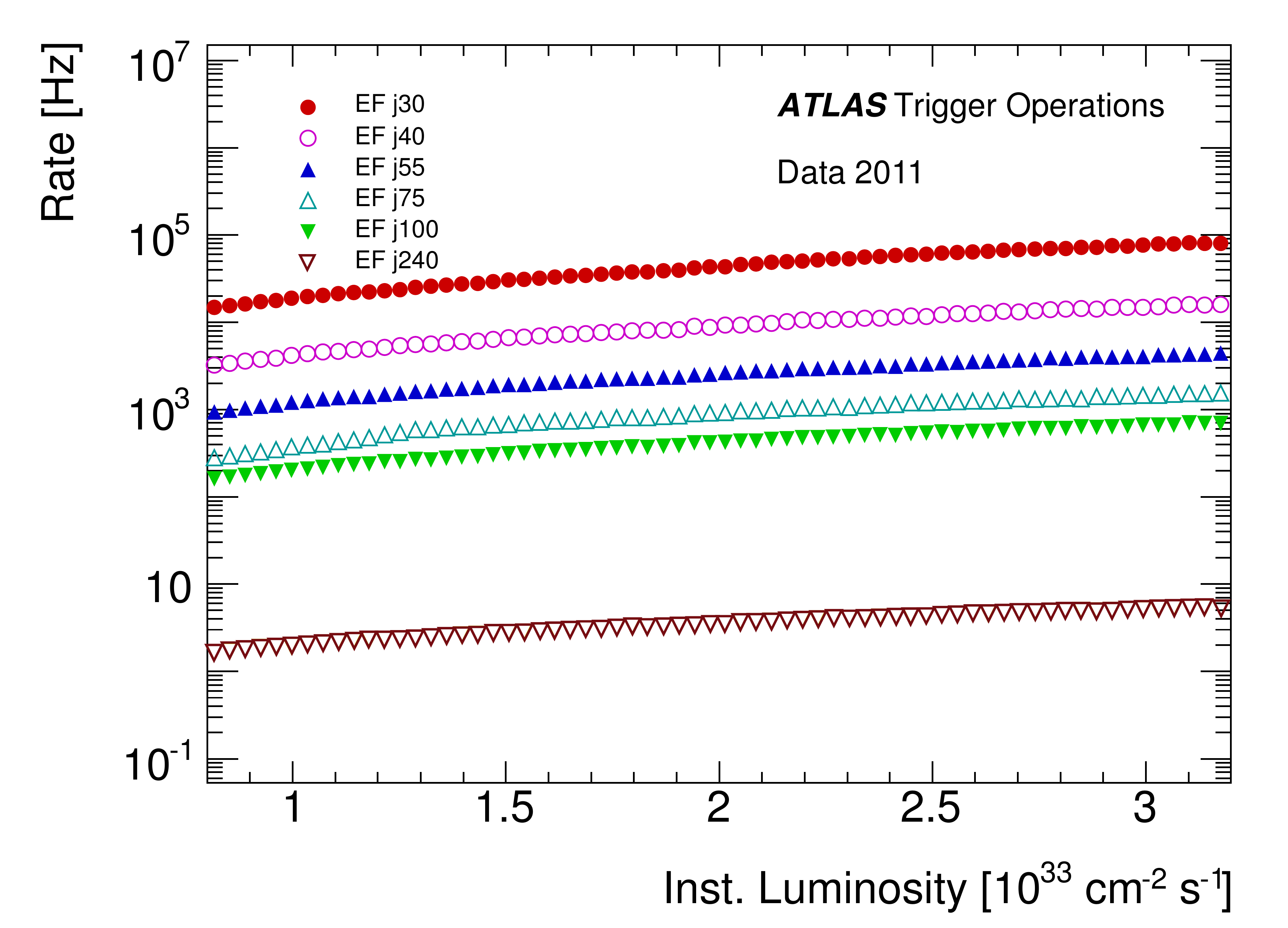}}
  \subfigure[]{
    \includegraphics[width=0.49\textwidth]{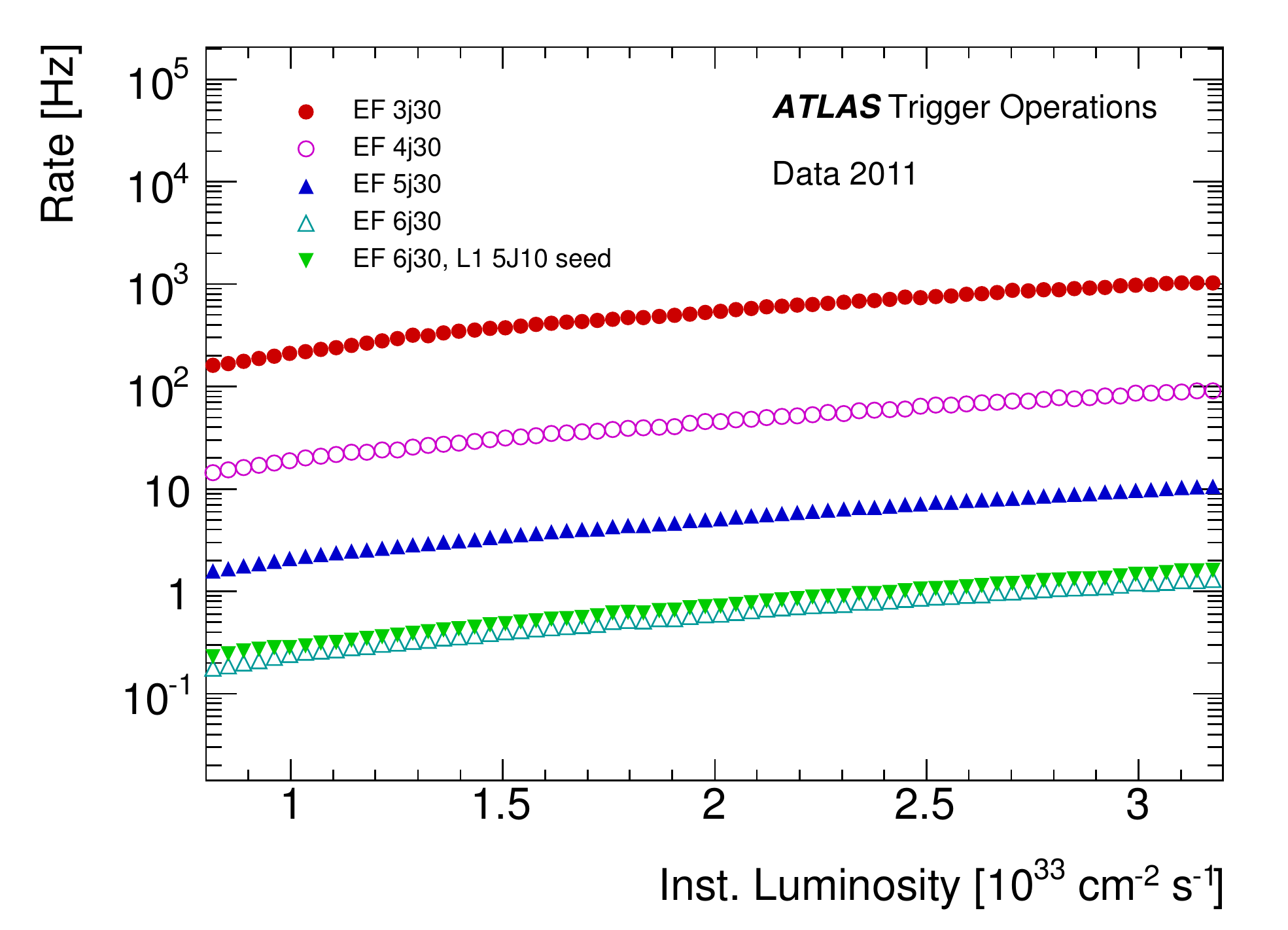}}
  \caption{The jet trigger rates, before application of prescale factors, for triggers operating in 2011: (a, c, e) for
    several single inclusive jet triggers; (b, d, f) for \multijet\ triggers. Shown are the rates for L1, L2 and EF signatures.  }
  \label{fig:inclusive_jet_rates}
\end{figure*}

In order to keep the rate from the jet trigger within the allowed
bandwidth, {\em prescale} factors are used to suppress the rates from
signatures with lower thresholds.  A prescaled trigger selects only a
fraction, 1/{\em prescale}, of events that would otherwise pass the
trigger. 
For the best expected statistical significance, wherever possible, 
triggers intended for searches or analyses requiring the highest possible 
number of data events, should not be prescaled.
As the luminosity 
increased during 2011 data taking, the prescale factors applied 
to the triggers with lower thresholds were increased accordingly, to ensure 
that the output rate remained within the available bandwidth 
for writing to offline storage.  
Figure \ref{fig:inclusive_jet_rates} illustrates the
evolution of jet trigger rates with
instantaneous luminosity
for a selection of single inclusive jet
and \multijet\ triggers operating in 2011 at each of the three trigger 
levels. The rates shown are  before application of any prescales. 
Typical prescale factors for the inclusive jet signatures applied on 
two separate dates during 2011 can be seen in Table~\ref{tab:prescales_for_2011}.

\begin{table*}
\caption{Typical values for the L1 and HLT prescales for the inclusive jet signatures, here 
denoted by the EF signatures, on two dates from different running periods. Also shown is the effective full chain 
prescale obtained by multiplying the L1 and HLT prescales. The three lowest \et\ signatures are seeded
by a random trigger at L1 with the same prescale, but have separate prescales at the HLT to control the rate.
The remaining signatures are seeded by a jet trigger at both L1 and L2.}
\label{tab:prescales_for_2011}
\begin{tabular}{lS[table-format=6.0]S[table-format=5.1]S[table-format=6.0]S[table-format=6.0]S[table-format=4.1]S[table-format=6.0]}
\hline   	       
                 &  \multicolumn{3}{c}{  Apr 28$^{\rm th}$}	&  \multicolumn{3}{c}{  Oct  22$^{\rm nd}$ } \\
Trigger	         &  \multicolumn{1}{c}{L1 prescale}  &     \multicolumn{1}{c}{HLT  prescale} &  \multicolumn{1}{c}{Combined} &   \multicolumn{1}{c}{L1 prescale}  &  \multicolumn{1}{c}{HLT prescale}  &  \multicolumn{1}{c}{Combined}   \\ \hline

\ \\
EF\_j10\_a4tc$^\dagger$   & ~2710 &   ~~60.9 & ~165039   &  ~58600 & ~~18.6 &  ~1089960~~   \\
EF\_j15\_a4tc$^\dagger$   & ~2710 &   ~~12.4 & ~~33604   &  ~58600 & ~~~4.3 &  ~~251980~~   \\
EF\_j20\_a4tc$^\dagger$   & ~2710 &   ~~~~3.8 & ~~10298   &  ~58600 & ~~~1.2 &  ~~~70320~~   \\
\ \\
EF\_j30\_a4tc             & ~7550 &   ~~~~1~~   & ~~~7550   &  ~39300 & ~~~1~~ &  ~~~39300~~   \\
EF\_j40\_a4tc             & ~5080 &   ~~~~1~~   & ~~~5080   &  ~25300 & ~~~1~~ &  ~~~25300~~   \\
EF\_j55\_a4tc             & ~1110 &   ~~~~1.3 & ~~~1443   &  ~~3940 & ~~~1.8 &  ~~~~7092~~   \\
EF\_j75\_a4tc             & ~~404 &   ~~~~1~~   & ~~~~404   &  ~~1910 & ~~~1~~ &  ~~~~1910~~   \\
EF\_j100\_a4tc            & ~~~~~1 &   ~116~~   & ~~~~116   &  ~~~~~~1 & ~529~~ &  ~~~~~529~~   \\
EF\_j135\_a4tc            & ~~~~~1 &   ~~~~3~~   & ~~~~~~~3   &  ~~~~~~1 & ~135~~ &  ~~~~~135~~   \\
EF\_j180\_a4tc            & ~~~~~1 &   ~~~~1~~   & ~~~~~~~1   &  ~~~~~~1 & ~~31.6 &  ~~~~~~31.6   \\
EF\_j240\_a4tc            & ~~~~~1 &   ~~~~1~~   & ~~~~~~~1   &  ~~~~~~1 & ~~~1~~ &  ~~~~~~~1~~   \\
EF\_j320\_a4tc$^\ddagger$ &       &           &           &  ~~~~~~1 & ~~~1~~ &  ~~~~~~~1~~   \\ 
EF\_j425\_a4tc$^\ddagger$ &       &           &           &  ~~~~~~1 & ~~~1~~ &  ~~~~~~~1~~   \\ 
\\ \hline  
\end{tabular}\\
$^\dagger$ Randomly seeded at L1, passthrough at L2\\
$^\ddagger$ Not active during early running
\end{table*}

\begin{table*}[htp]
    \caption{
     \label{tab:thresholds_for_2011}
     The evolution of the lowest \ET, unprescaled EF threshold for
     single-jet triggers during 2011 data taking. }
    \begin{center}
      \begin{tabular}{cc}
        \hline
         \begin{tabular}{l}
         Instantaneous Luminosity  \\ $[10^{33} \textrm{cm}^{-2}\textrm{s}^{-1}]$ \end{tabular} &
         \begin{tabular}{l}
           Lowest unprescaled trigger \\ \ET\ threshold [\GeV] 
           \end{tabular} \\
        \hline
 0~~~~~ \ -- \ 0.16 & 100 \\
 0.16 \ -- \ 0.25 &  135\\
 0.25 \ -- \ 1.1~~ &  180\\
 1.1~~ \ -- \ 3.6~~ &  240\\
        \hline
      \end{tabular}
    \end{center}
    \vspace{-5mm}
\end{table*}

In addition to applying prescale factors to low-threshold triggers,
the EF \ET\ threshold of the lowest-\ET\ unprescaled single inclusive
jet trigger was raised on three occasions to accommodate the
increasing instantaneous luminosity.  The evolution of the minimum
unprescaled EF threshold is detailed in
Table~\ref{tab:thresholds_for_2011} and effectively determines the
lowest trigger threshold which can be used in several physics
analyses.  
Technical improvements were implemented to improve trigger
rejection and cope with the increasing luminosity and varying LHC 
conditions.  From May 2011, 
calorimeter noise suppression and
\pileup\ corrections were applied in the L2 calorimeter data
preparation in order to reduce sudden increases in the trigger rate
due to bad detector conditions, as well as maintaining performance
under higher \pileup conditions.


\section{Timing}
\label{section:timing}

As a hardware system, the L1 trigger operates with a fixed latency,
whereas the L2 and EF systems operate with a variable processing time, and
must complete their respective processing within the constraints
provided by the L1 rate, the rate at which events can be recorded
offline, and the number of available CPU nodes in each HLT farm.  In this
section, the time taken to process events for the L1 system and the
HLT is discussed.

\subsection{Level 1}

The L1 jet trigger is a fixed latency, hardware based trigger operating 
synchronously with the LHC bunch clock and the rest of the L1 system. 
The  pipelines in the detector front-end electronics are
typically 120 bunch crossings deep and as such the latency from the 
complete L1 processing must fit within the corresponding time.
Throughout the L1 system each step is handled in
parallel with other steps.  Data transfers between parts of the system
are performed concurrently with the processing of the data that has
already been transferred. The analogue data are digitised and sent as
input to a jet algorithm, and the final decision is sent from the L1
calorimeter system to the Central Trigger Processor (CTP).  The jet
algorithm processing itself is very fast and takes only approximately
50~ns, but represents only part of the processing necessary to reconstruct
jet candidates, the rest being in formatting the input and output data such that
the algorithm can execute quickly. The overall time for all these
stages including the transmission of the results of the calorimeter
trigger reconstruction to the CTP is approximately 1.5\,\mus.  The
additional time required for the subsequent CTP processing to
determine the global L1 decision, and the time taken for transmission
of this decision back to the detector front-end is approximately
0.5\,\mus\ so that the full latency of the entire L1 system is within
the required  maximum 2.5\,\mus.

\subsection{High level trigger}

\begin{figure}[thp]
  \subfigure[]{\includegraphics[width=0.49\textwidth]{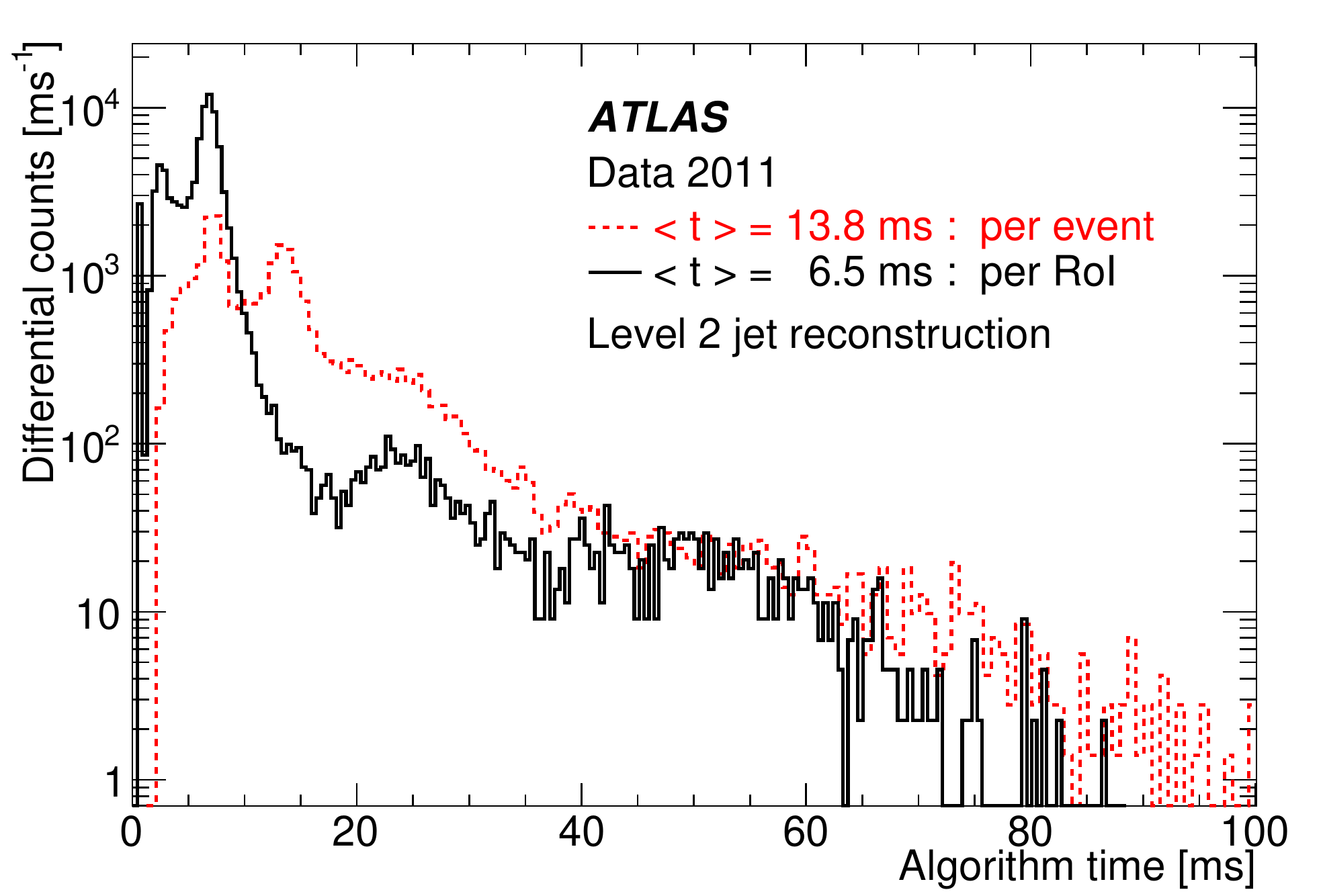}\label{figs:timingleveltwoa}}
  \subfigure[]{\includegraphics[width=0.49\textwidth]{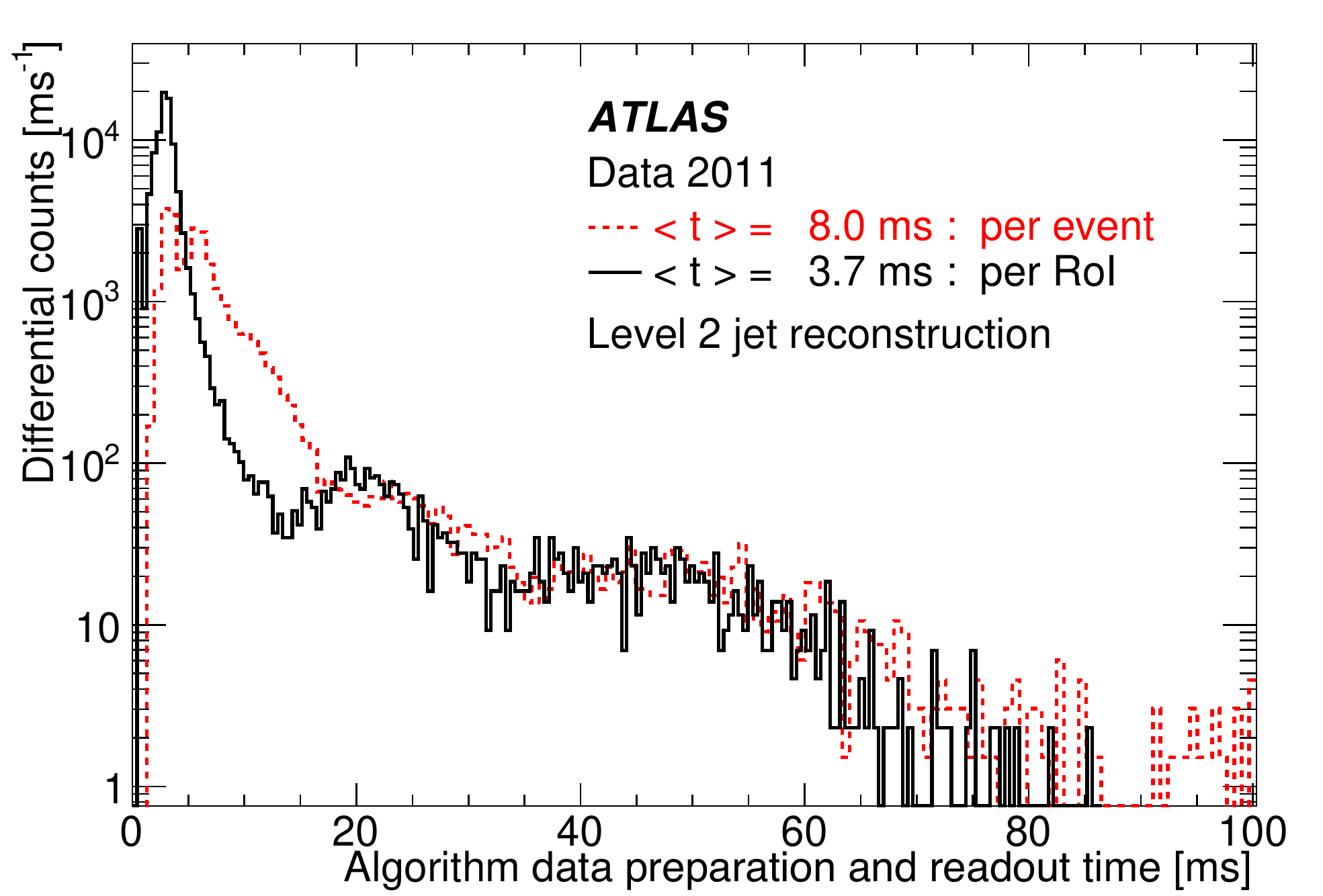}\label{figs:timingleveltwob}}
  \subfigure[]{\includegraphics[width=0.49\textwidth]{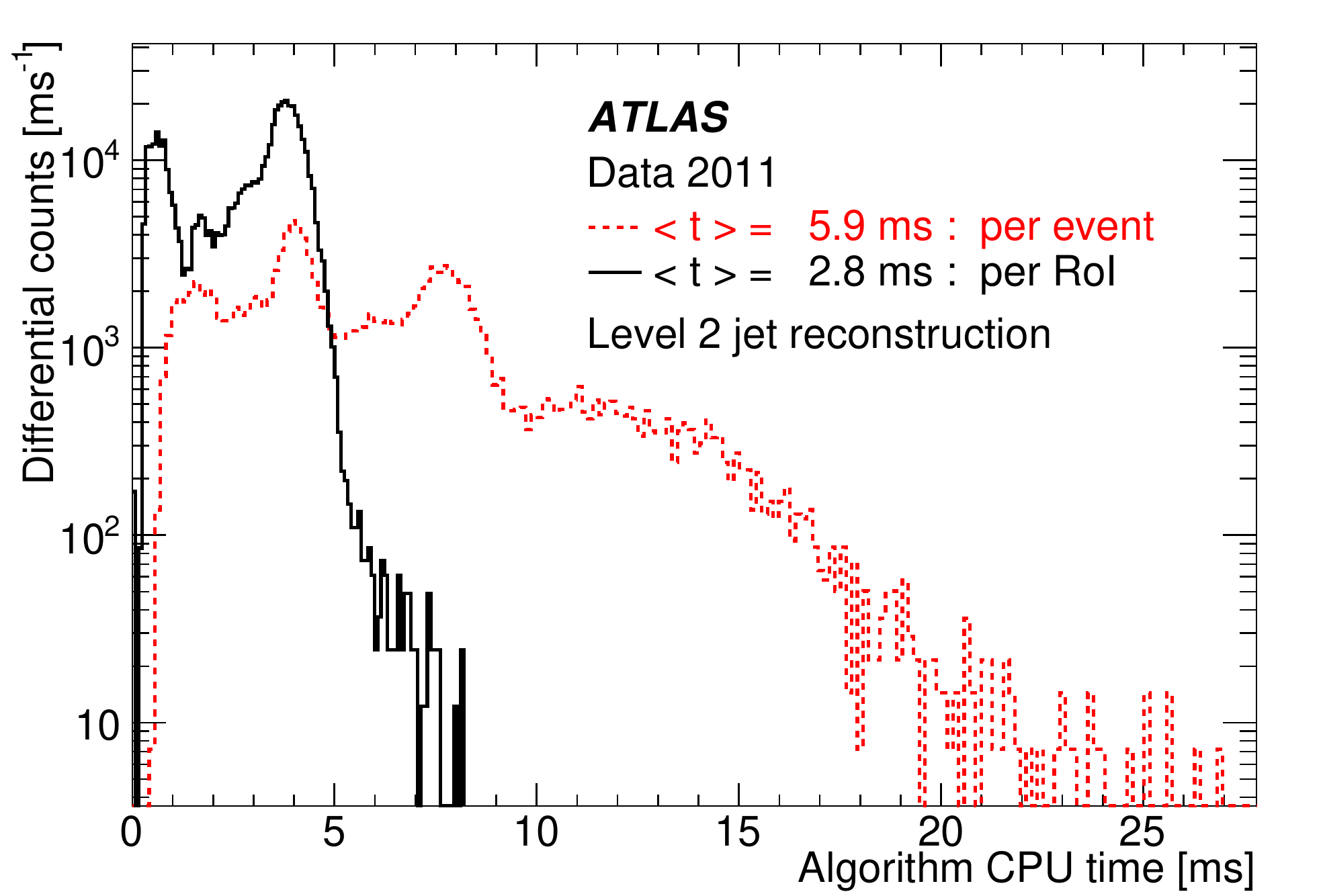}\label{figs:timingleveltwoc}}
  \caption{The processing time for the L2 jet trigger: (a)~the full
    algorithm time; (b)~the data preparation
    time; (c)~the algorithm processing CPU time. The full algorithm time includes 
    both the data preparation and algorithm processing times. The solid
    lines show the processing time per call, and the dashed lines
    show the processing times per event. }
  \label{figs:timingleveltwo}
\end{figure}

In this section, timing distributions are presented for a physics run
taken during October 2011. During this run, the peak
instantaneous luminosity was 3.5$\times 10^{33} $\,cm$^{-2}$s$^{-1}$
with a mean of 17 interactions per bunch crossing at the start of the
fill.  The total L2 processing time is shown in
Figure~\ref{figs:timingleveltwoa}. This includes the data
preparation time for the extraction of the data from the readout
buffers, shown in Figure~\ref{figs:timingleveltwob}, and the
algorithmic CPU time, shown in Figure~\ref{figs:timingleveltwoc}.
Since the L2 algorithm executes on a per RoI basis, the time per event
is determined by the time for processing a single RoI and the number
of RoIs in the event.  The full algorithm time for a single RoI has a
mean of 6.5\,ms and a long tail extending to approximately 
80\,ms, corresponding to an algorithm processing mean time of 
2.8\,ms and a combined data preparation and readout time with a mean of 3.7\,ms which also provides the 
long tail. 
\begin{figure*}[th]
  \subfigure[]{\includegraphics[width=0.49\textwidth]{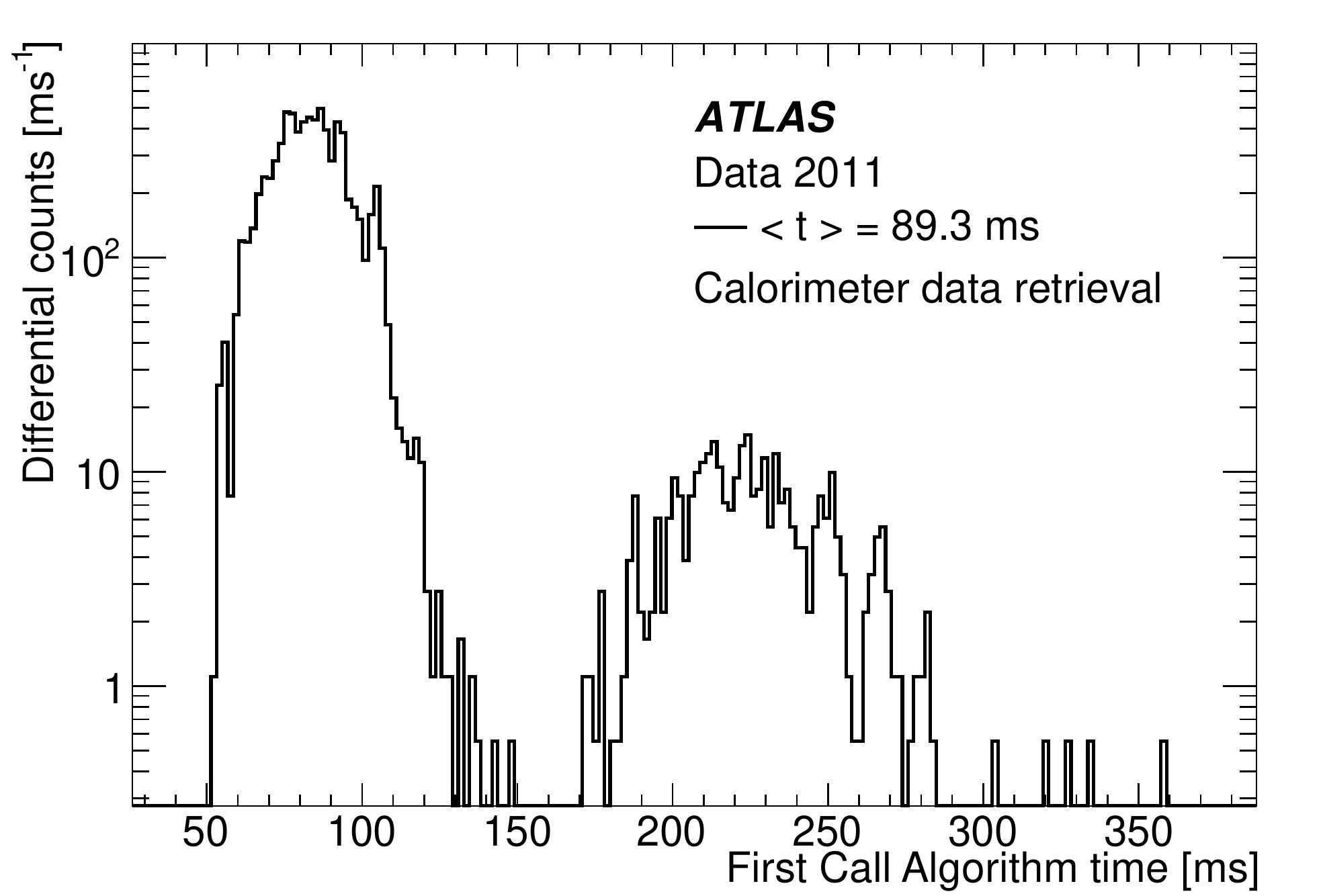}\label{figs:timingEFa}}
  \subfigure[]{\includegraphics[width=0.49\textwidth]{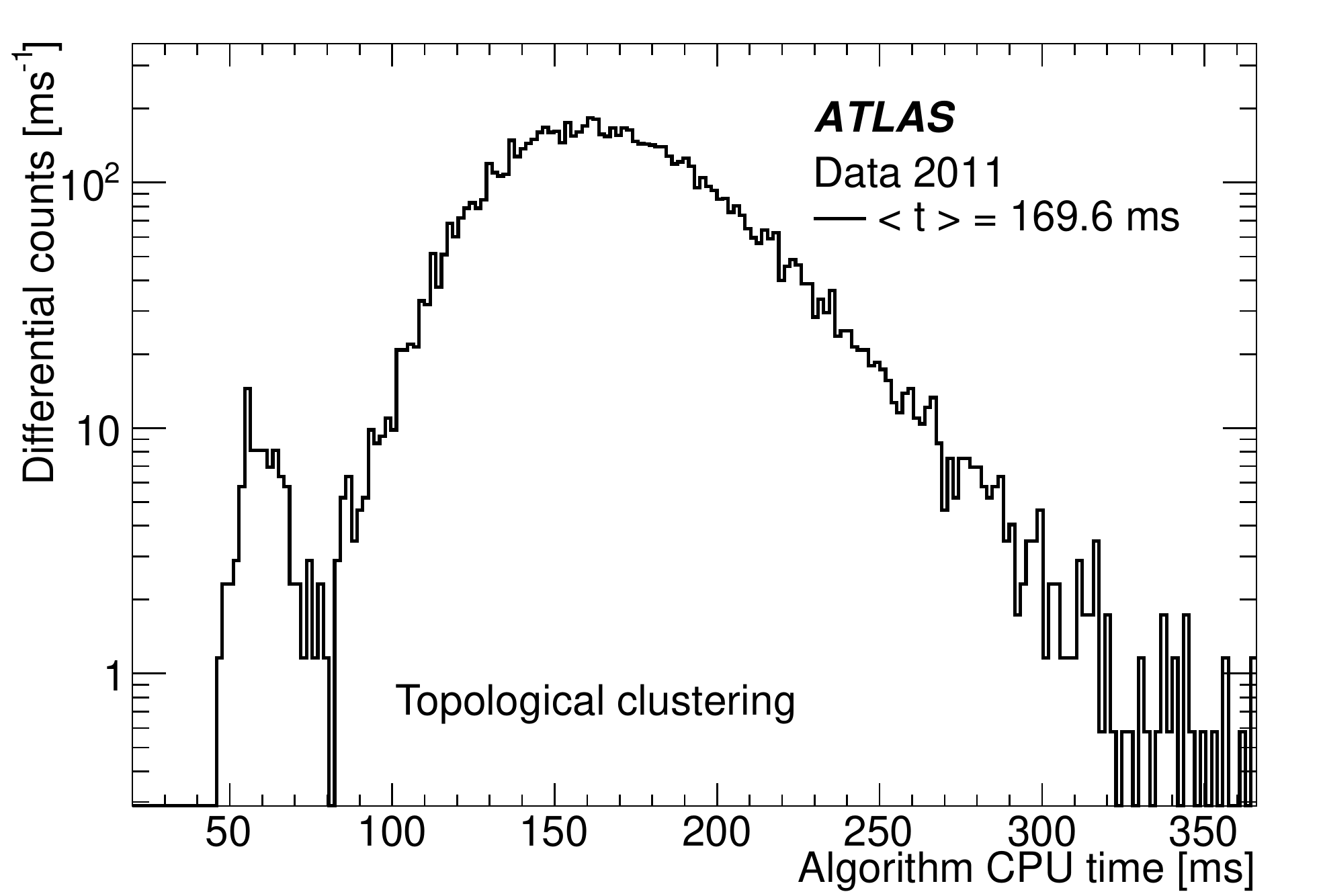}\label{figs:timingEFb}}
  \subfigure[]{\includegraphics[width=0.49\textwidth]{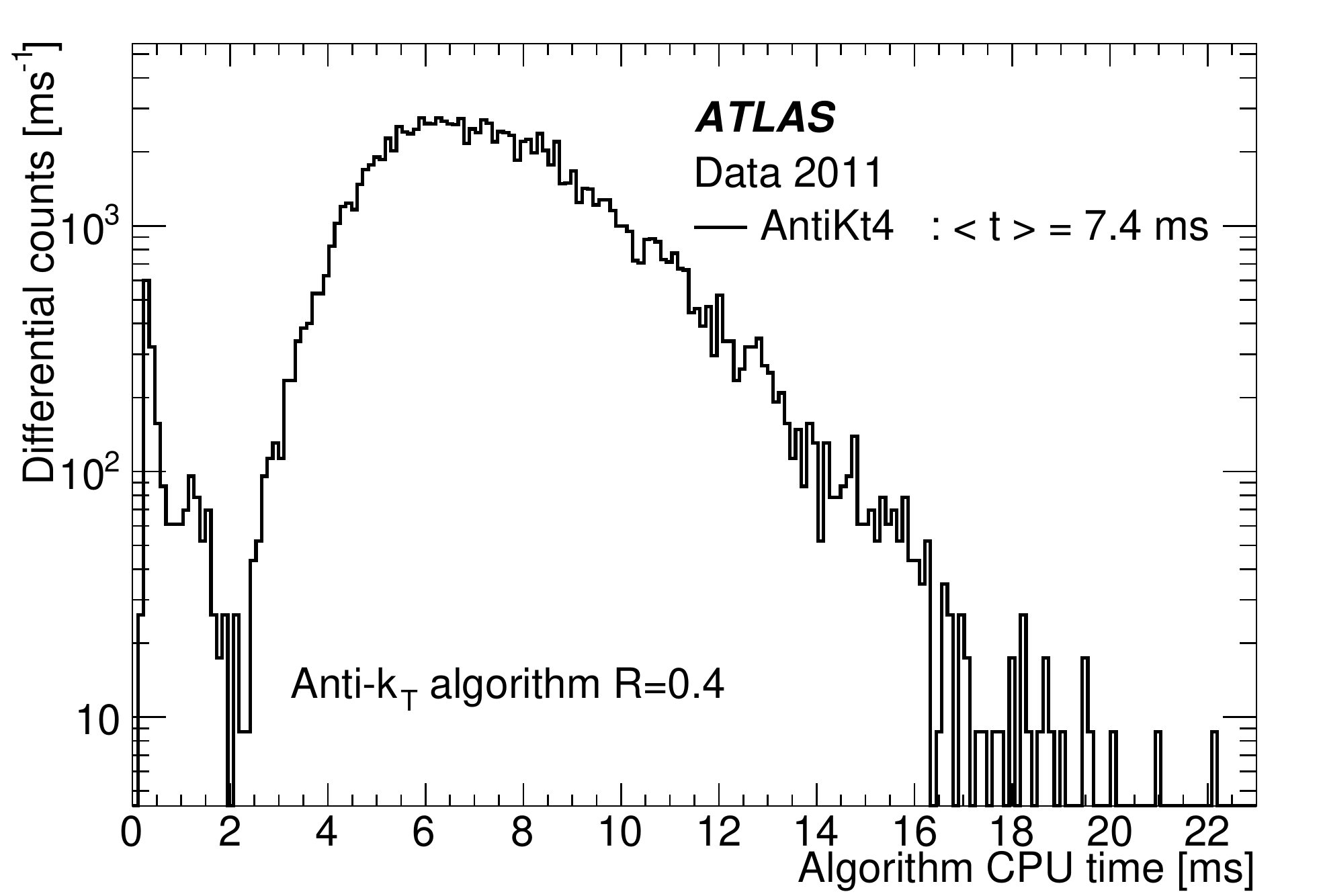}\label{figs:timingEFc}}
  \subfigure[]{\includegraphics[width=0.49\textwidth]{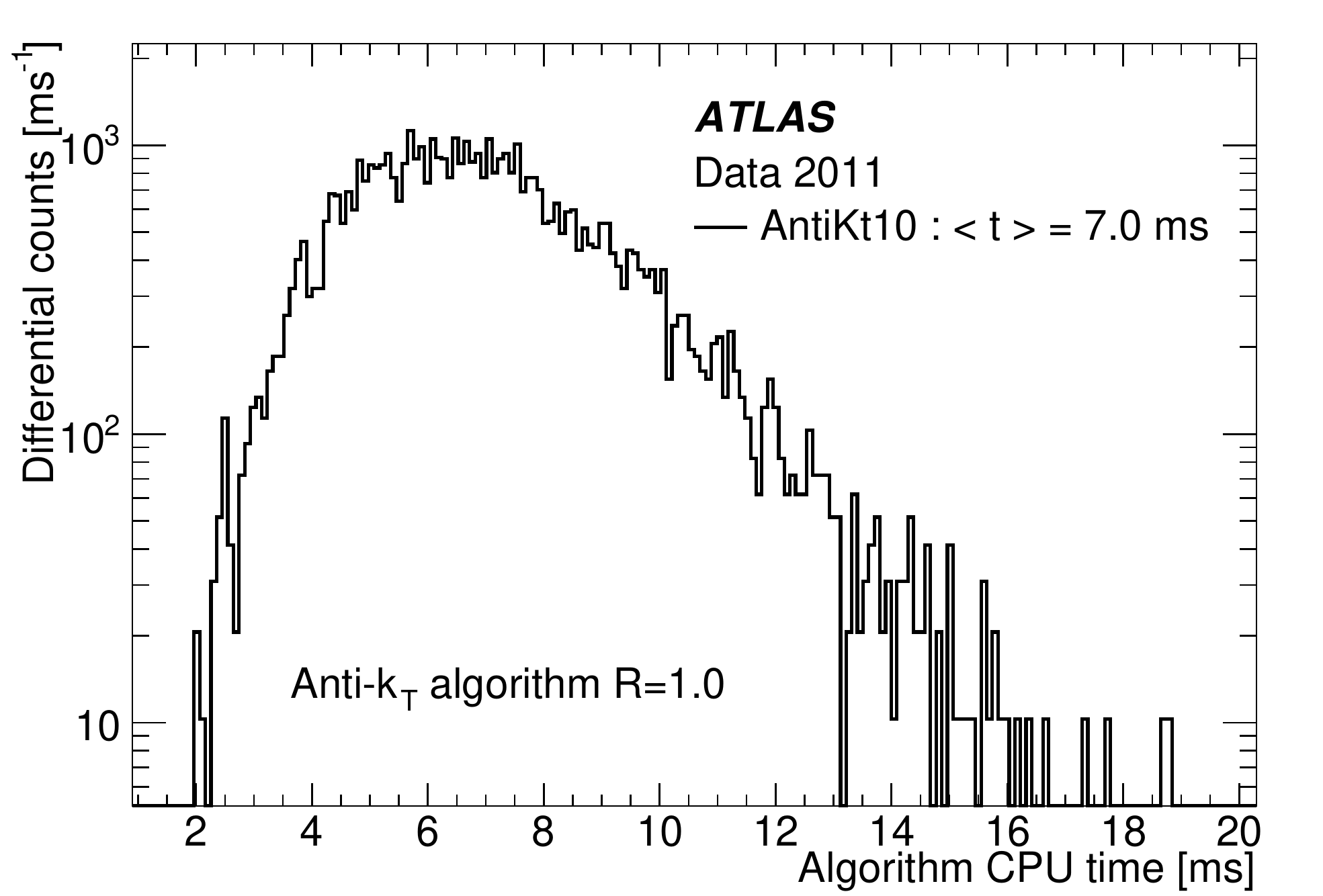}\label{figs:timingEFd}}
  \caption{The processing times for the Event Filter jet trigger:
    (a)~the time for the data preparation for the full calorimeter
    data; (b)~the processing time for the topological clustering;
    ~(c, d) the times for the actual jet finding for the \antikt\ algorithm,
    for instances with radius parameters (c)~$R=0.4$, and
    (d)~$R=1.0$.}
  \label{figs:timingEF}
\end{figure*}
The execution time of the algorithm
per event, rather than per RoI shows a clear peak around 6\,ms and a
second peak around 12\,ms, due to events containing two RoIs. The data
preparation time for the full event has a peak at around 3\,ms for
single RoI events and another from the two-RoI events around 6\,ms. The
algorithm CPU time has corresponding peaks around 4\,ms and
8\,ms.

At the EF, following the data preparation steps -- the retrieval
of calorimeter cell information from memory and building of  topological clusters -- 
a single instance of the \antikt\ jet
finder is executed for each of the required jet radii, $R=0.4$ and
$R=1.0$.  The times for each of the data preparation stages and the
two jet radii are shown in Figure~\ref{figs:timingEF}, for the same
2011 run.  For the data retrieval stage, before the topological
clustering, two distinct peaks are observed.  The first, with a mean
around 80\,ms, represents the processing of the complete event.  The
second broader peak, with a mean of approximately 220\,ms, is due to
an artefact of the trigger processing by the EF farm: each CPU in the
EF farm runs a separate instance of the EF software and performs some
additional initialisation for the first event each receives,
increasing the processing time for the first event processed by each
CPU node.  The number of events in this second peak then corresponds
to the number of individual CPUs in the farm.  The most time consuming
part of the full EF processing is the topological clustering, with a
mean of approximately 170\,ms. The jet finding itself is comparatively
fast, requiring approximately 7\,ms per instance. Due to the different
prescales and thresholds used for the triggers for the different jet
radii, the event topology and \et\ spectrum is slightly different 
for the events processed by each instance of the \antikt\ algorithm,
resulting in the slightly different distributions seen in
Figures~\ref{figs:timingEFc} and~\ref{figs:timingEFd}. The peaks
seen at short times in the topological clustering and $R=0.4$
\antikt\ jet finding are due to the low threshold EF triggers seeded
by the random trigger at L1 which therefore may contain fewer calorimeter 
cells with significant energy and so do not take as long to
process. The \antikt\ jet finding using $R=1.0$ only processes events
seeded by jets found both at L1 and L2, where these jets pass the
95\,\GeV\ L2 threshold, so this peak is largely absent in
Figure~\ref{figs:timingEFd}.

After the jet finding has completed,  
the selection hypothesis algorithms are executed
both at L2 and the EF. For the single inclusive and \multijet\ triggers
the hypothesis algorithm typically executes in approximately 
10~\mus\ for each signature.

\section{Comparison of trigger and offline performance}
\label{section:performance}

\newcommand{\fullerror}{Statistical uncertainties only are shown: the data are shown as the solid points with error bars, and
the \herwig\ simulated sample as the hatched band.}

\newcommand{\statonly}{Statistical uncertainties only are shown.}

An important concern for the trigger reconstruction is the producion of objects 
resembling as closely as possible those later reconstructed offline,
to allow informed event selection with high efficiency while minimising 
any increase in the trigger rate. This is achieved  
by reducing any finite trigger--offline resolution or bias so that any 
selection of objects on the basis of trigger quantities more closely 
corresponds to the offline selection used for physics analyses.  
For this reason,
the performance of the jet trigger during 2011 data
taking has been evaluated with respect to the offline jet
reconstruction.  The offline reference jets have been reconstructed
using the infrared and collinear safe
\antikt\ algorithm~\cite{antikt} implemented in the
FASTJET~\cite{fastjet} package. The same values of the radius 
parameter are used offline: $R = 0.4$ for the standard analyses, and
$R=1.0$ used for the analysis of boosted objects.

The trigger performance is defined in terms of specific metrics,
comparing offline and trigger reconstructed jets, such as jet
selection efficiency, and the transverse energy or angular resolution
with respect to offline jets. Comparisons of the same metrics with
Monte Carlo simulated samples are shown in this and the following
sections to illustrate how well the simulation describes the data and
where disagreements appear. It should be emphasised, however, that the
focus is on performance indicators determined from collision data, and
detailed comparisons of different simulation configurations are beyond
the scope of this paper.

\subsection{Data samples and event selection}
\label{section:dataset}

It is informative to evaluate the performance of the trigger in
simulated events and compare it to the real trigger running in collision
data. The ATLAS trigger simulation runs exactly the same code for the HLT 
as that run online, and a very precise emulation for L1. The
differences observed between collision data and simulation are due either to
differences in the underlying physics, such as the composition and
internal topology of the jets themselves, or to the  kinematics,
hadronisation, treatment of underlying event, or to differences in
the simulation of the detector response or the detector
conditions.

Because of these potential sources of differences between data and
simulation, for jet physics analyses, trigger selection and trigger
related calibrations are generally obtained using the data rather than
relying on the trigger performance from the Monte Carlo
simulation. Therefore, while it is desirable for the simulation to
accurately reproduce the behaviour of the trigger, it is by no means
essential.

For the evaluation of the trigger performance, events are selected
from those written offline that are free from known problems with the
detector or beam conditions. From these events, offline reconstructed
jets which satisfy standard ATLAS jet selection criteria used in physics
analyses~\cite{atlasjets1,atlasjets2,atlasjets3} are selected to
provide a reference jet sample.  Besides the kinematic selection, 
these criteria also include jet-quality 
selections~\cite{atlasjets7,atlasjetquality,atlasjetquality2}
to reject fake jets 
reconstructed from non-collision signals, such as beam-related background, 
cosmic rays or detector noise. Similar jet quality criteria are applied 
online to the trigger jets.

The efficiency for each specific chain has been evaluated using events 
selected by an alternative chain which is unbiased by the selection of 
the chain being evaluated. 
Therefore, wherever possible, the efficiencies 
have been evaluated using trigger
chains seeded by a random trigger at L1, passing through L2 and EF
without additional trigger selection. Where this was not possible, the
standard chains have been used, but selecting only those 
{\em pass through} events, where the trigger accepted the event
irrespective of the trigger jet selection, as discussed in
Section~\ref{section:menu}. 

There are a number of general purpose event generators for LHC
physics: For more complete review, see elsewhere~\cite{MCforLHC}.  In
the following studies, data are compared with simulated events produced
using either the \herwig~\cite{herwig} or 
\pythia~\cite{pythia6} Monte Carlo generators. Each simulates
complete physics events using a hard subprocess with a leading
logarithmic parton shower 
followed by a soft
hadronisation model to generate the outgoing hadrons. 
Both include models for the underlying event: In
\herwig, the formation of hadrons from the final state quarks and gluons produced
in the parton shower is described using a cluster hadronisation
model~\cite{clusterHad}, whereas the \pythia\ generator uses the Lund
string fragmentation model~\cite{lundstring,lundmodel}.

In the following discussion the central and the forward jets triggers
are discussed separately.
For central jet triggers in the range \mbox{$|\eta_{\mathrm {jet}}|<3.2$},
offline jets are required to lie in the range $|\eta_{\mathrm {jet}}|<2.8$
in order to completely contain jets with radius parameter $0.4$.
Similarly, for the forward jet triggers, which lie in the range
\mbox{$3.2<|\eta_{\mathrm{jet}}|<4.9$}, offline jets satisfying
$3.6<|\eta_{\mathrm {jet}}|<4.5$ are required.

For offline analyses, jets are corrected for the difference between
electromagnetic and hadronic responses in the calorimeter. Therefore
jets can be defined either at the electromagnetic (EM) scale, which 
correctly measures the energy deposited by electromagnetic showers 
in the calorimeter, 
or after the full hadronic jet energy scale (JES) calibration~\cite{emscale1,atlasjetquality2}.
In the trigger,
the JES calibration was not applied in 2011 since the full calibration
was not available during data taking. The standard calibration of the
reference offline jets is referred to as EM+JES, meaning jets built from
(electromagnetic-scale) topological clusters, with jet energy
corrected by the application of the JES calibration.

\subsection{Jet trigger performance metrics} 
\label{section:metrics}

Descriptions of the metrics used to assess the jet trigger 
performance can be found in this section: specifically for the efficiency 
measurement, and the evaluation of the resolution and bias arising from  
any offset between the trigger and offline reconstructed quantities.

\subsubsection{Efficiency definition}

Unless otherwise stated, the inclusive single jet efficiencies
presented in this paper are of the form of {\em per jet} efficiencies
with respect to the corresponding jet reconstructed offline.  
This represents the probability that an offline jet will have a corresponding 
jet reconstructed in the trigger that satisfies the trigger selection.
Efficiencies {\em per event} can also be defined, in terms of global
event properties, such as the \et\ of the leading jet in the
event. These are more sensitive to the event topology and more
difficult to interpret, since, for example, any other jet might cause
the event to be accepted, even if the leading offline jet does
not. For a multi-jet trigger however, where the selection depends on
the properties of many jets, these {\em per event} selections may be
very informative; this is discussed further in
Section~\ref{section:multijet_efficiencies}.

The jet reconstruction efficiency, $\epsilon$, for a sample of jets
can be defined as the ratio of the number of offline jets, $N$,
passing some selection which defines the sample, and the number of
those jets, $m$, which are also reconstructed in the trigger to within
some appropriate matching criteria, such that $\epsilon \equiv m/N$.

\begin{figure}[th!]
  \subfigure[]{\includegraphics[width=0.48\textwidth]{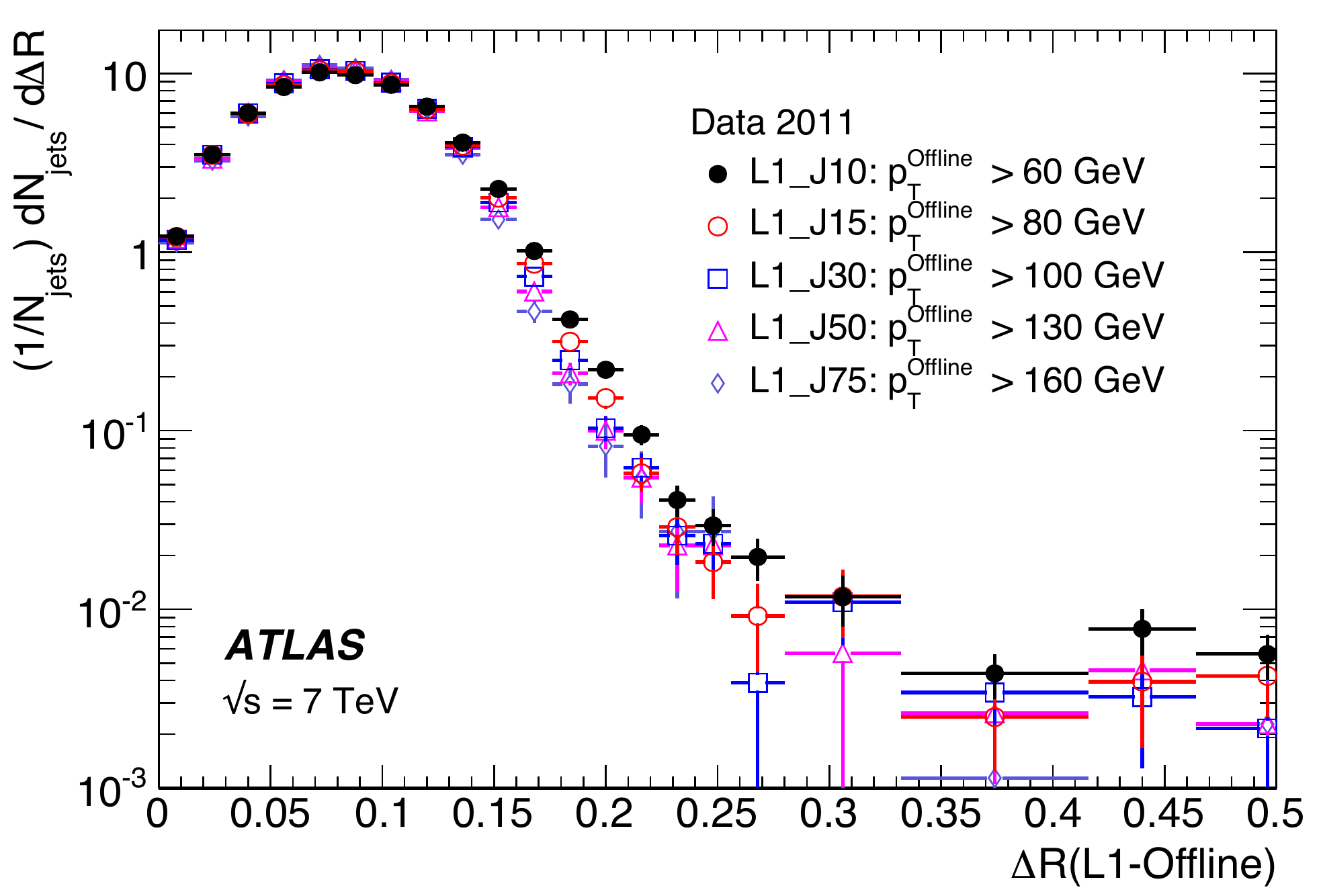}\label{fig:deltaRa}}
  \subfigure[]{\includegraphics[width=0.48\textwidth]{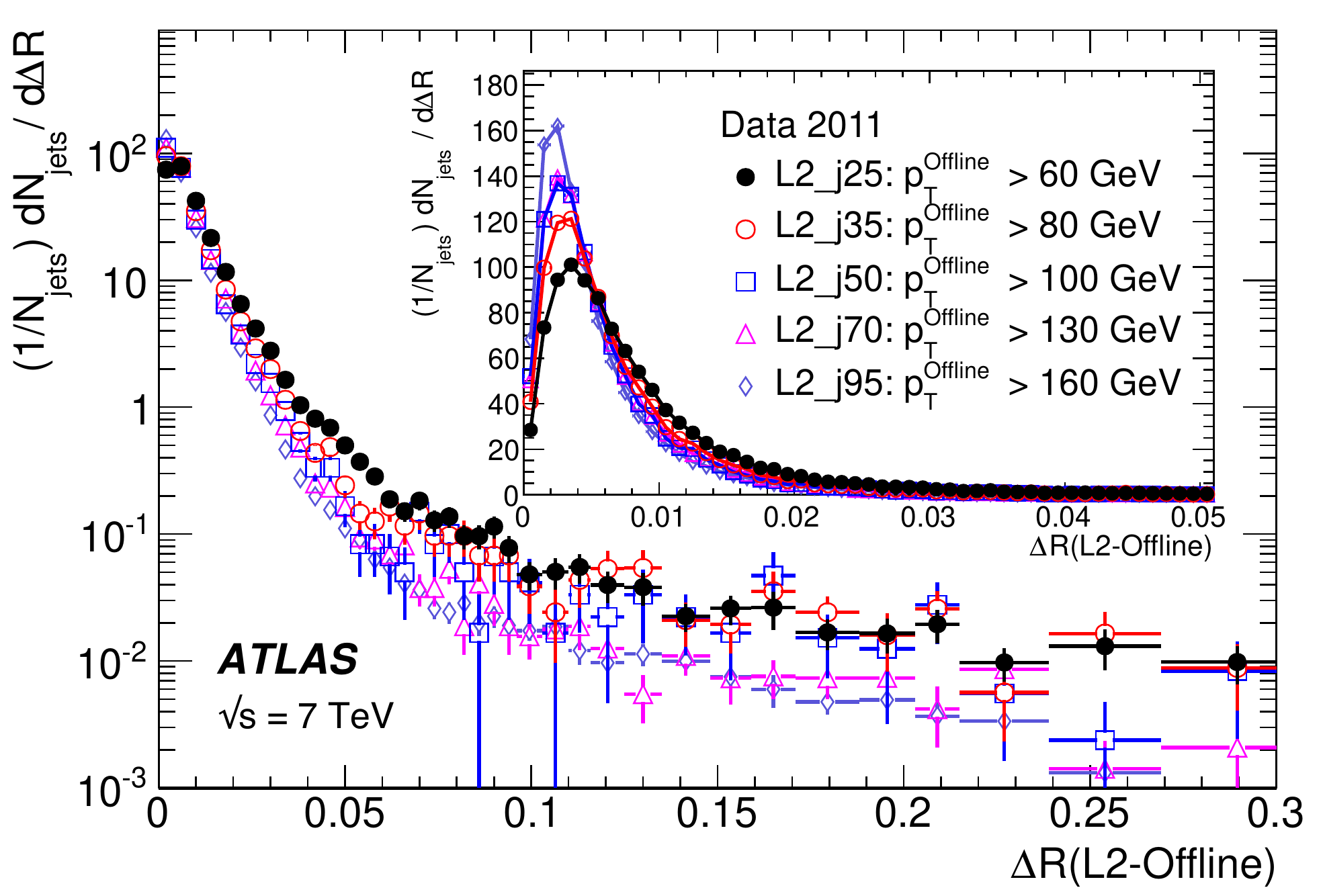}\label{fig:deltaRb}}
  \subfigure[]{\includegraphics[width=0.48\textwidth]{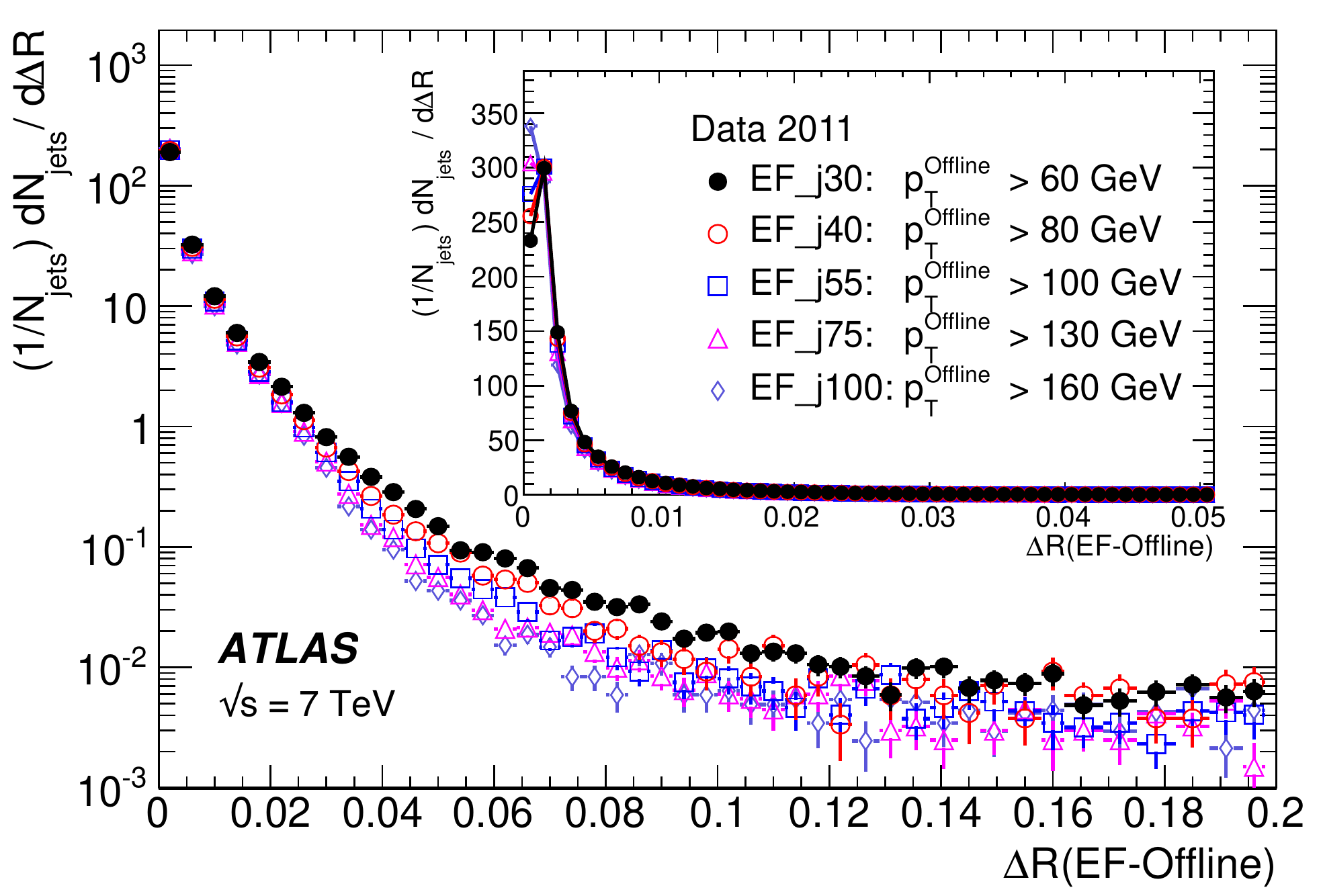}\label{fig:deltaRc}}
  \caption{ The distribution of \deltaR\ between the offline jets and the
    closest matching trigger jet: (a) for L1; (b) for L2; (c) for the EF.
    In each case the differences are shown with respect to offline
    jets above the \pt\ thresholds indicated such that the trigger for
    each threshold is fully efficient. Statistical uncertainties only 
    are shown.}
  \label{fig:deltaR}
\end{figure}

The choice of matching criteria must be considered as an important aspect of the
definition of the efficiency, since a tighter matching will
necessarily result in a lower efficiency and {\em vice versa}.  This
is important since the correspondence of offline jets to trigger jets
is not one-to-one.

The binned differential efficiency, $\epsilon_i$, in some  generic variable
$x_{\mathrm {jet}}$, where $x_i \leq x_{\mathrm{jet}} < ( x_i+\Delta x )$, is
defined analogously,
\begin{equation}
 \epsilon_i = \frac{m(x_i \leq x_{\mathrm{jet}} < (x_i+\Delta x) )}{N(x_i \leq x_{\mathrm{jet}} < ( x_i+\Delta x) ) }.  
\label{eqn:object_trigger_efficiency}
\end{equation}
The criterion applied here for matching online and
offline jets is based on the distance $\Delta R = \sqrt{
  (\Delta\eta)^2+(\Delta\phi)^2}$ in the $\eta$ -- $\phi$ plane between
the offline jet and the closest matching trigger jet.

Figure~\ref{fig:deltaR} shows the distribution of \delR\ for trigger
jets from L1, L2 and the EF. Distributions are shown for each trigger
for several different \pt\ ranges such that the trigger in each case
is fully efficient.  For the L2 and EF \delR\ distributions, the
agreement between online and offline clearly improves with increasing
jet \et.  Since the L1 trigger uses only coarse granularity
calorimeter information, and quantises the $\eta$ and $\phi$
directions to the nearest 0.2, the resolution in $\eta$ or $\phi$ from
L1 would be expected to be approximately 0.06.  
For similar Gaussian distributed residuals in $\eta$
and $\phi$ this would result in a maximum in the \delR\ distribution
of around 0.08, as observed in Figure~\ref{fig:deltaRa}. Although
the L2 trigger operates only within each RoI, it uses calorimeter information
at the full detector granularity. Therefore the jet $\eta$ and $\phi$
reconstruction in Figure~\ref{fig:deltaRb} for L2 is significantly improved 
with respect to that seen in Figure~\ref{fig:deltaRa} for L1. 
The EF
uses the same topological clustering algorithm and the same jet
finding algorithm as the offline reconstruction. This leads to a
further improvement in the resolution between the trigger and offline jets
for the EF with respect to what is already acheived at L2, and can be
seen in Figure~\ref{fig:deltaRc}.

For the matching used to define the resolution and efficiency, 
criteria in \delR\ have 
been chosen which allow high efficiency for genuine matches while 
reducing the contribution from random matches that may 
degrade the resolution.
For the analyses of the efficiency and resolution for single jets
presented here, trigger jets are required to match with the closest
offline jet to within $\delR<0.4$ at L1, and to within $\delR<0.2$ for
L2 and the EF.

\subsubsection{Trigger efficiency behaviour near threshold}

The trigger system selects jets based largely on the \et\ and
\pseudorapidity\ of the jets reconstructed at the three trigger
levels. The principle source of difference between the \et\ of offline and trigger
jets in 2011 was the hadronic calibration, which was not applied
online in this period.  Smaller differences at the different levels arise
from the detector granularity at L1, the input objects to the jet
algorithms and the L2 algorithms.  These differences give rise to
shifts and additional resolution smearing of the \ET\ reconstructed in
the trigger with respect to that reconstructed offline.  The selection
efficiencies for the various trigger levels resulting from these
shifts and smearing are therefore not step functions when measured as
a function of the offline \ET.  Instead, the
efficiency as a function of \et\ will exhibit a more
slowly rising edge as the trigger turns on.  This has an impact on
ATLAS physics analyses; in general, a steeply rising efficiency near
the \ET\ threshold which rapidly approaches a plateau near 100\%
efficiency indicates good performance of the trigger.  A more slowly
rising efficiency, or one which does not saturate near 100\% can be
problematic for offline data analysis as it has the potential to
introduce large systematic uncertainties in the selection efficiency
and background estimates.

A more slowly rising edge is expected at L1 due to the poorer
\ET\ resolution, arising from the coarse granularity data. Care must
therefore be taken to ensure that the L1 efficiency reaches its
plateau for \ET\ values below the onset of the rising edges of the L2
and EF efficiency curves. The thresholds for the higher trigger levels 
therefore impose upper limits on the corresponding L1 thresholds, 
and so reduce the efficacy of raising
these thresholds in order to reduce the rate of L1 accepted
events. Because of the steeply falling \pt\ spectrum this implies that
more events need to be accepted at L1 to avoid reducing the EF
efficiency significantly at higher \ET.  A more steeply rising efficiency
is expected at the EF due to the improved \ET\ resolution and the greater similarity
in the reconstruction algorithms used online and offline.  To minimise
systematic uncertainties associated with the trigger, most ATLAS
physics analyses relying on jet triggers require that the offline
\ET\ for selected jets lie in the {\em efficiency plateau} region,
where the efficiency is above 99\%.

\subsubsection{Definition of transverse energy resolutions and offsets}

The transverse energy resolutions and offsets are computed from the 
distributions of the residuals between the quantities computed offline 
and at trigger level.

To provide a single statistic to parameterise the resolution, the
root-mean-square (RMS) deviation of the central 95\% of the residual
distribution is used.  This is further divided by the RMS for the
central 95\% of a Gaussian distribution with unit standard deviation. In this way,
if the distribution were Gaussian, the normal Gaussian resolution
would be obtained.  This measure was chosen because the RMS of the
full distribution can be strongly biased by significant non-Gaussian
tails.  Similarly, a measure for the resolution based on the width of
a Gaussian distribution fitted over the central region of the distribution will
fail to take into account a significant fraction of the distribution
if there are large tails and will not be representative of the actual
performance.

Offsets and resolutions between jets reconstructed in the trigger and
reconstructed offline are obtained from the distribution of the
quantity
\begin{eqnarray}
\frac{E_{\mathrm T}^{\mathrm{Trigger}}-\etoffline}{\etoffline}. 
\end{eqnarray}

For a comparison of offline, fully calibrated, jets with the trigger 
jets reconstructed at the electromagnetic scale, the transverse energy 
offset will be large. Therefore, results are first presented in 
terms of the jet definitions actually used by offline analyses and 
the online systems, and then also shown for the case of comparison 
of the online jets with the offline jets, both reconstructed at the 
electromagnetic scale, which more closely resemble each other.

\subsection{Transverse energy offsets and resolutions}
\label{sec:hltres}

Understanding of the offsets and resolutions in the data is useful for the 
determination of the behaviour of the trigger efficiencies near the
\et\ thresholds, since the offset and resolution will, respectively,
have an impact on the position and gradient of the rising edge.

As the resolutions and offsets presented in this section are with
respect to offline jets at the EM+JES scale, large offsets are
expected. For brevity, only the performance of the EF trigger is 
presented as this corresponds most closely to the offline reconstruction.
Since physics analyses generally use \pt, where applicable, the 
offset and resolution of the quantity actually reconstructed in the 
trigger -- namely \et\ -- are shown as a function of the offline jet \pt.

\begin{figure*}[thp]
  \subfigure[]{\includegraphics[width=0.5\textwidth]{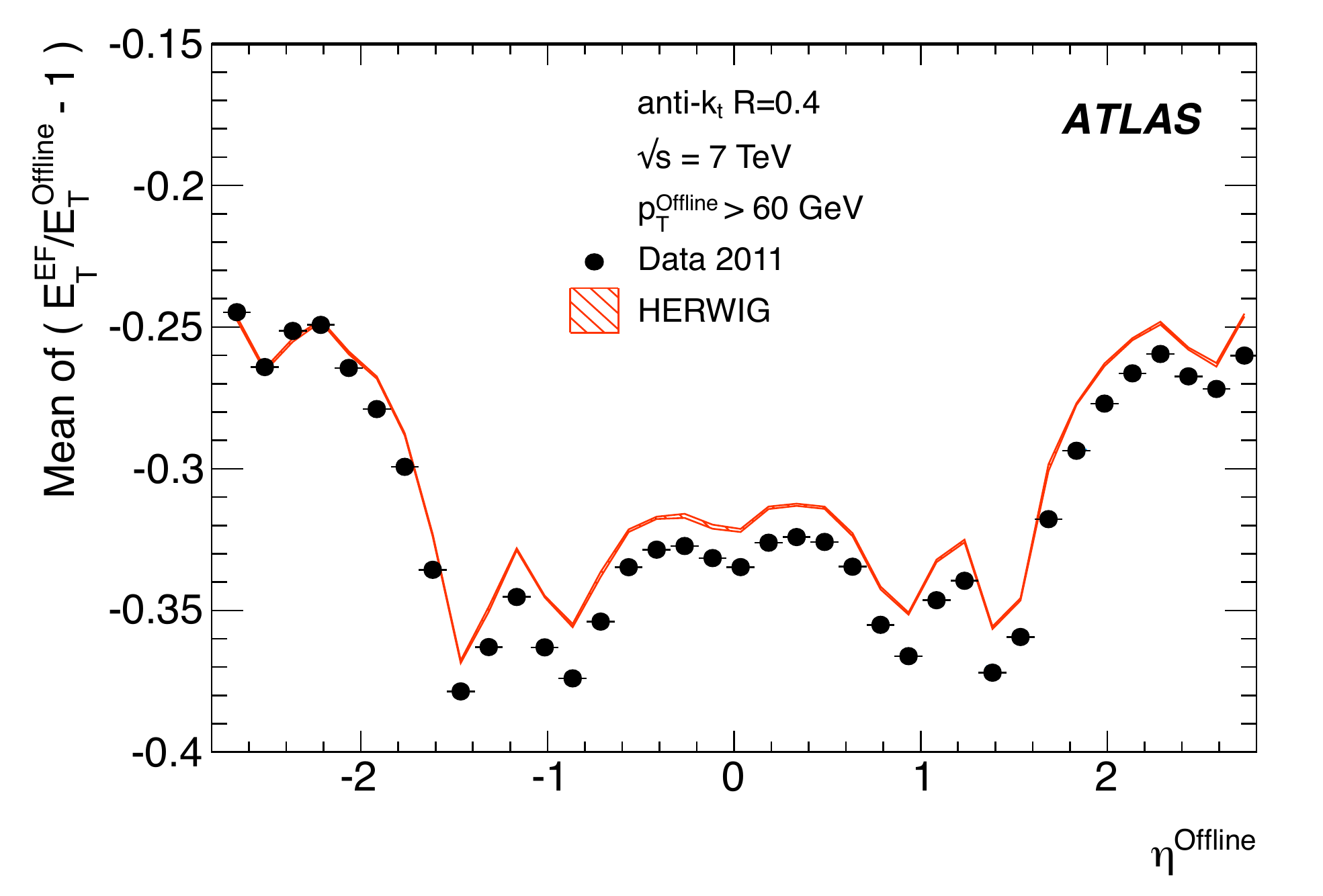}}
  \subfigure[]{\includegraphics[width=0.5\textwidth]{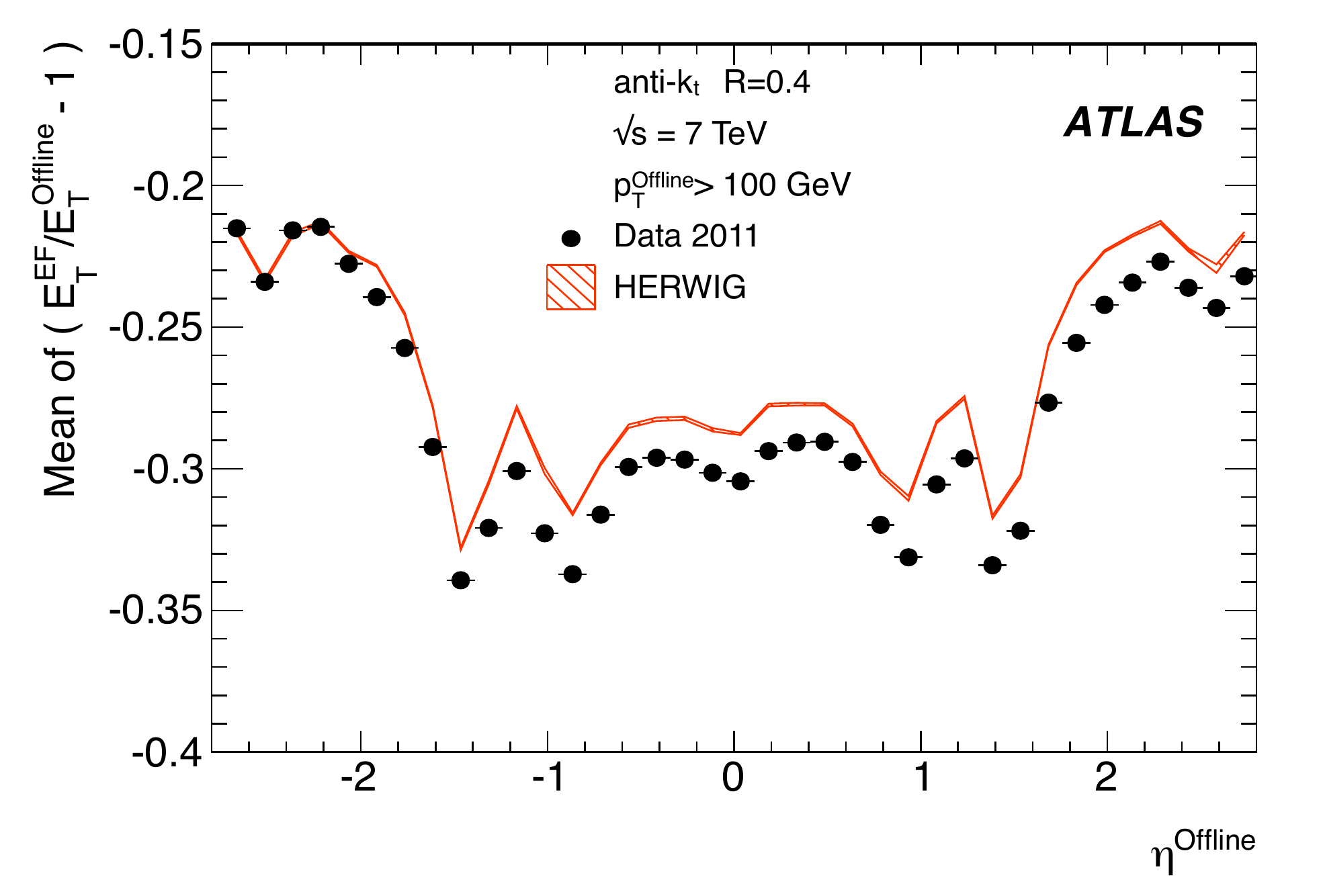}}
  \subfigure[]{\includegraphics[width=0.5\textwidth]{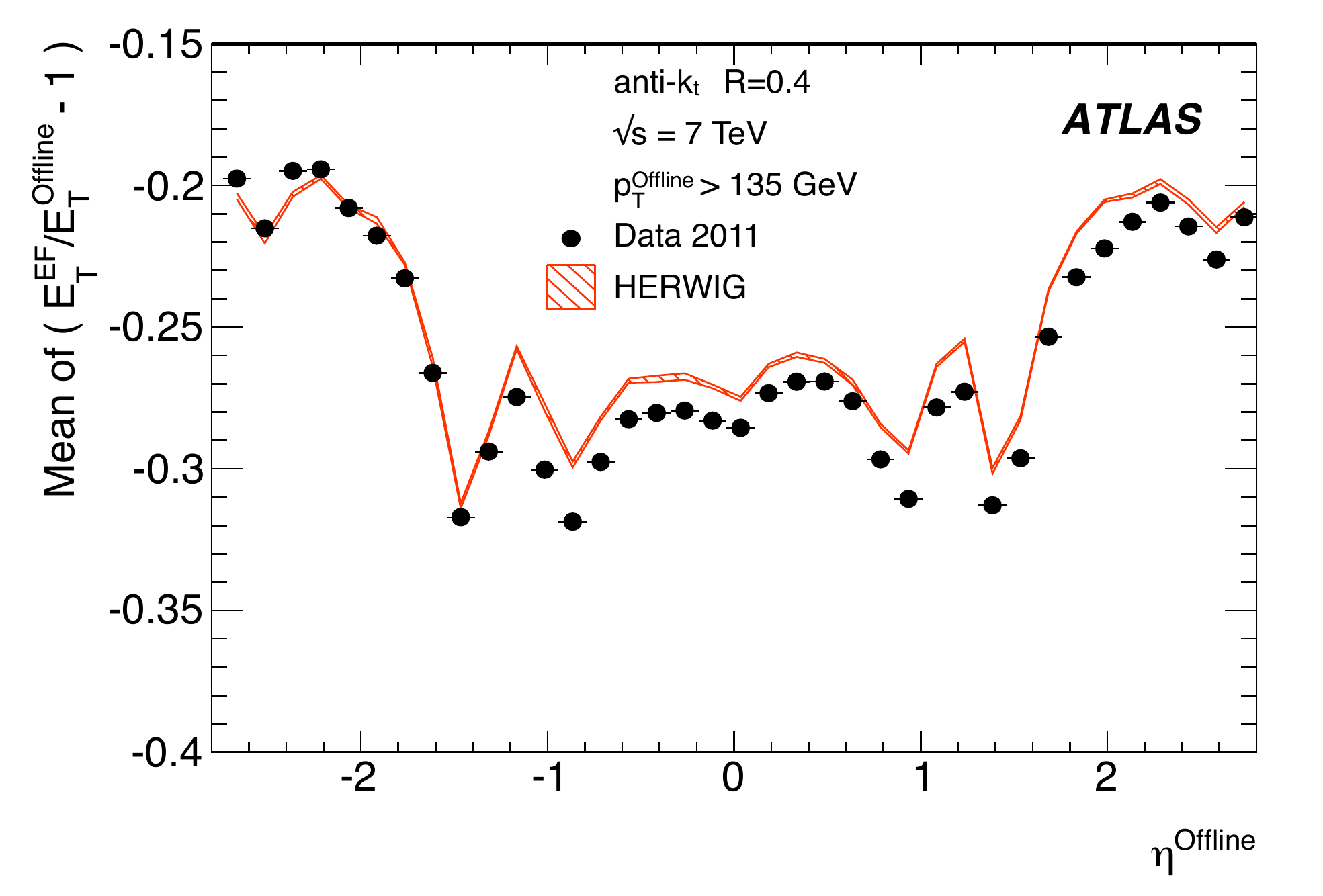}}
  \subfigure[]{\includegraphics[width=0.5\textwidth]{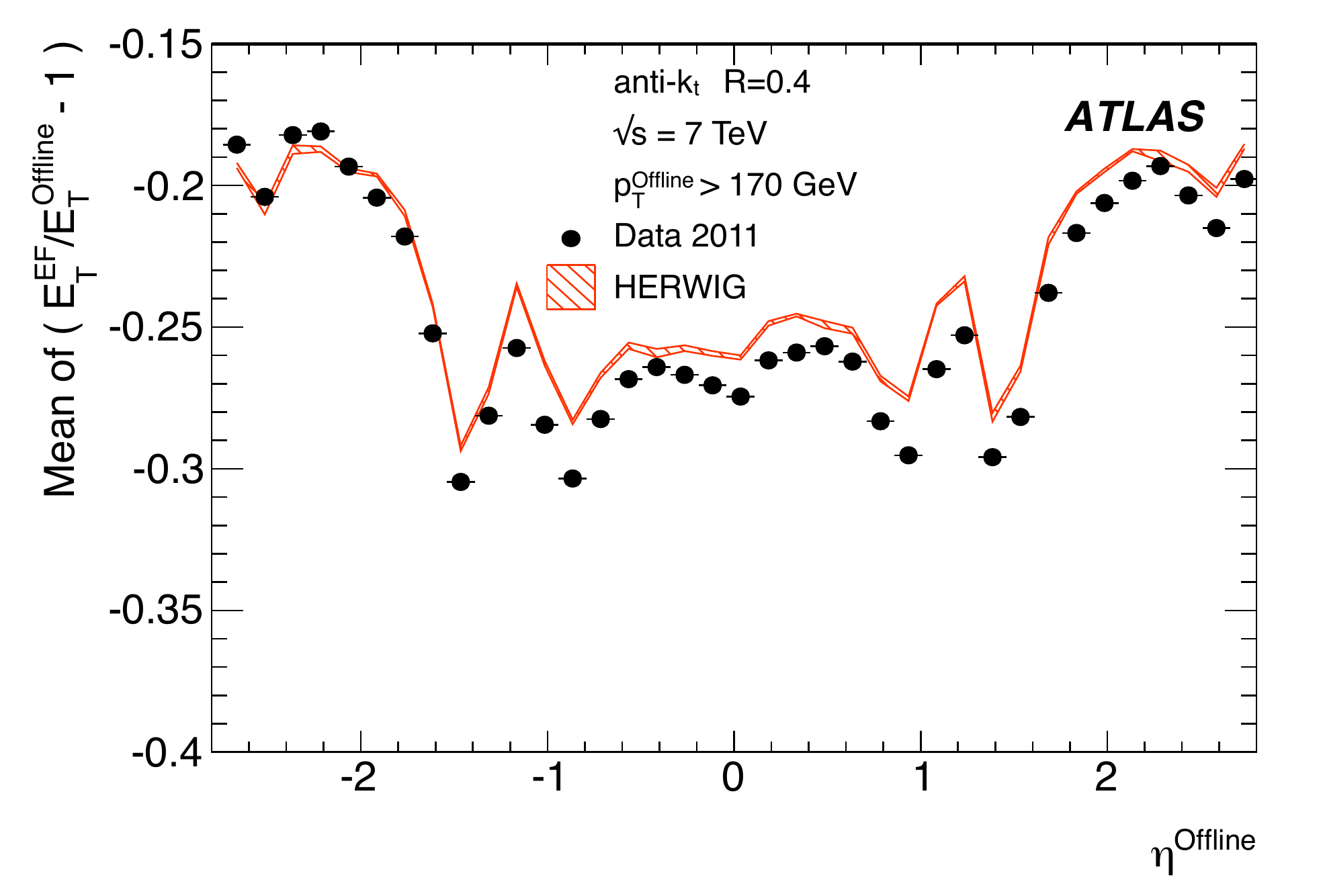}}
  \caption{ The mean relative offset for the EF trigger jets with
    respect to offline jets at the EM+JES energy scale as a function
    of the offline jet $\eta$ in four different ranges of jet \pt:
    (a)~\ptoffline$>60$\,\GeV; (b)~\ptoffline$>100$\,\GeV;
    (c)~\ptoffline$>135$\,\GeV; and (d)~\ptoffline$>170$\,\GeV.  
    \fullerror }

  \label{fig:efresfigone}
\end{figure*}

In order to ensure that the trigger has reached plateau efficiency for
the lowest L1 jet \et\ threshold, the performance in terms of
reconstruction of the jet transverse energy and \pseudorapidity\ is
presented for offline jets with $\pt>60$\,\GeV\ when evaluating the central
jet triggers seeded by the  L1 jet trigger, and $\pt>50$\,\GeV\ when 
evaluating the forward jet trigger.

\subsubsection{Central jets} 

Figure~\ref{fig:efresfigone} shows the mean relative offset 
as a function of the offline jet
$\eta$ for both the data and the \herwig\ simulated sample
integrated over \pt\ above four  different thresholds, indicated in 
the figure.
The general trend of the data is reasonably  well
reproduced by the Monte Carlo simulation, with small differences at
the percent level. 
As discussed previously, differences between the simulation and data result from
inaccuracies or approximations in the simulation of the detector response or 
in the application of the detector conditions, 
but also from differences in the underlying kinematics and \pt\ spectrum 
of the Monte Carlo sample.

A large $\eta$ dependence can be discerned: at low \pt, negative offsets of 
between 24\% and 27\% in the endcaps, and 
between 32\% and 35\% in the barrel are observed.
This variation with $\eta$ is
largely determined by the detector geometry and the different performance 
of the respective calorimeter subsystems  
-- notably with
larger differences in the transition (crack) regions between the barrel and endcap
subsystems, around $|\eta|\sim 1.5$, which are populated with
detector services and around $|\eta|\sim 0.8$ where there is a 20\% 
reduction in the depth of active material in the LAr calorimeter with 
respect to more central pseudorapidities. 
For the same minimum offline \pt\ requirement, jets at higher $\eta$
values also have higher energy, which may also contribute to the
observed differences in the endcap response when compared to more central 
pseudorapidities.  These effects are largely 
accounted for in full calibration for the offline jets, but
not for the trigger jets, where this correction was not applied in 2011. 
Differences in the
detector conditions between online and offline reconstruction, such
as information on masked, or inactive front end boards, which is only
obtained following the offline calibration, also play a
r\^{o}le. This can be seen in the small asymmetry observed between the
forward and rear barrel regions for $|\eta|<0.6$, where   
the simulation, which includes these effects, broadly
follows the trend seen in data, albeit with small quantitative
differences.  Larger offsets are seen in the crack regions due to the
greater energy loss in the additional dead material in front of the
calorimeter. These effects, including changes in the detector
conditions occurring during data taking, are corrected for in the
offline reconstruction using the full calibration.

The relative offset in the data is in general
slightly more negative than that shown by the Monte Carlo
simulation. The size of the offset of the EF trigger jets with respect
to offline jets tends to decrease for the higher \pt\ selections. This 
is largely attributable to the comparitively reduced energy loss in 
inactive material for jets of high energy when compared to those 
with lower energy. This trend is also fairly well reproduced by the 
simulation

\begin{figure*}[thp]
  \subfigure[]{\includegraphics[width=0.5\textwidth]{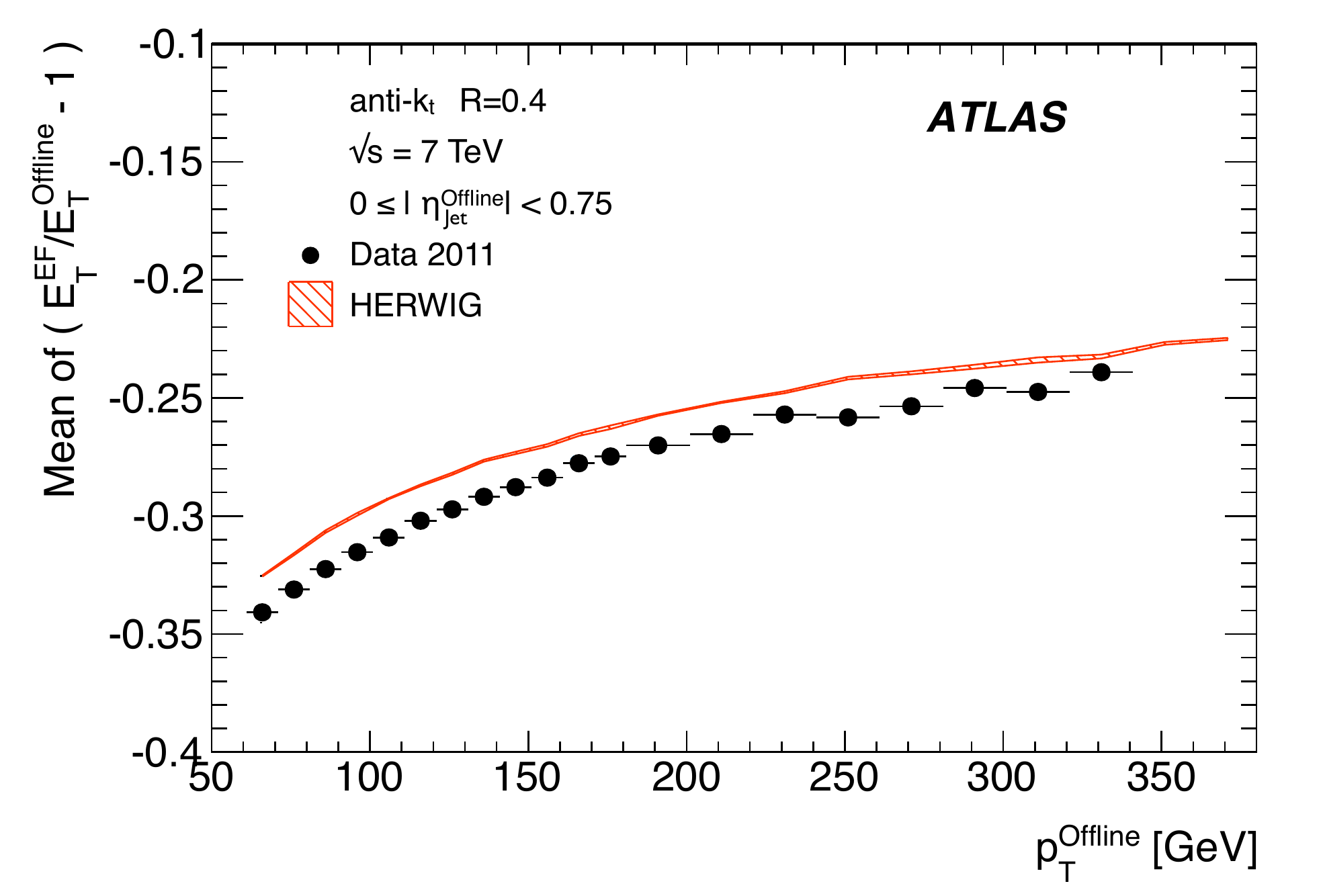}}
  \subfigure[]{\includegraphics[width=0.5\textwidth]{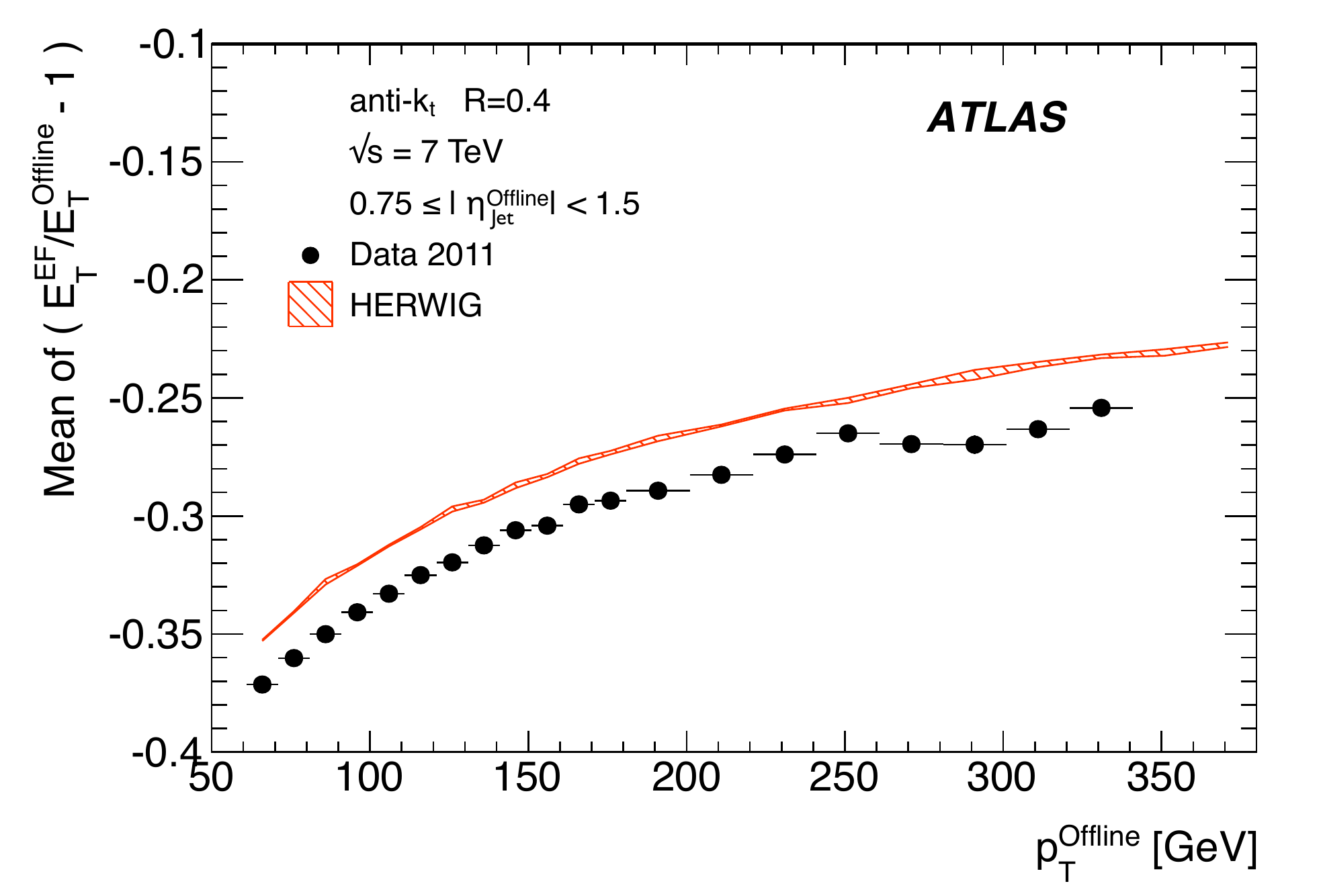}}
  \subfigure[]{\includegraphics[width=0.5\textwidth]{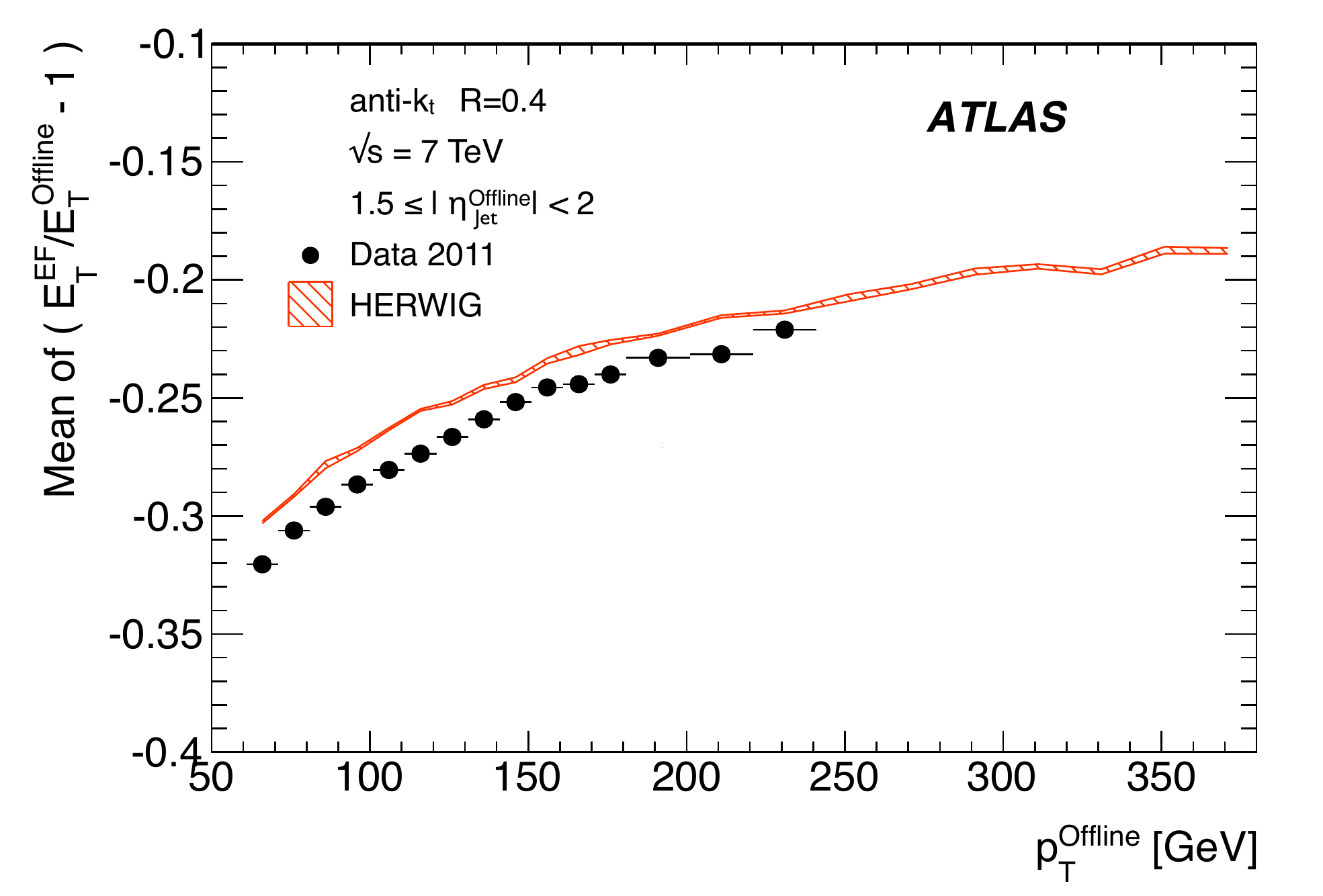}}
  \subfigure[]{\includegraphics[width=0.5\textwidth]{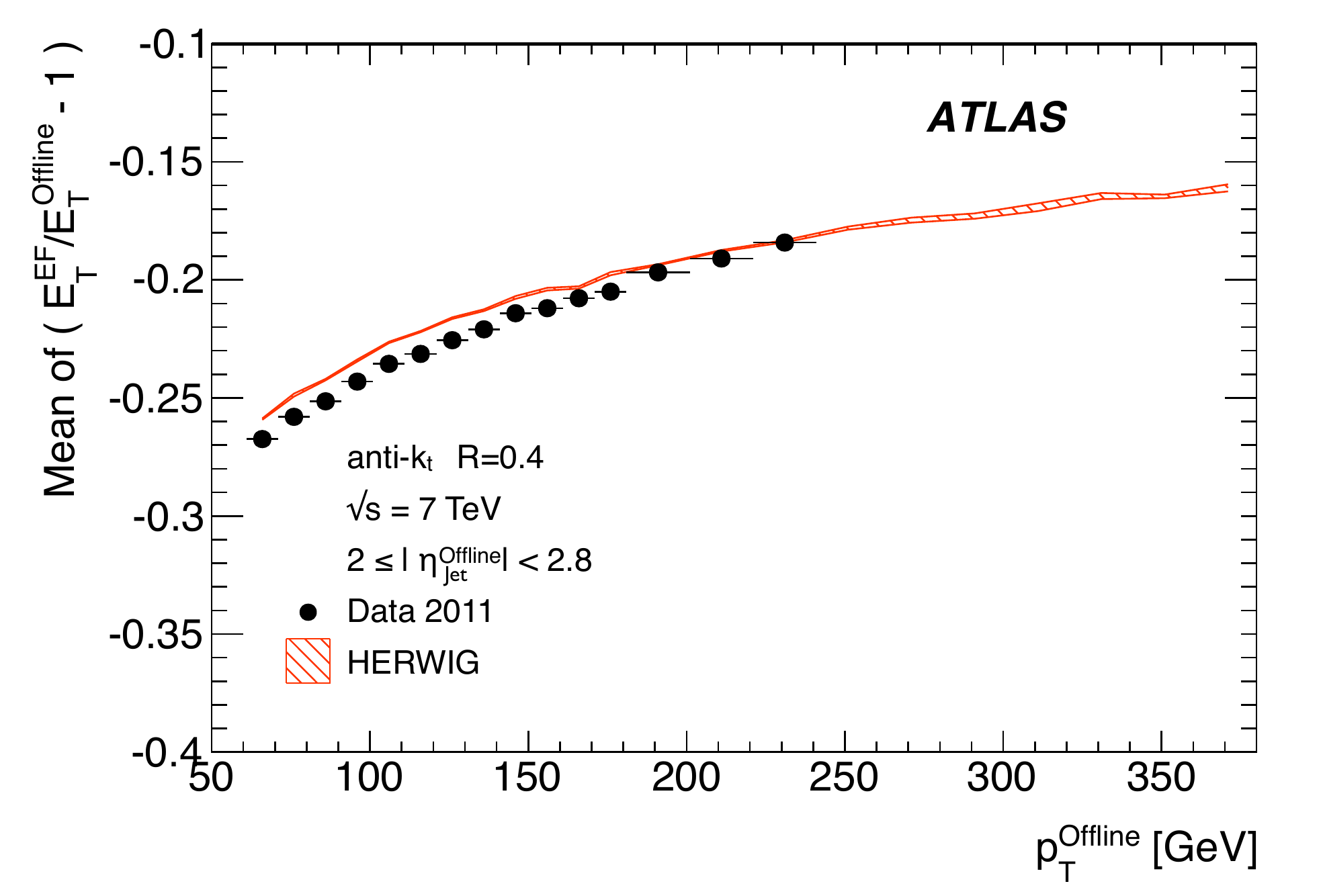}}
  \caption{ The mean relative offset for the EF trigger jets with
    respect to offline jets at the EM+JES energy scale as a function
    of the offline jet \pt, in four different regions of offline jet
    pseudorapidity: (a)~$|\eta\offline| < 0.75$; (b)~$0.75\leq
    |\eta\offline| < 1.5$, (c)~$1.5\leq |\eta\offline| < 2$; and
    (d)~$2\leq |\eta\offline| < 2.8$. 
    \fullerror }

  \label{fig:efresfigtwo}
\end{figure*}

In Figure~\ref{fig:efresfigone} the offset in transverse energy from the 
Event Filter, $\et^{\mathrm {EF}}$, with respect to the offline jets
is seen to vary over a range of approximately 10\% with \eta, with the 
offsets themselves being only two or three times larger than this range.  
Therefore
the widths of the distributions of residuals obtained when integrating 
over the entire $\eta$ range
would result from a convolution of the true resolution and the variation 
of the offset with $\eta$.  The consequent resolution would appear
artificially large, smeared by this additional factor of 10\%.

As a result, the mean offset and resolution in \et\ as a function of
\pt\ has been measured in four separate regions of $|\eta\offline|$;
\begin{eqnarray}
0~\leq &|\eta\offline|& < ~0.75 \\
0.75~\leq &|\eta\offline|& < ~1.5 \\
1.5~\leq &|\eta\offline|& < ~2 \\
2~\leq &|\eta\offline|& < ~2.8,
\end{eqnarray}
with the first and last corresponding to those regions where the offset is
approximately constant in the barrel and endcap regions, respectively.
In the remaining two regions the offset varies rapidly due to the
crack regions in the calorimeters.  The resulting offsets are shown as
a function of \ptoffline\ in Figure~\ref{fig:efresfigtwo}.  It is seen
that the mean offset decreases as the offline jet \pt\ increases, with
the smallest offset in the endcap regions, as expected from the
variation of the offset with $\eta\offline$ seen in Figure~\ref{fig:efresfigone}.
Again, as can be seen in Figure~\ref{fig:efresfigtwo}, the simulation
underestimates the offsets for all
$|\eta\offline|$, by aproximately 1\% for $|\eta\offline|<0.75$ and 2\% 
for the regions $0.75<|\eta\offline|<2$.
This is also true, albeit to a lesser degree, 
for the range $2\leq|\eta\offline|<2.8$, since the positive and
negative \pseudorapidity\ regions seen in Figure~\ref{fig:efresfigone} have been
combined.

\begin{figure*}[th!]
  \subfigure[]{\includegraphics[width=0.5\textwidth]{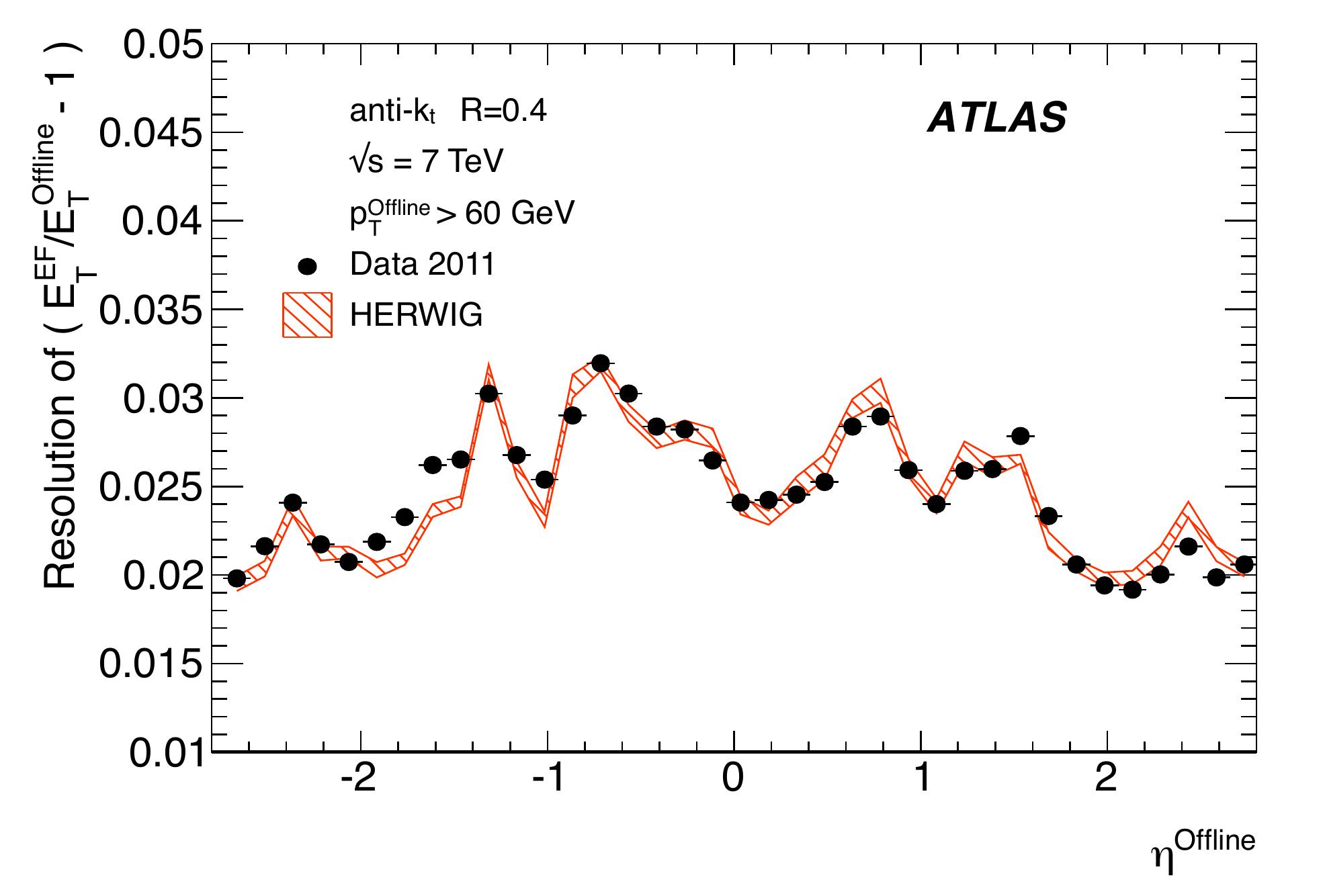}}
  \subfigure[]{\includegraphics[width=0.5\textwidth]{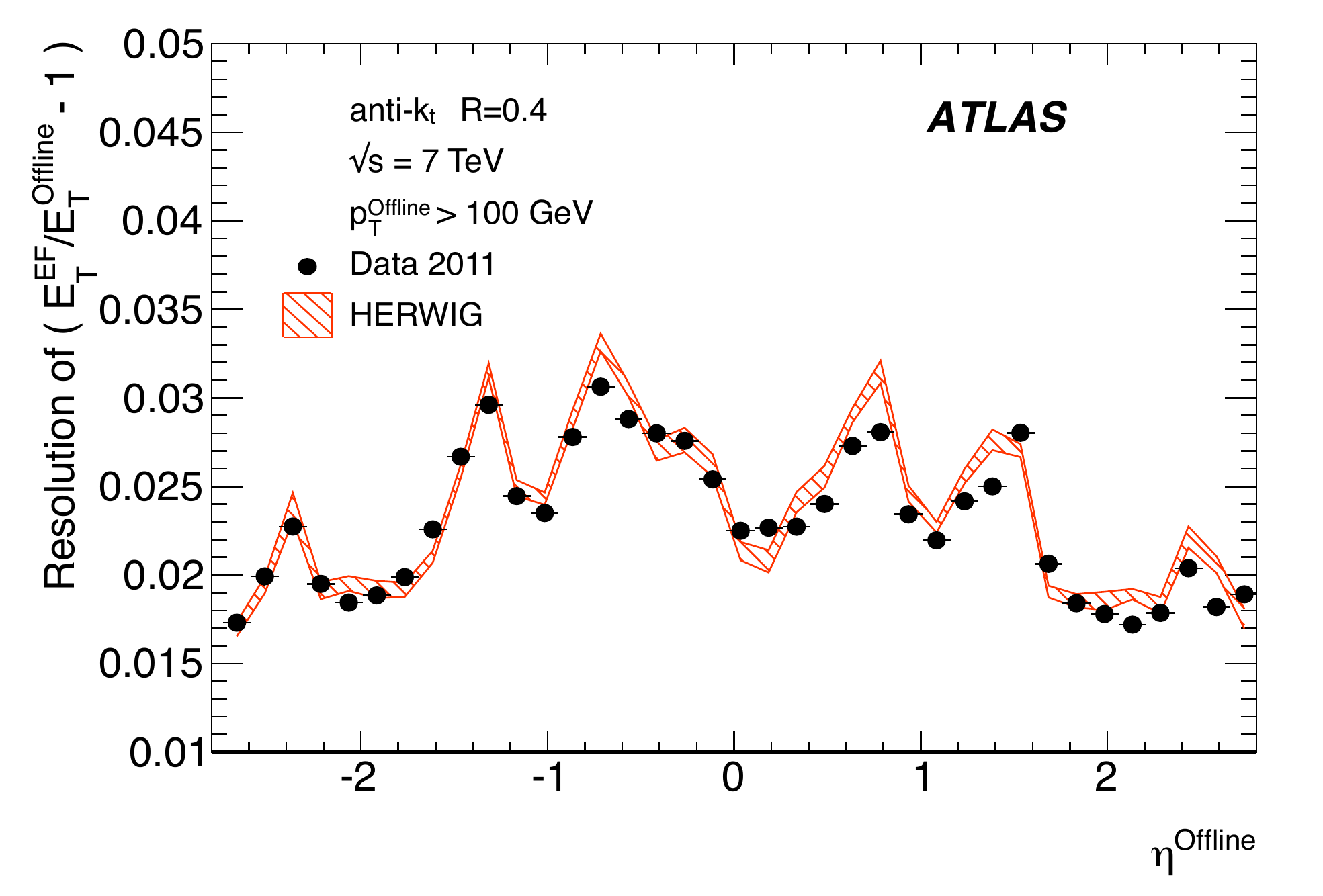}}
  \subfigure[]{\includegraphics[width=0.5\textwidth]{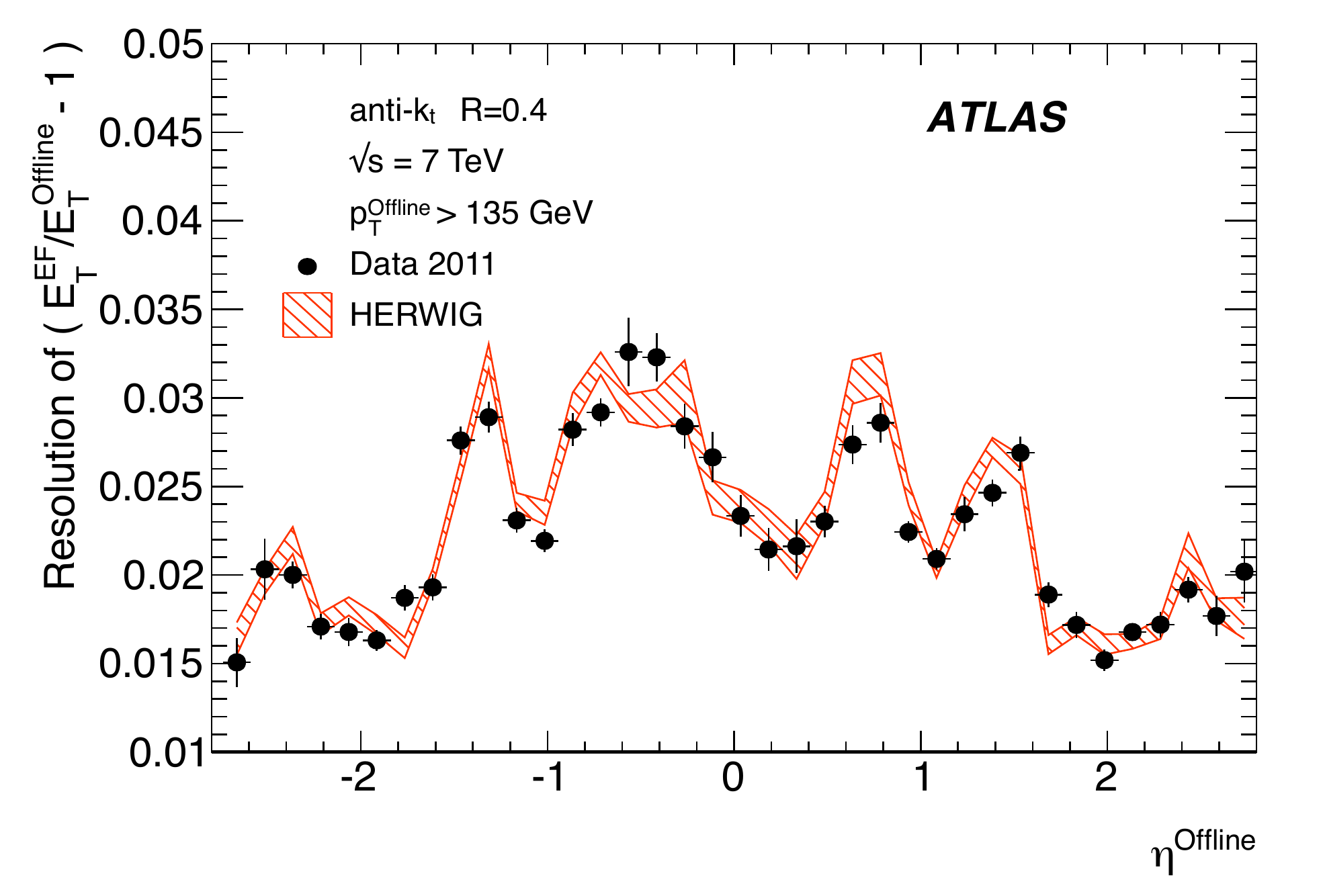}}
  \subfigure[]{\includegraphics[width=0.5\textwidth]{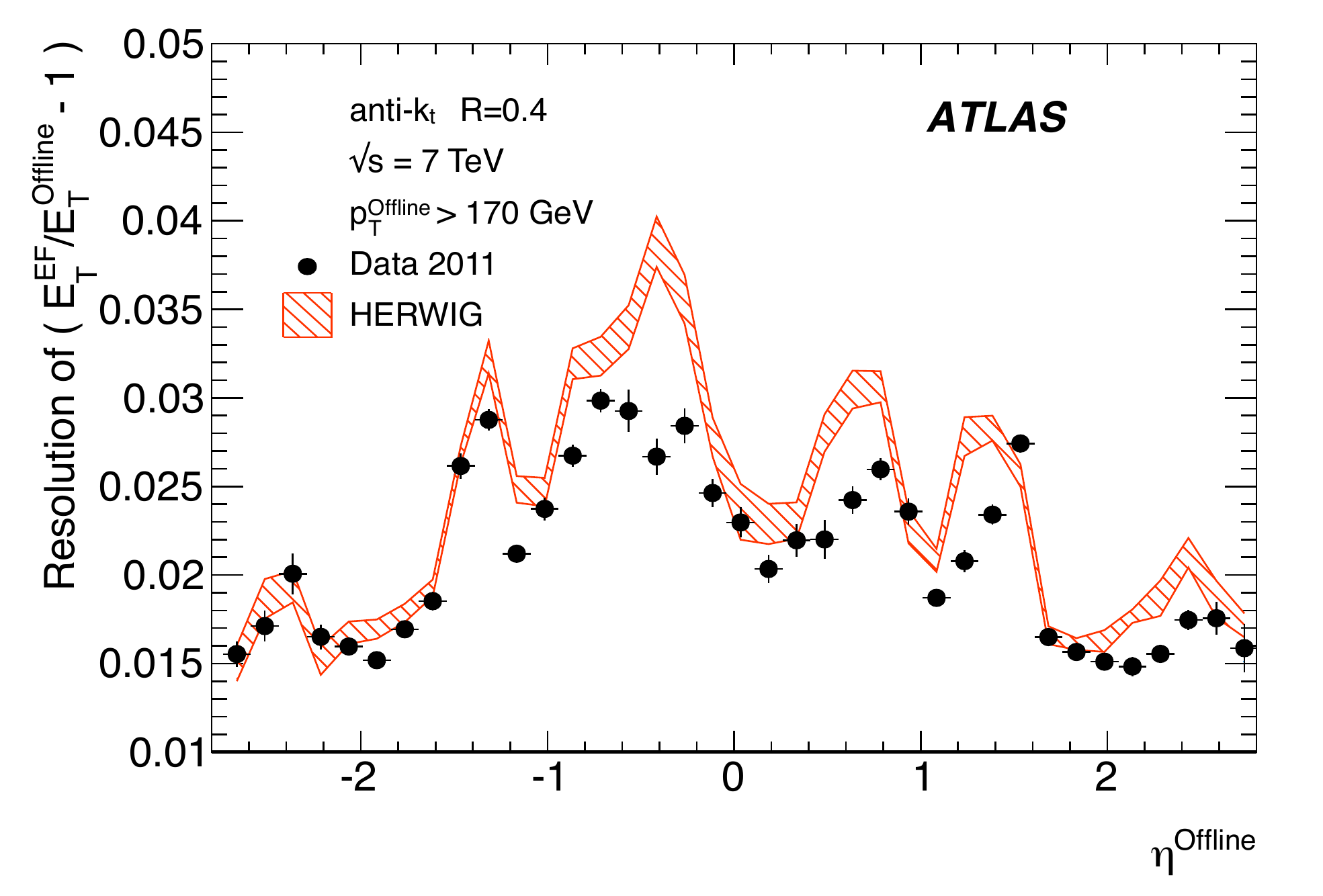}}
  \caption{ The resolution for the EF trigger jets with respect to
    offline jets at the EM+JES energy scale as a function of the
    offline jet $\eta$ in four different ranges of jet offline \pt:
    (a)~\ptoffline$>60$\,\GeV; (b)~\ptoffline$>100$\,\GeV;
    (c)~\ptoffline$>135$\,\GeV; and (d)~\ptoffline$>170$\,\GeV.  
    \fullerror }

  \label{fig:efresfigthree}
\end{figure*}

Because of the large dependence of the offset on \pt, when integrating
the distribution over \ptoffline\ to show for example the variation 
of the resolution as a function of $\eta\offline$, this will include the 
convolution of the resolution from the detector response with the variation 
of the offset with \ptoffline\ itself. This should  be taken into 
account when estimating the resolution. 
This effect will, however, not be as pronounced as in the case of the
similar variation of the offset with $\eta\offline$, because of the steeply 
falling \pt\ spectrum, so the shape of the residual distributions
will be largely determined by the jets near the \ptoffline\ threshold.

Figure~\ref{fig:efresfigthree} shows the resolution versus
$\eta\offline$ for the four $\pt$ ranges shown earlier. The resolution 
is generally better in the endcap regions than in the barrel and does 
not vary greatly between the four \pt\ ranges, although it varies sharply as 
a function of $\eta\offline$. The resolution is quite well described by the
simulation where, as in data, it does not vary greatly with \pt.  The large
asymmetry in the resolution between the barrel regions at positive and
negative $\eta$ due to the detector conditions is approximately
reproduced by the simulation. At high \pt, the simulation predicts
a worse resolution than seen in the data, by up to 0.5\% or slightly
higher in some regions.  In the more forward directions, the better
\et\ resolution partly results from the larger jet energy relative to 
jets in the barrel with the same \et.

\begin{figure*}[th!]
  \subfigure[]{\includegraphics[width=0.5\textwidth]{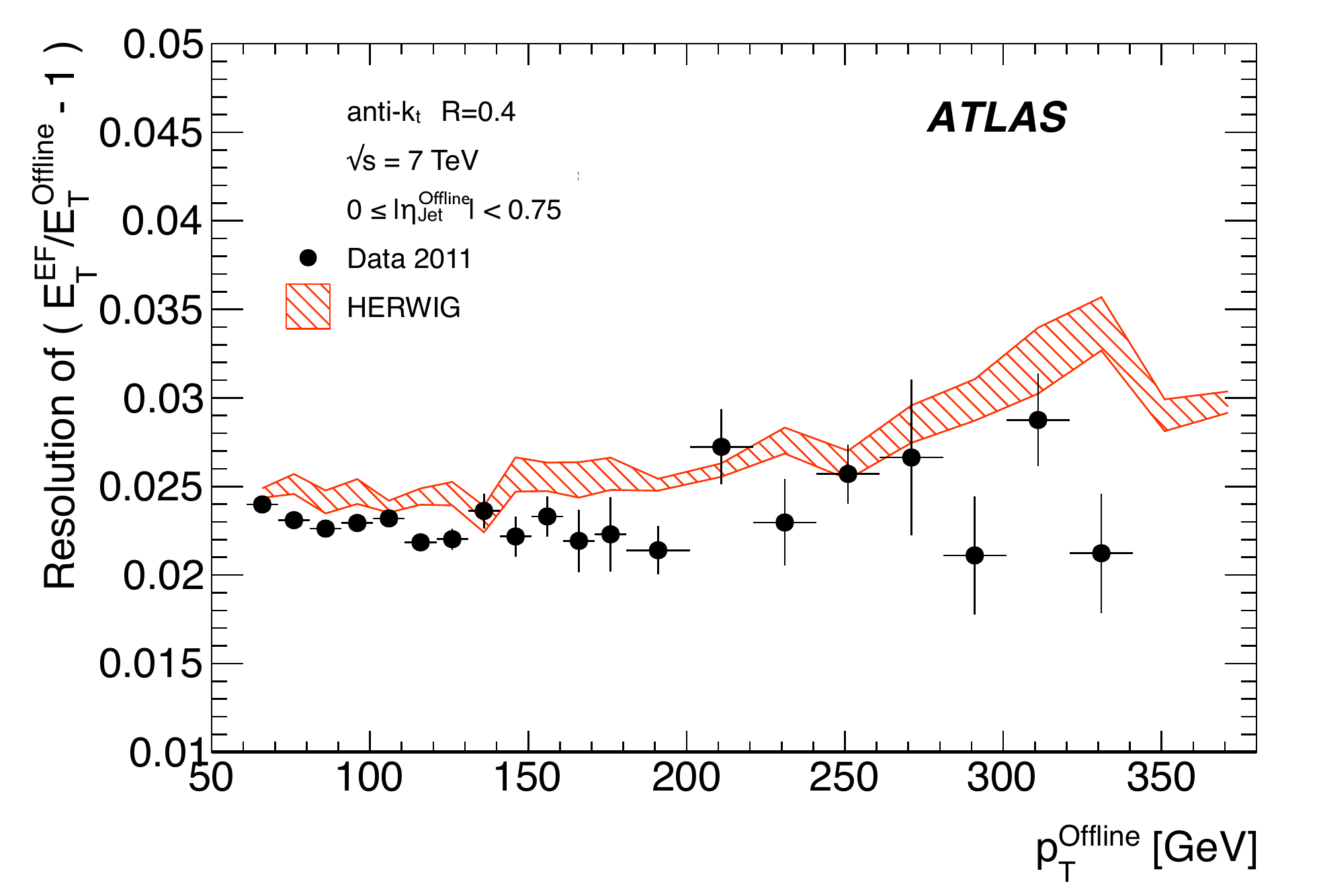}}
  \subfigure[]{\includegraphics[width=0.5\textwidth]{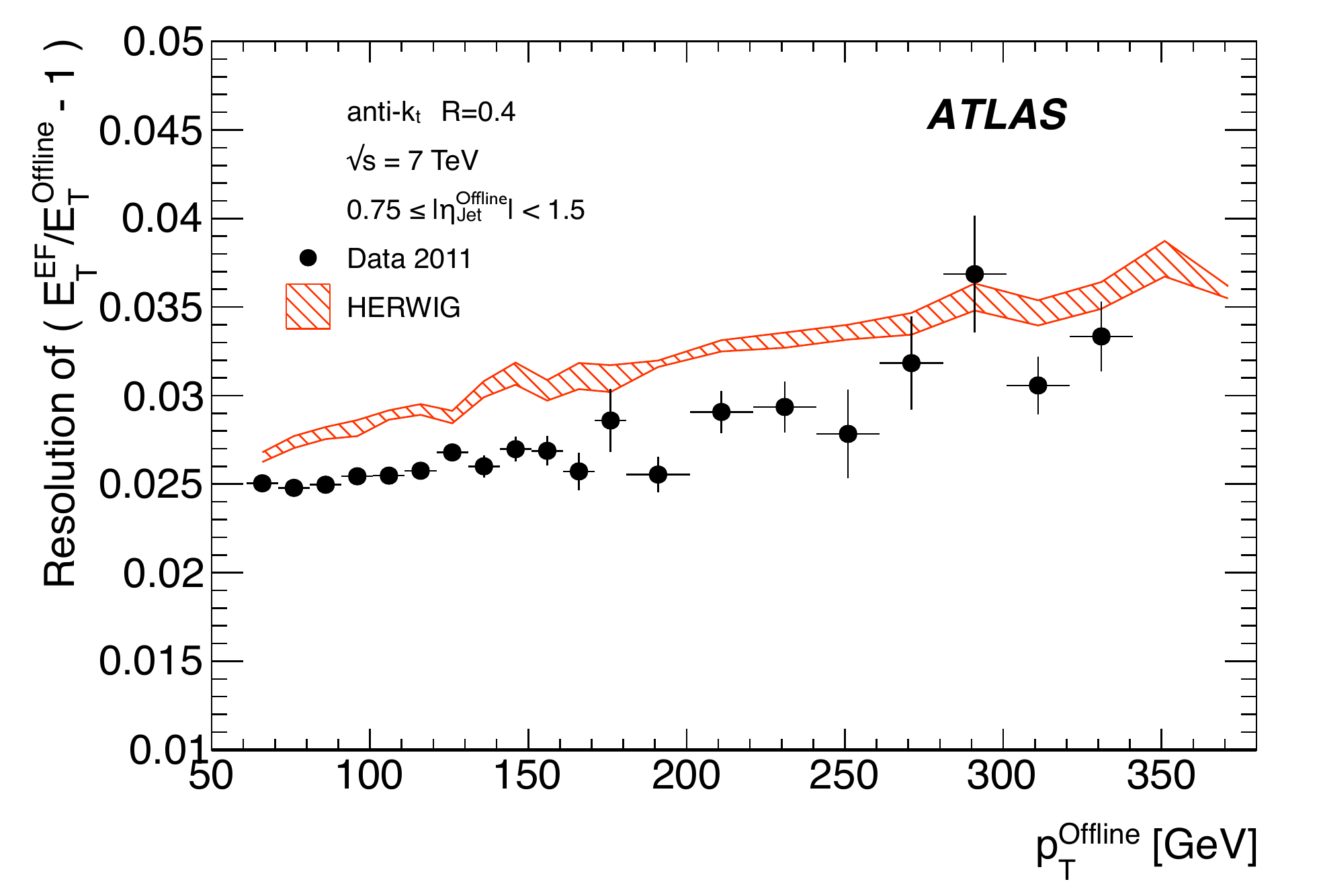}}
  \subfigure[]{\includegraphics[width=0.5\textwidth]{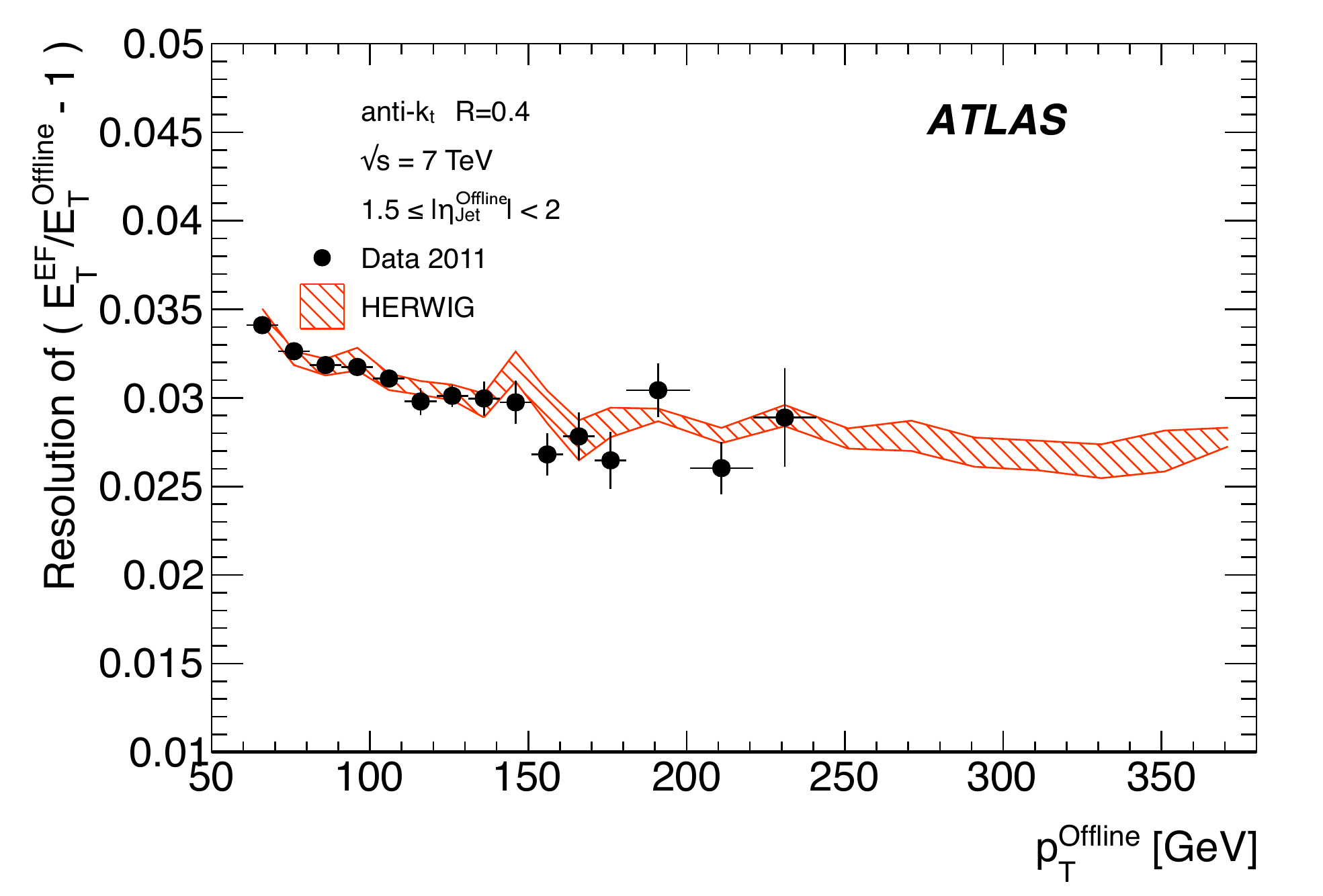}}
  \subfigure[]{\includegraphics[width=0.5\textwidth]{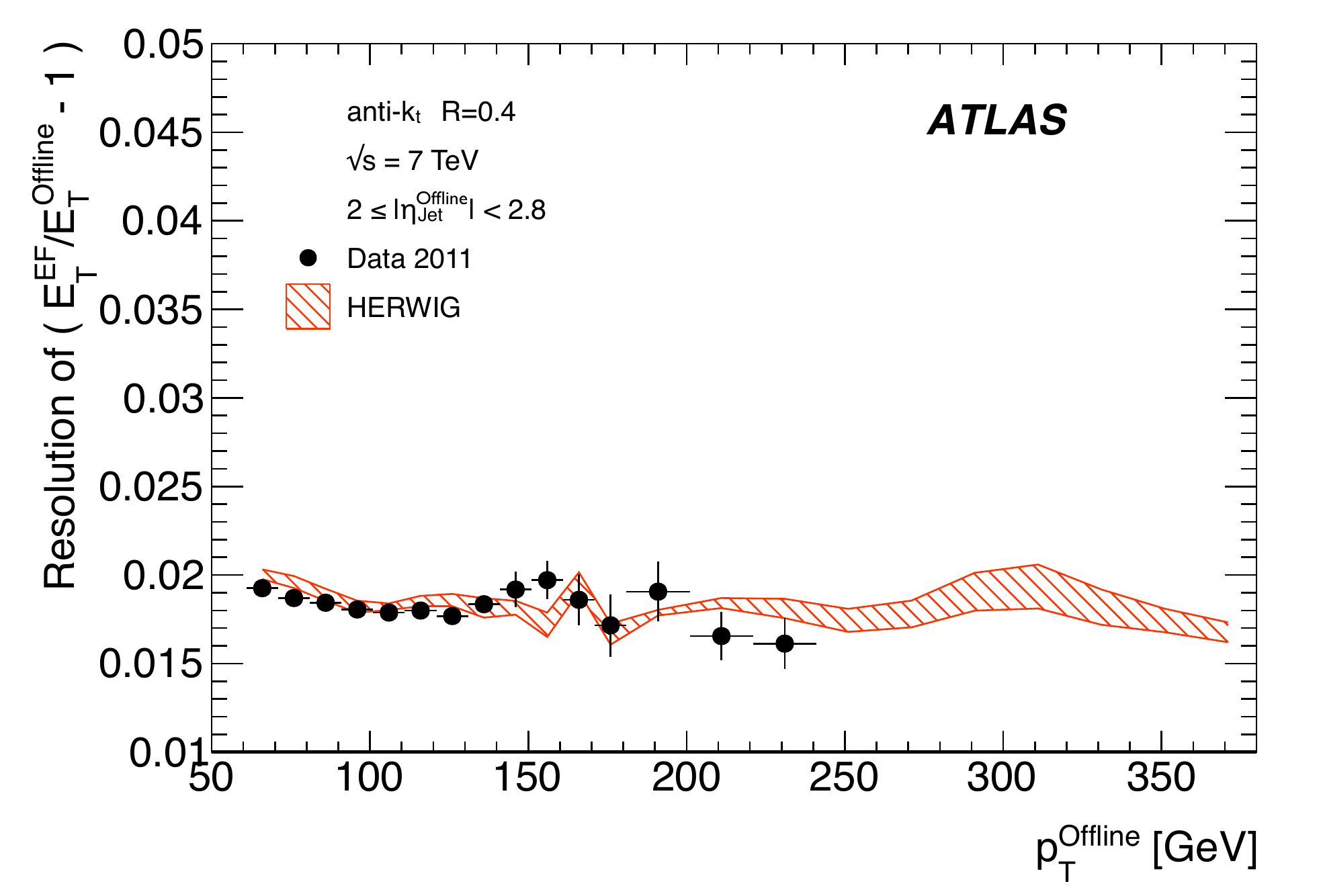}}
  \caption{ The resolution for the EF trigger jets with respect to
    offline jets at the EM+JES energy scale, as a function of the
    offline jet \pt, in four different regions of offline jet
    pseudorapidity: (a)~$|\eta\offline| < 0.75$; (b)~$0.75\leq
    |\eta\offline| < 1.5$; (c)~$1.5\leq |\eta\offline| < 2$;
    and (d)~$2\leq |\eta\offline| < 2.8$. 
    \fullerror }

  \label{fig:efresfigfour}
\end{figure*}

Figure~\ref{fig:efresfigfour} illustrates the resolution as a function
of transverse energy in the same four $\eta\offline$ ranges discussed
previously. In the barrel region, at low \ptoffline, the resolution is
approximately constant. The resolution found in collision data is
again not fully reproduced by simulation, although the differences
found are small, being only around 0.5\% at most. In the crack
regions, as might be expected from the larger energy loss indicated by
the larger offsets observed earlier, the resolutions are worse than in the
barrel or endcaps and show a larger dependence on the \pt.

\subsubsection{Forward jets}
\label{sec:efresfj}

\begin{figure*}[tp]
  \spacer
  \subfigure[]{\includegraphics[width=0.505\textwidth]{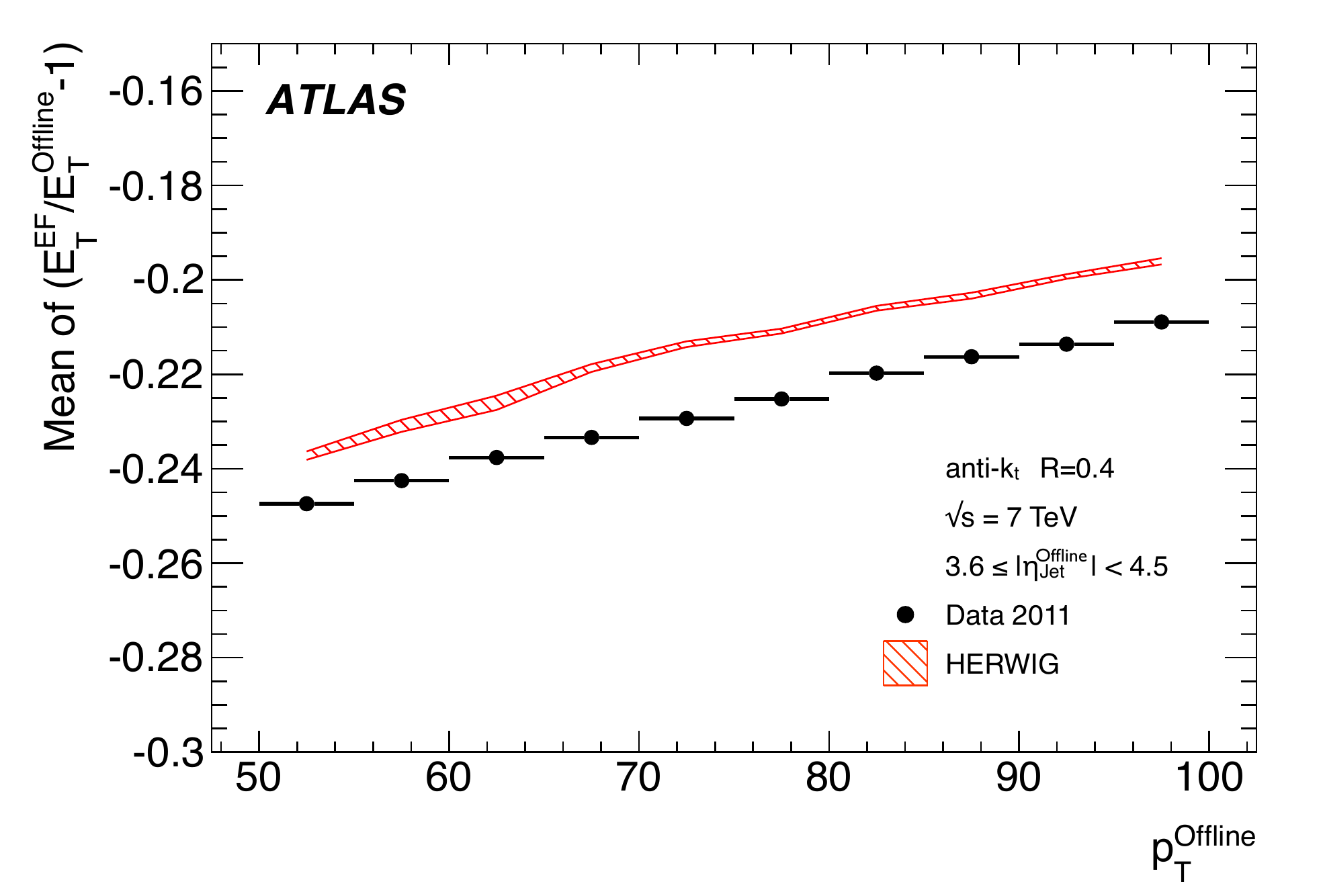}\label{fig:effjresfigtwoa}}
  \subfigure[]{\includegraphics[width=0.505\textwidth]{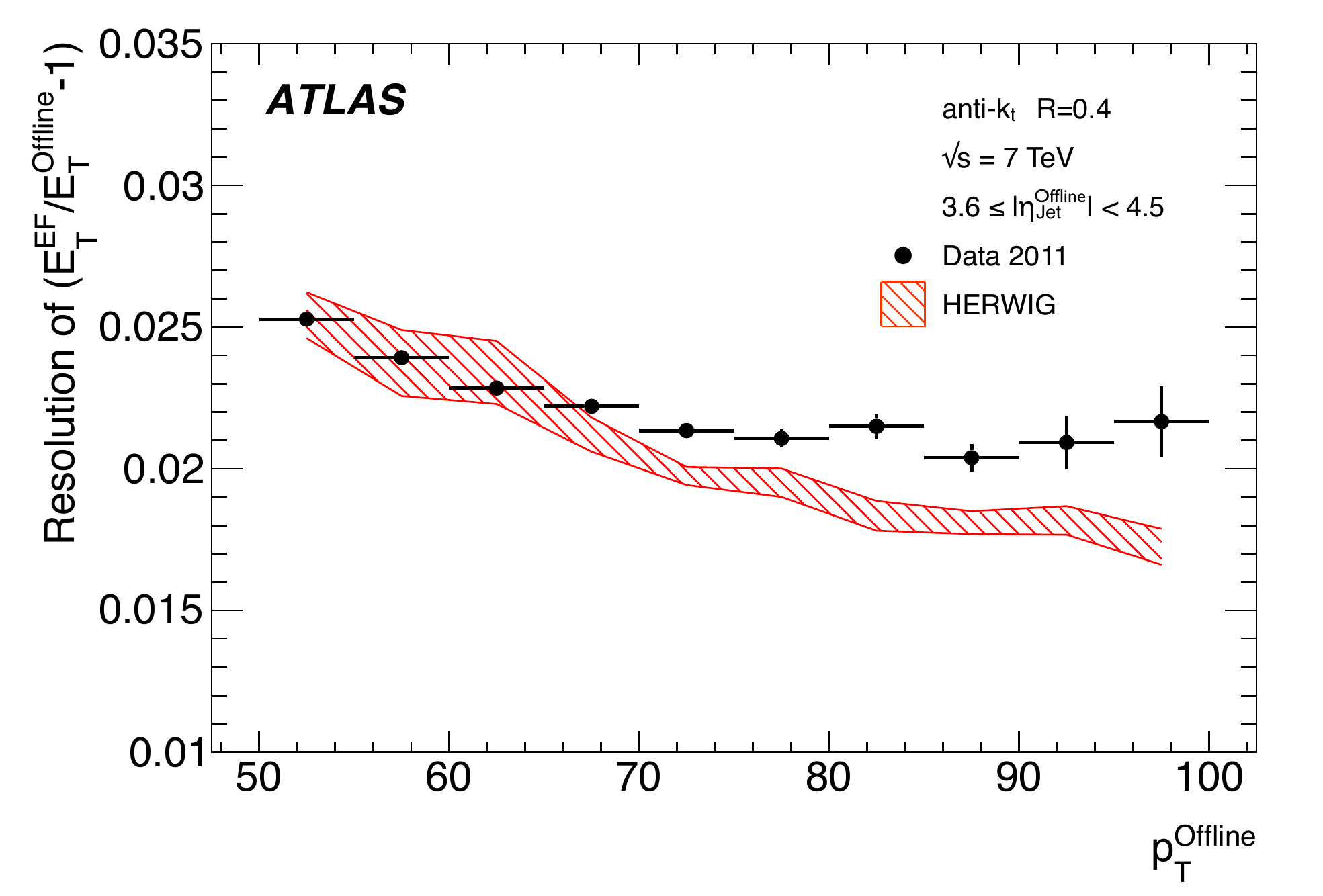}\label{fig:effjresfigtwob}}
  \caption{ The performance for 
    the EF forward trigger jets with respect to offline jets at the
    EM+JES energy scale as a function of the offline jet \pt\ for jets
    with \mbox{$3.6<|\eta\offline|<4.4$}: (a) The mean relative offset; and (b) the resolution.
  \fullerror }
  \label{fig:effjresfigtwo}
\end{figure*}

\newcommand{\excludefjtwo}{
The offsets and resolutions with respect to offline jets at the EM+JES
scale for the forward jets reconstructed by the EF can be seen in
Figures~\ref{fig:effjresfigone} and~\ref{fig:effjresfigtwo}. The
performance for jets with a radius parameter $R=0.4$ are shown. Jets
are required to be in the range $3.6< |\eta\offline| < 4.4$ which
ensures complete containment of jets with $R=0.4$ in the forward
calorimeter.  As in the case for the central jet trigger, at large
$\eta\offline$, offsets of approximately 20\% are observed, with
resolutions of approximately 2\% increasing to 3\% for
\pseudorapidity\ approaching 4.3.}

The offsets and resolutions with respect to offline jets
as a function of offline \pt
for jets from the forward jet trigger are shown in 
Figure~\ref{fig:effjresfigtwoa} and Figure~\ref{fig:effjresfigtwob} respectively.
The offline jets are produced
with a radius parameter $R=0.4$ and are required to be within the
range $3.6< |\eta\offline| < 4.4$.
These show a dependence of the offset on the jet \pt\ which improves 
towards high \pt\ as for the central jets, from approximately 
24\% at lower \pt to only 20\% at high \pt, broadly consistent with 
the behaviour seen in the central jet trigger.  The simulation shows approximately 
1\% smaller offsets than seen in the data over the full \pt\ range. 
The jet resolution is reasonably well described by the simulation at low \pt, with 
larger differences at high \pt, of less than 0.5\%. The data show only a 
small variation of the resolution with offline \pt, of between 2.0\% and 2.5\%, 
whereas the simulation shows a slightly stronger dependence at higher \pt.

\subsubsection{The performance with respect to offline EM scale jets}
\label{sec:emscale}

All of the above results compare electromagnetic-scale jets, as
measured by the trigger, with offline reconstructed jets at the EM+JES
scale. In this section, trigger jets are compared to offline jets
reconstructed at the EM scale only, to better illustrate the
correspondence between the reconstruction in the trigger and offline 
when the same calibrations are applied to both. 
\footnote{The
full calibration was not applied to the trigger jets since it was 
not available at the time of data taking, but only after the 
offline reconstruction of the data.}

\begin{figure}[th!]
  \subfigure[]{\includegraphics[width=0.5\textwidth]{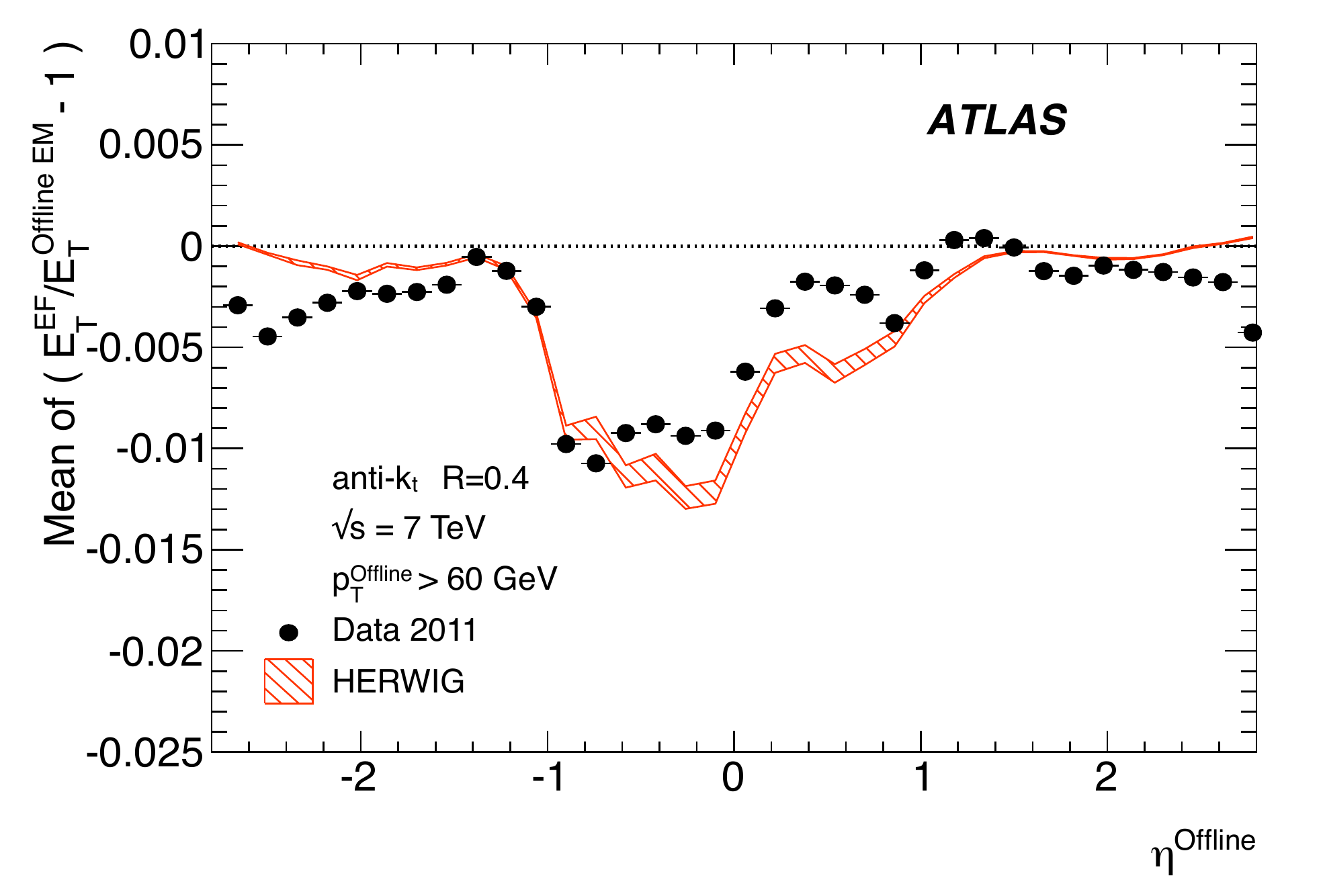}}
  \subfigure[]{\includegraphics[width=0.5\textwidth]{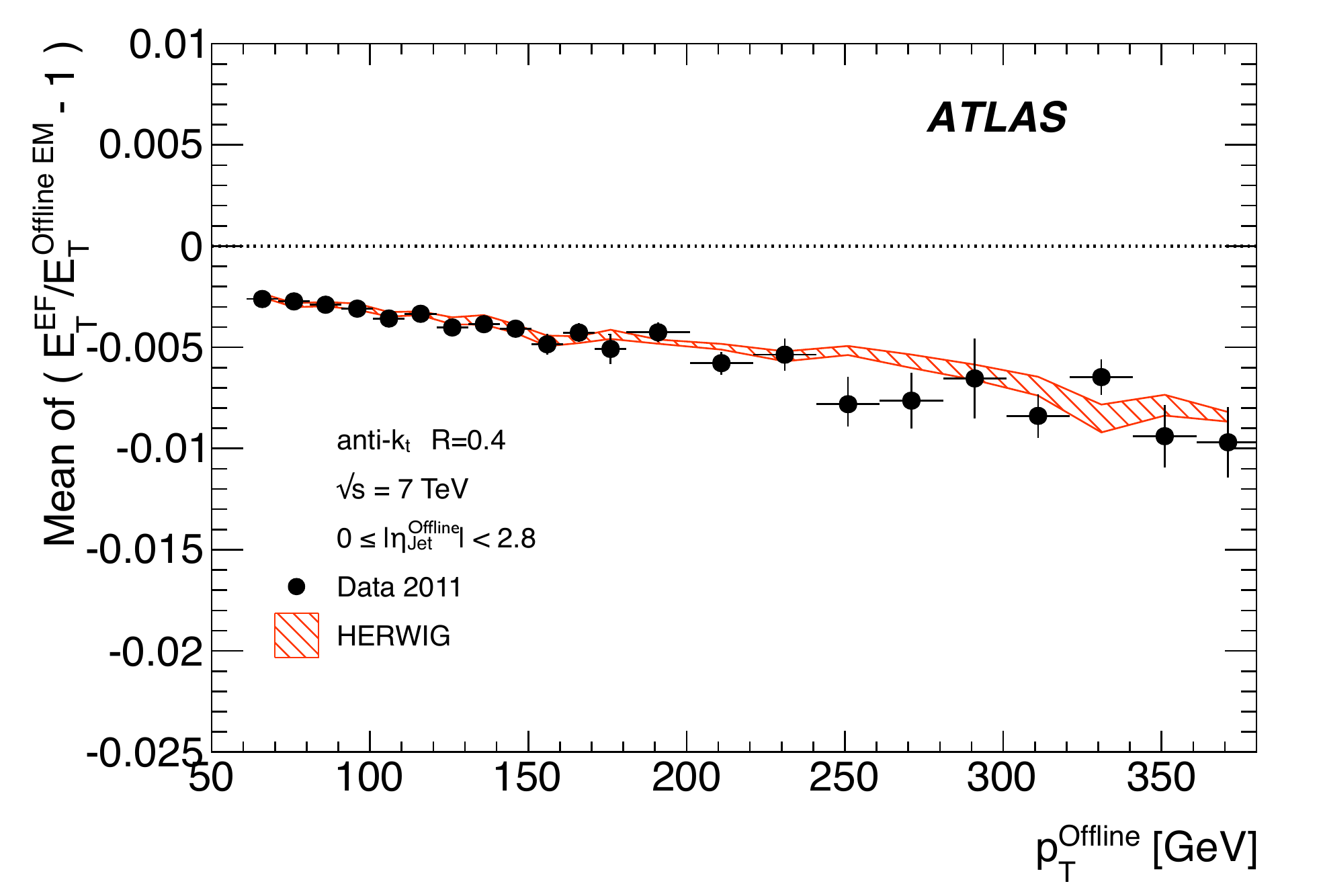}}
  \caption{ The mean relative offset for the EF trigger jets with
    respect to offline jets at the EM energy scale as a function:
    (a)~of the offline jet $\eta$ for jets with \pt$>60$\,\GeV;
    (b)~as a function of the offline jet \pt\ at the hadronic energy
    scale for jets in the range $|\eta^{\mathrm {Offline}}|<2.8$. \fullerror }
  \label{fig:efresemfigone}
\end{figure}

Figure~\ref{fig:efresemfigone} shows the mean relative offset of the
trigger jets at the EF with respect to offline jets at the EM scale as
functions of the offline jet \pseudorapidity and \pt\ at the EM+JES
scale.  A mean offset of less than 0.5\% is observed at low \pt,
increasing to around 1\% at higher \pt. This is in contrast to the
much stronger dependence on \pt\ seen for the offset when comparing to
fully corrected offline jets from Figure~\ref{fig:efresfigtwo}, where
the offset varies from 35\% at low \pt\ to 15\% at
higher \pt. The offset as a function of offline jet \pseudorapidity,
integrated over the range \pt$>60$\,\GeV, has a value of better than
0.5\% everywhere, except for the barrel region in the range $-1<\eta<0$,
where the offset is approximately 1\%. This is again in stark contrast
to the offsets found between EM scale trigger jets and EM+JES offline
jets shown in Figure~\ref{fig:efresfigone}, which extend to 35\%.

\begin{figure}[th!]
  \subfigure[]{\includegraphics[width=0.5\textwidth]{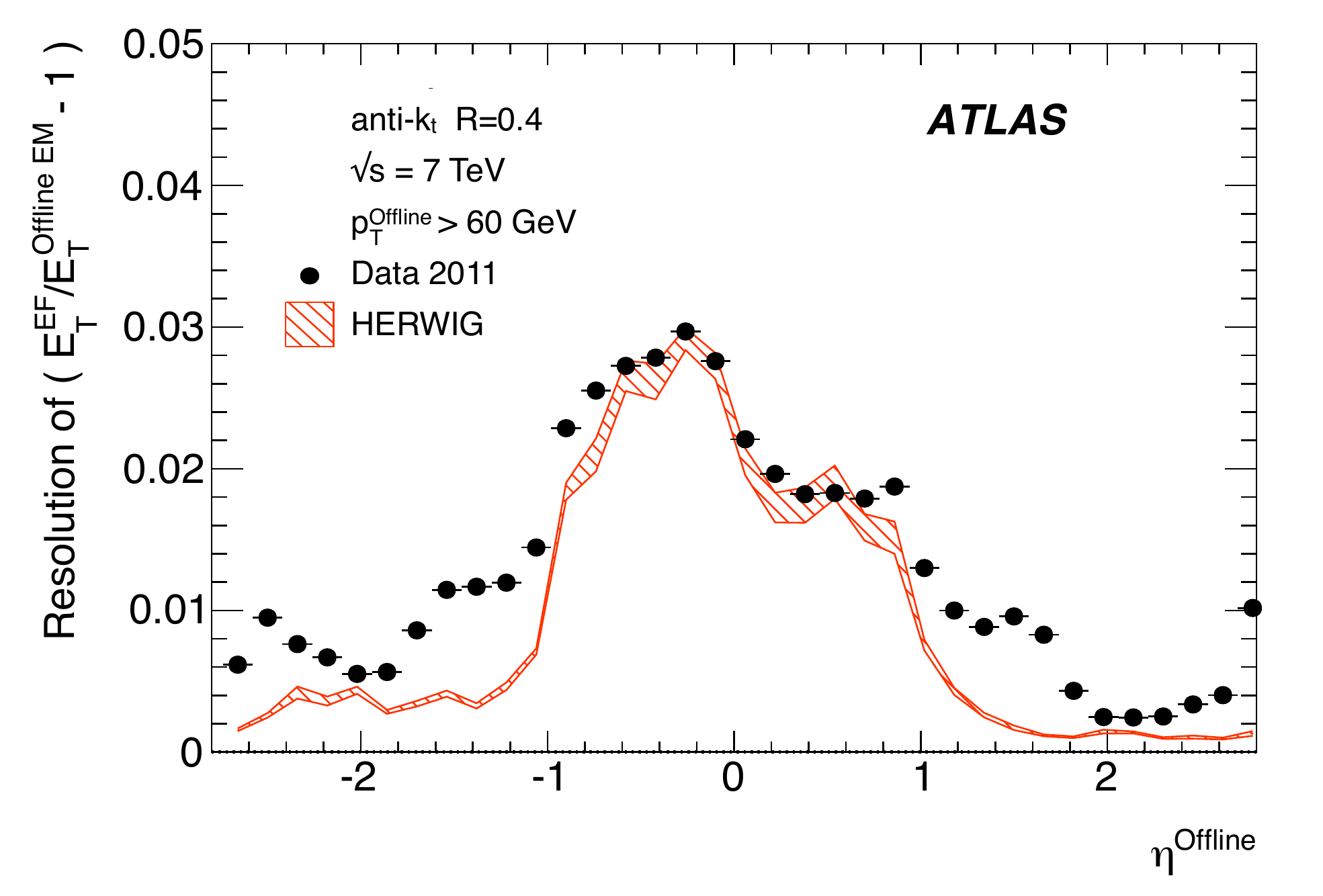}}
  \subfigure[]{\includegraphics[width=0.5\textwidth]{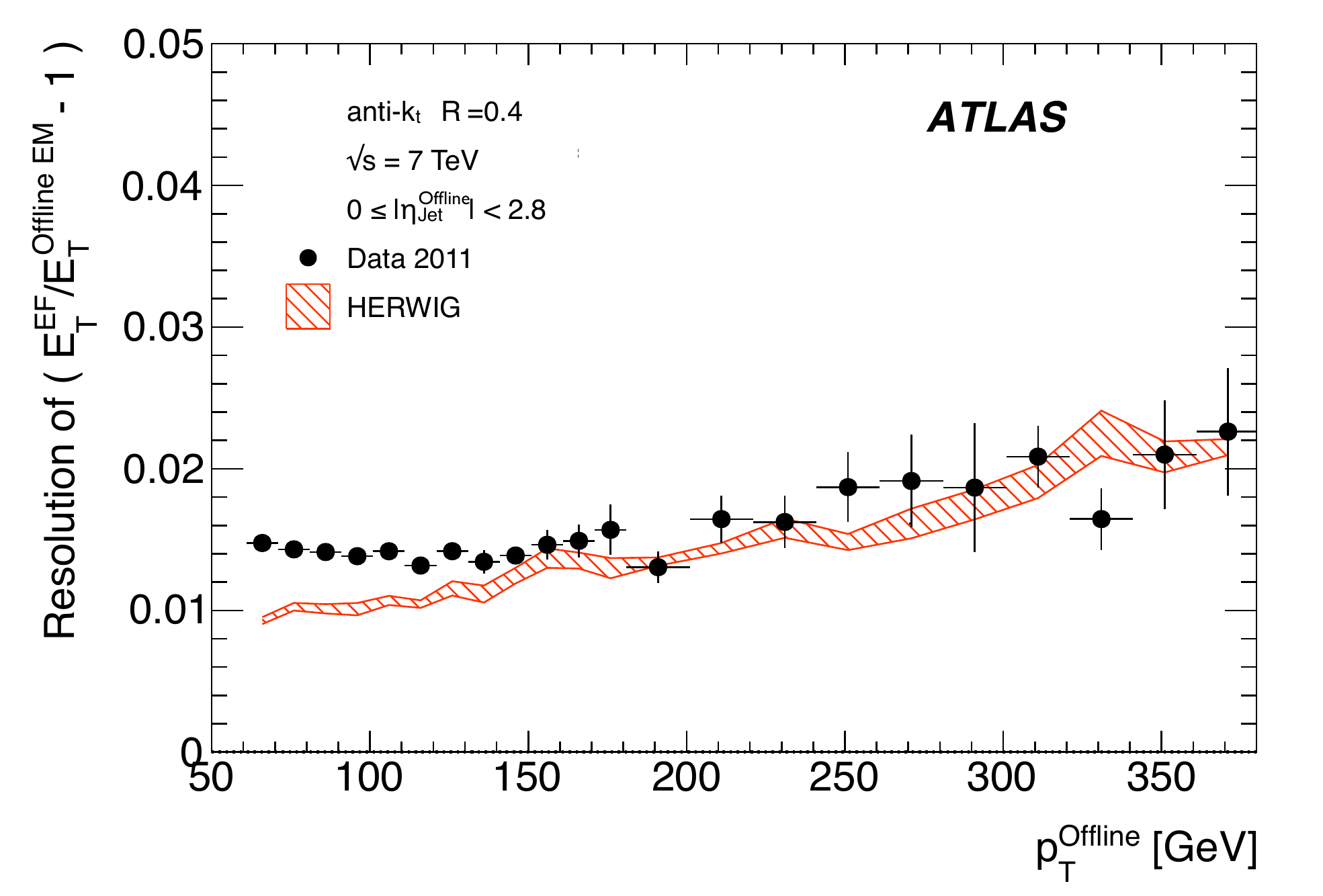}}
  \caption{ The resolution for the EF trigger jets with respect to
    offline jets at the EM energy scale: (a)~as a function of the
    offline jet $\eta$ for jets with offline \pt$>60$\,\GeV; (b)~as a function
    of the offline jet \pt\ at the hadronic energy scale for jets in
    the range $|\eta^{\mathrm {Offline}}|<2.8$. \fullerror }
  \label{fig:efresemfigtwo}
\end{figure}

The resolutions as a function of the offline jet \pseudorapidity\ and
\pt\ are presented in Figure~\ref{fig:efresemfigtwo} for all jets with
$\ptoffline>60$\,\GeV.  Resolutions of approximately 1\%, 2\% and 3\%
are seen for $|\eta|>1$, and the positive and negative
\pseudorapidity\ regions in the barrel respectively. The resolution in
the central barrel region is approximately the same as that with respect to
fully corrected offline jets, but around 1\% better in the region
$1<|\eta|<2$. As a function of \pt, the resolution degrades slightly
from approximately 1.5\% at low \pt\ to 2\% at higher \pt. The
simulation describes the resolution reasonably well in the barrel
region, but predicts significantly better resolution than in the data for
the regions with $|\eta|>1$. The asymmetry between the positive and
negative $\eta$ barrel regions observed when comparing with EM+JES jets is also
observed here.

The difference in the offsets observed between the EM and EM+JES jets
serves to illustrate the size of the correction applied during the 
JES calibration and suggests
that, should the same correction be applied online, the correspondence
between offline and trigger reconstructed jets would be better than a
few percent with resolutions of better than 3\%.  Since 2012
the JES calibration has been applied to the trigger jets online.

\subsection{Jet trigger reconstruction efficiency}
\label{section:efficiencies}

To understand the performance of the trigger in more detail, the
trigger efficiency versus \ET\ has been studied for all the major
inclusive trigger chains, comparing once again to fully calibrated
offline jets at the EM+JES scale. Here \ET\ is used as the variable 
of merit, since the trigger selects jets based on \ET\ rather than the 
\pt\ used in physics analyses.

Measuring the efficiency in data requires events to be selected with
an independent reference trigger which is unbiased with respect to the
trigger being studied.  The reference trigger is usually chosen to
have a lower \ET\ threshold than the specific trigger being evaluated, 
for which the \ET\ region to be studied must lie well within the plateau
region of the reference trigger.  To study the very low \ET\ triggers,
triggers selecting events randomly at L1 or pass through events
without additional trigger selection have been used.

\subsubsection{The single inclusive jet trigger efficiency}

\begin{figure}[thp]
 
  \subfigure[]{
    \includegraphics[width=0.48\textwidth]{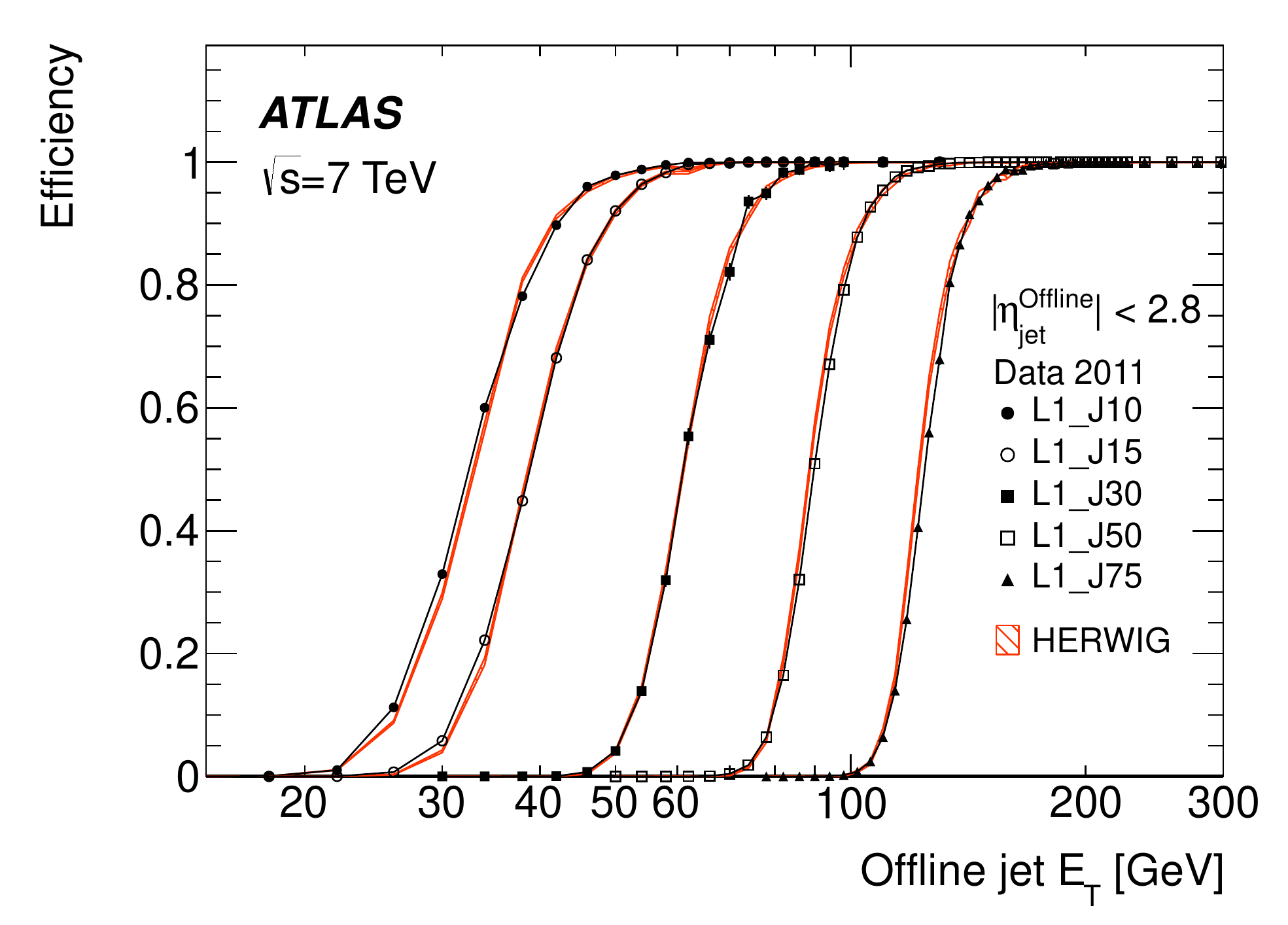}}
  \subfigure[]{
    \includegraphics[width=0.48\textwidth]{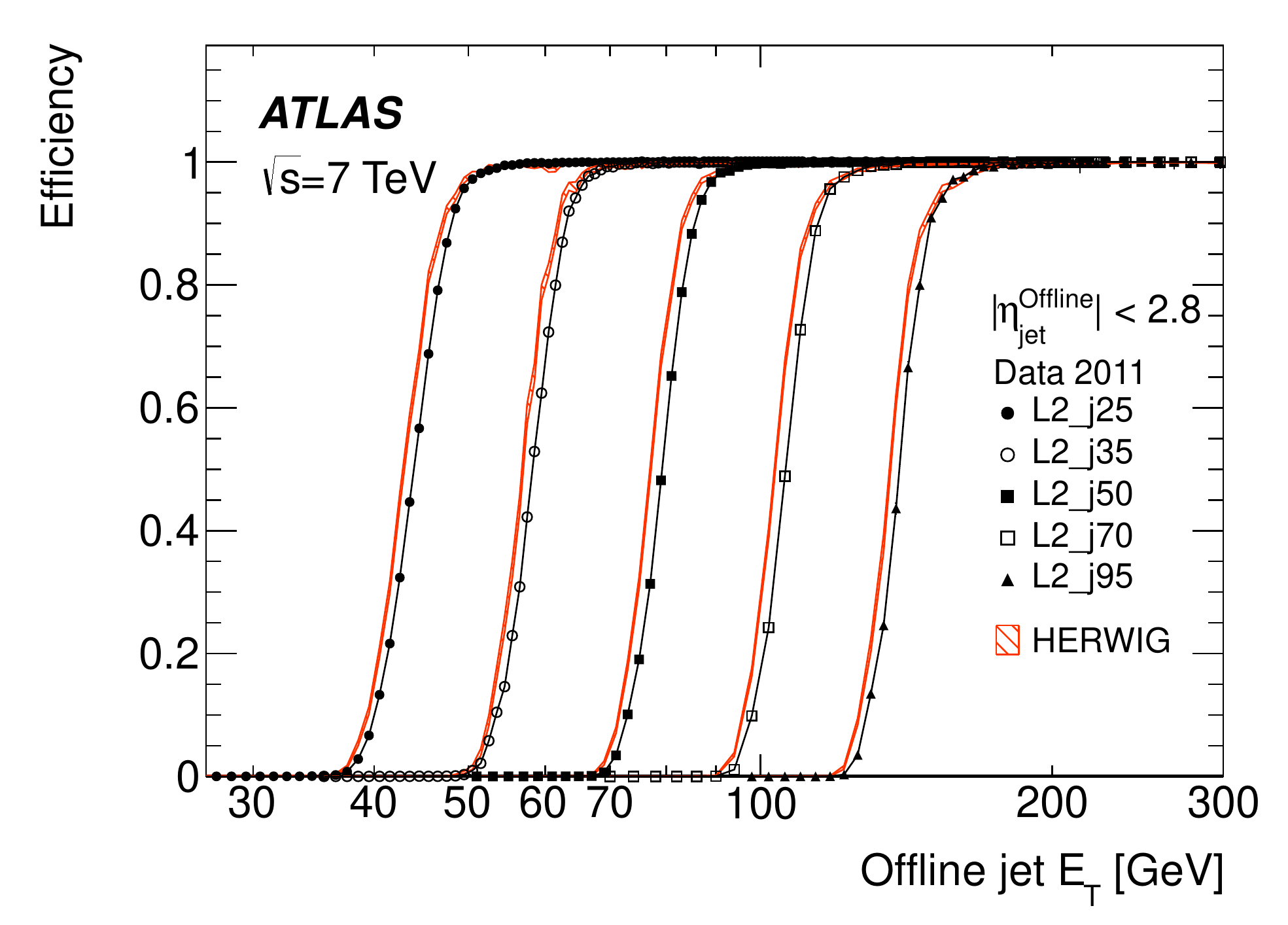}}
  \subfigure[]{
    \includegraphics[width=0.48\textwidth]{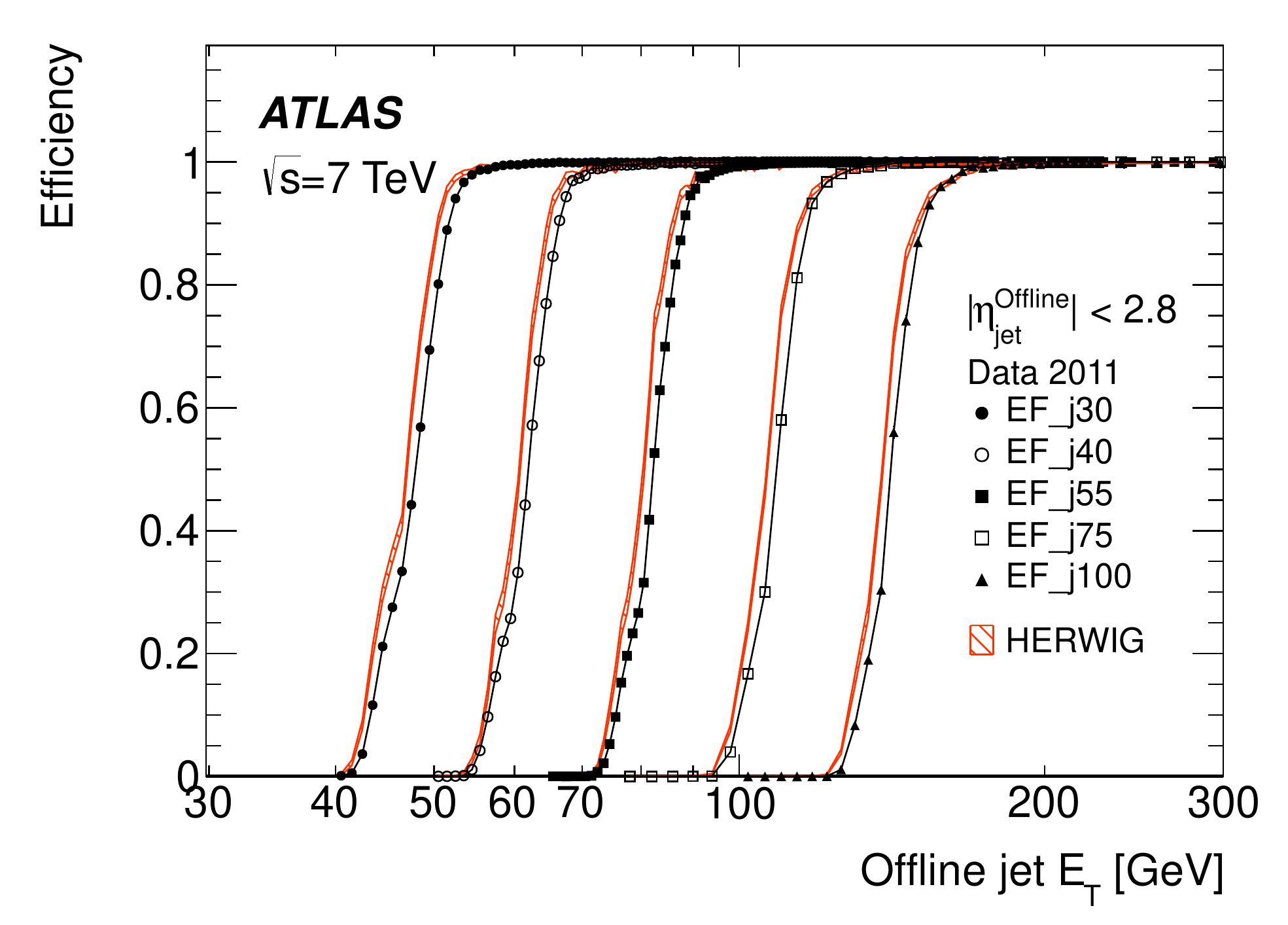}}
  \caption{ The efficiency as a function of offline jet \et\ for various
    single inclusive jet triggers. Shown are the efficiency for data,
    and for the \herwig\ simulated sample for: (a)~L1; (b)~L2; and (c)~the EF
    triggers. For data, the efficiency is computed with respect to
    events taken by an independent trigger that is 100\% efficient in
    the relevant region. \fullerror }
  \label{fig:eff}
\end{figure}

The efficiency curves for a selection of single inclusive jet triggers
as a function of \ET\ are shown in Figure~\ref{fig:eff} for data and
simulation, for each of the three trigger levels.  Relative trigger
efficiencies are shown: the L2 trigger requires that a jet has already
satisfied the L1 trigger in the chain; similarly an EF trigger
requires that L2 has been satisfied.  The rising edges for the L2 and
the EF selection are considerably sharper than for the corresponding
L1 selection due to the improved \ET\ resolution in the HLT.  At all
levels, any discrepancies between data and simulation are of the order
of a few percent close to the full efficiency region.

In Section~\ref{sec:hltres}, the Monte Carlo simulation was seen to
predict smaller offsets than the data at nearly all \pt. The result of
this is that the trigger jets in the \herwig\ sample would have a correspondingly
higher \et\ than those from the data, and so the trigger would be
expected to turn on earlier than the data. 

\begin{figure}[tp]
  \subfigure[]{
    \includegraphics[width=0.48\textwidth]{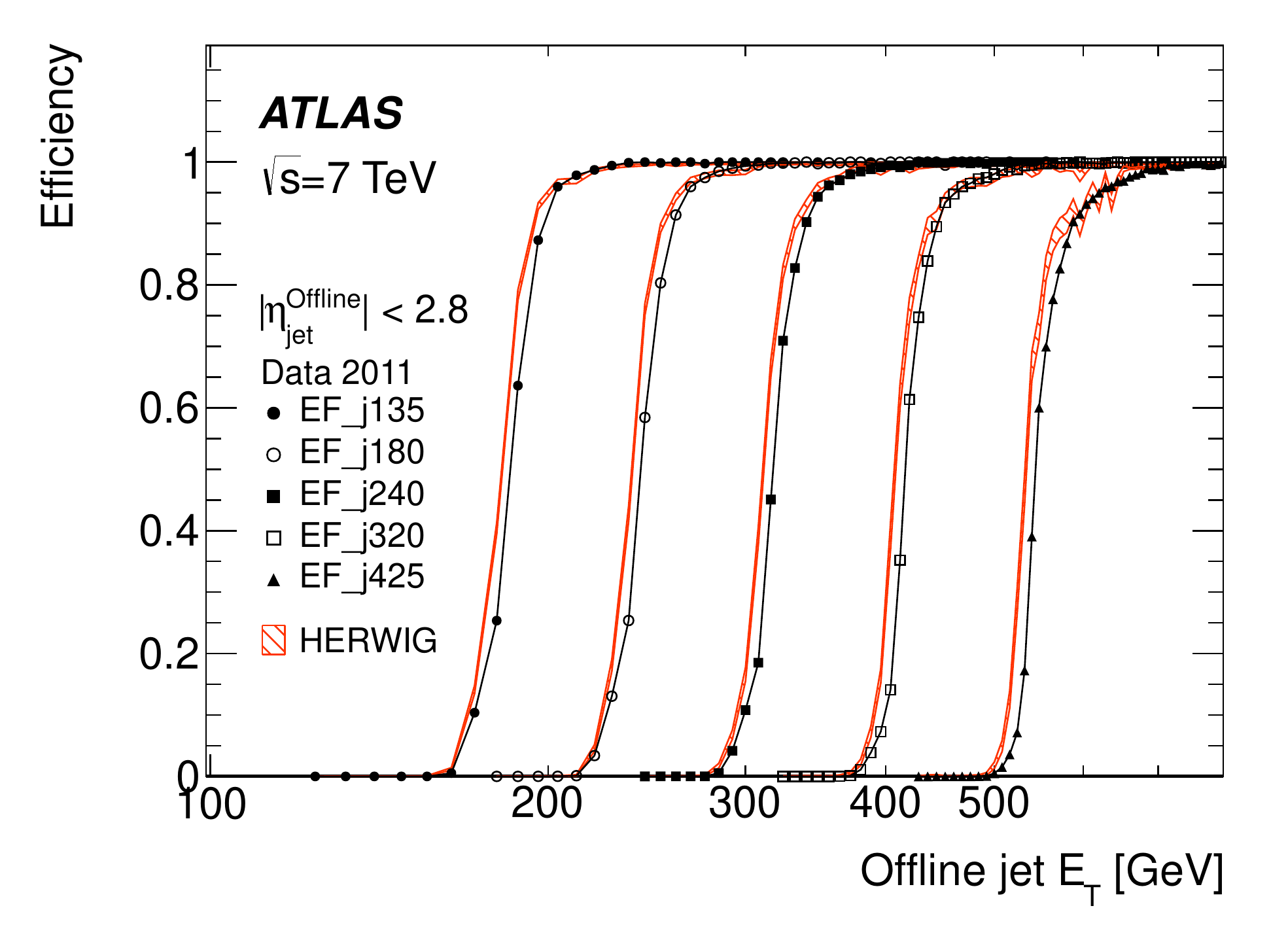}
    \label{fig:eff_morea}}
  \subfigure[]{
    \includegraphics[width=0.48\textwidth]{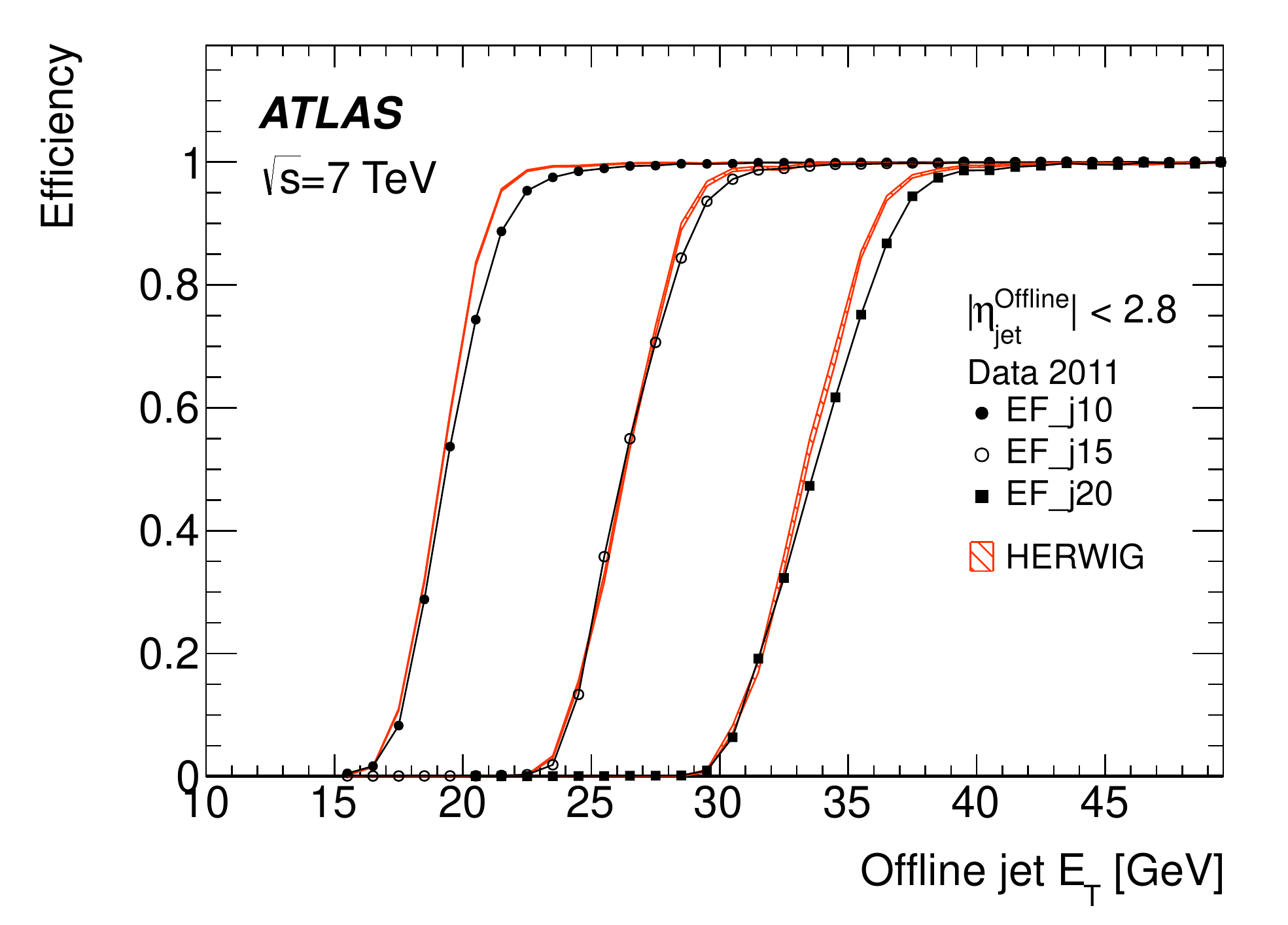}
    \label{fig:eff_moreb}}
  \caption{ The efficiency for various EF triggers as a function of
    offline jet \et. Shown are the efficiencies for data and the
    \herwig\ simulated sample for the: (a) EF triggers seeded by L2\_j95 and
    L1\_J75; (b) EF triggers seeded by a random trigger at L1 and
    passed through L2. \fullerror }
  \label{fig:eff_more}
\end{figure}

The efficiencies as a function of \ET\ for additional EF triggers which 
ran in 2011 are shown in Figure~\ref{fig:eff_more} for data and
simulation. The high \ET\ threshold
triggers are shown in Figure~\ref{fig:eff_morea}.  
The efficiencies as a function of \ET\ for
the EF triggers seeded by a random trigger at L1 which are passed 
through L2 are shown in Figure~\ref{fig:eff_moreb}.  
Since the random triggers
require no jet selection at either L1 or L2, these EF triggers are
unaffected by the coarse resolution and the less steep rising edge
seen for the low threshold jet triggers at L1. This allows the
triggers to reach their full efficiency at a lower \et\ than is possible
for the chains seeded by an L1 jet trigger. In this case, the lowest
threshold trigger, with a transverse energy requirement of 10\,\GeV, is fully
efficient by 25\,\GeV.

\begin{figure}[thp]
  
  \subfigure[]{
    \includegraphics[width=0.48\textwidth]{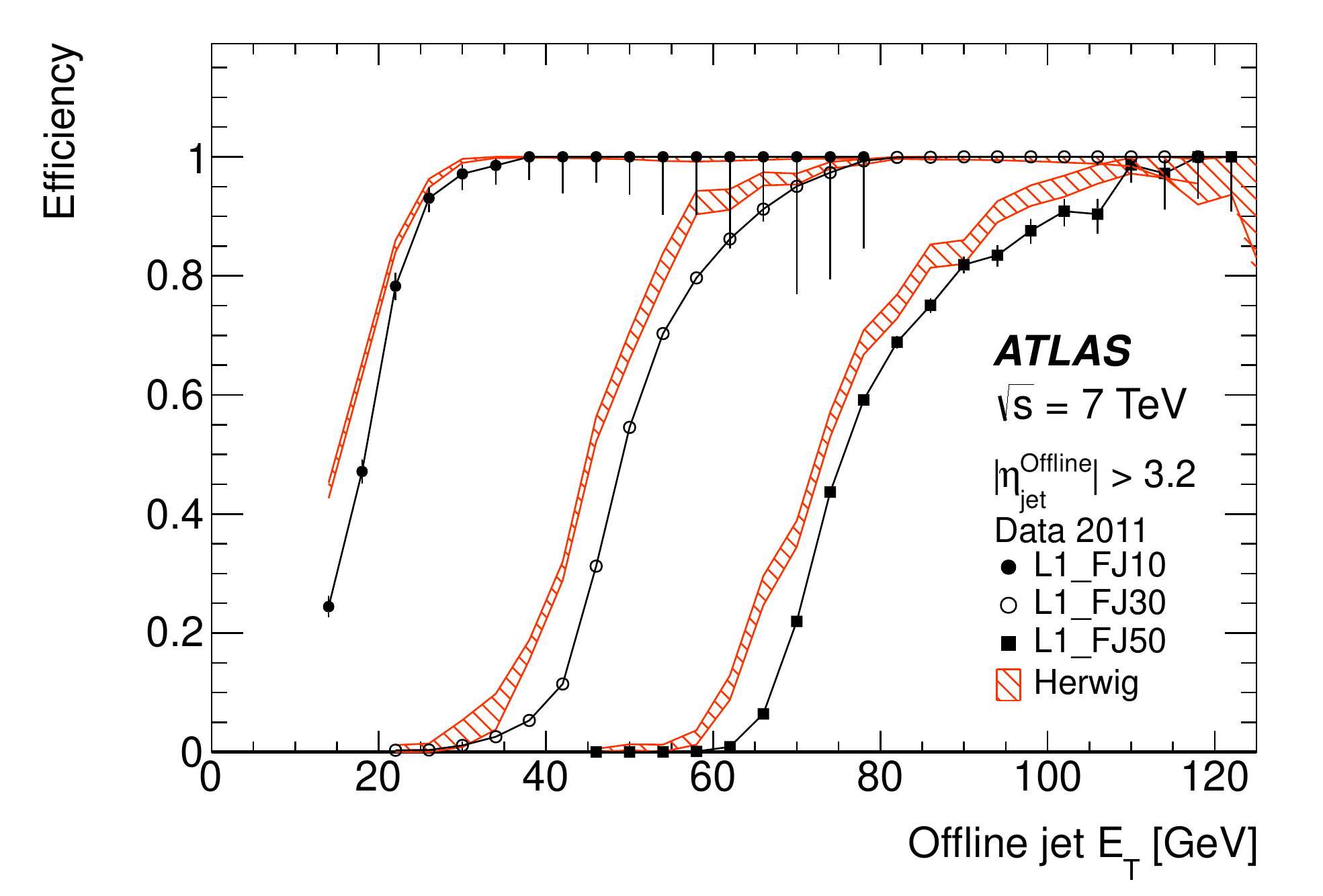}}
  \subfigure[]{
    \includegraphics[width=0.48\textwidth]{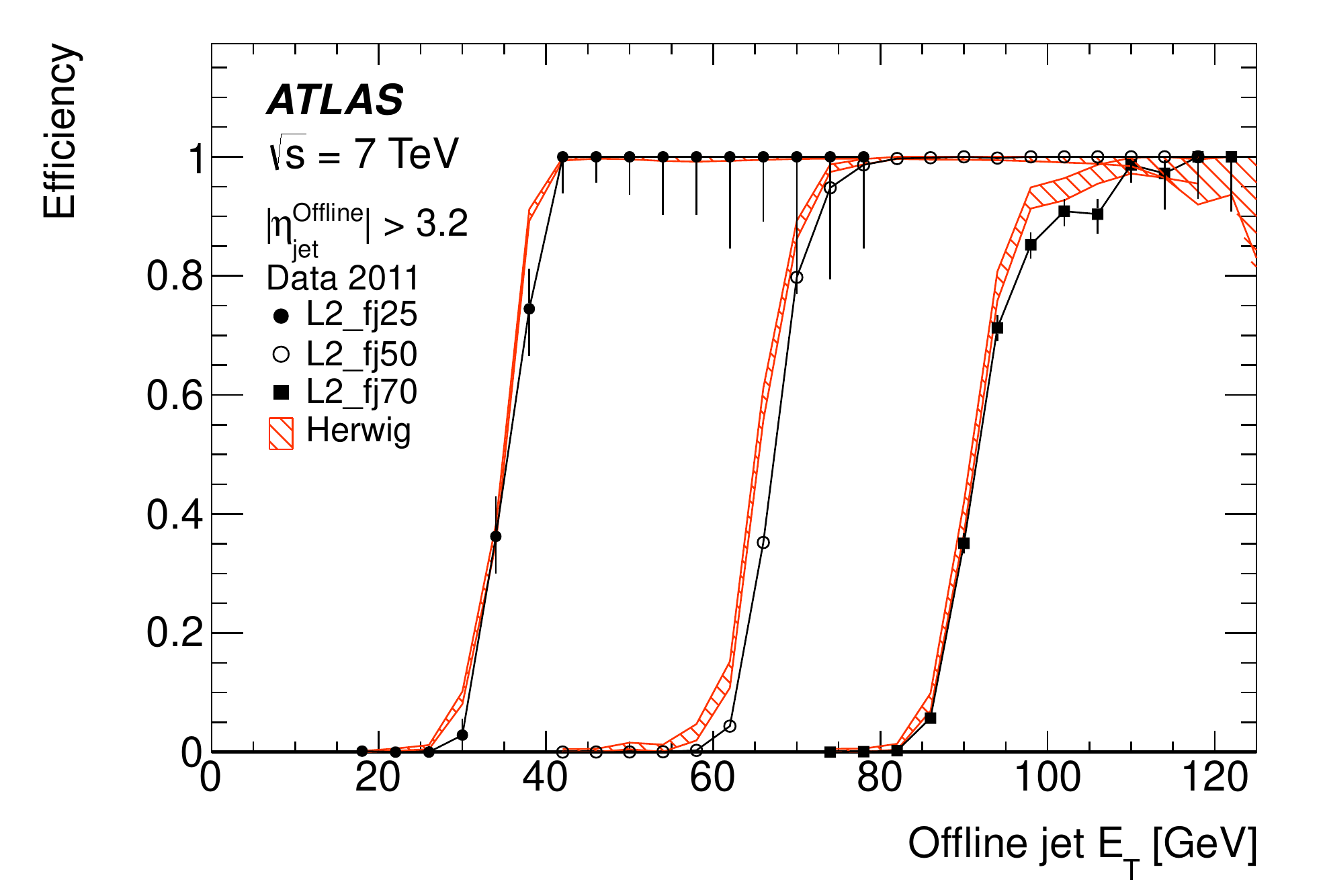}}
  \subfigure[]{
    \includegraphics[width=0.48\textwidth]{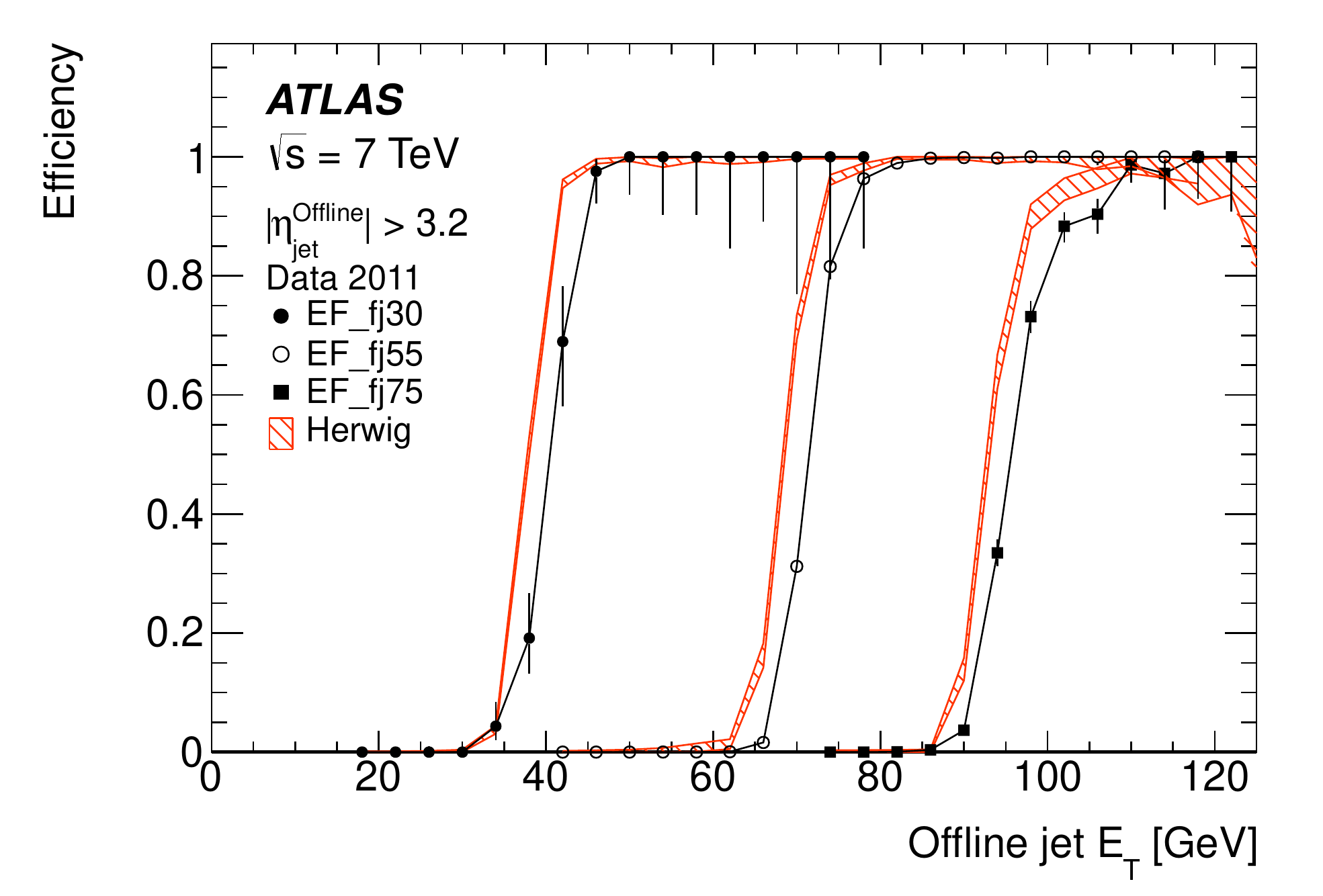}}
  \caption{ The efficiency for various forward jet triggers in data
    and the \herwig\ simulated sample as a function of offline jet
    \ET\ for: (a)~L1\_FJ10, L1\_FJ30 and L1\_FJ50; (b)~L2\_fj25, L2\_fj50 and L2\_fj70;
    (c)~EF\_fj30, EF\_fj55 and EF\_fj75. \fullerror }
  \label{fig:eff_fwd}
\end{figure}

Figure~\ref{fig:eff_fwd} shows the efficiency as a function of
\ET\ for L1, L2 and EF jets in the {\em forward region}, defined as having a
\pseudorapidity\ $|\eta|> 3.2$. However, in order for these jets to be
fully contained in the forward calorimeter the offline $|\eta|$ is
required to be in the range $3.6 \leq|\eta|\leq 4.8$.  The agreement between data and
simulation is worse in the forward region than for central jets. This is
related to the smaller offsets seen in simulation 
in Section~\ref{sec:efresfj} when compared to the data. 
This results in the trigger turning on at slightly lower \et\ in the 
simulation than in the data.  

\subsubsection{Trigger efficiency versus pseudorapidity}

The offset and resolution of the trigger, and the underlying
kinematics, each affect the rising edge of the trigger efficiency as
it increases towards plateau.

The resolution and offset of the trigger jets have been shown to vary
significantly with \pseudorapidity. This has a significant effect on
the trigger efficiency and introduces a strong dependence on the
\pseudorapidity, of both the position of the midpoint and the sharpness of the rising
edge of the trigger, and of the \ET\ at which the trigger reaches its maximal
plateau efficiency.

\begin{figure}[thp]
  
  \subfigure[]{
    \includegraphics[width=0.48\textwidth]{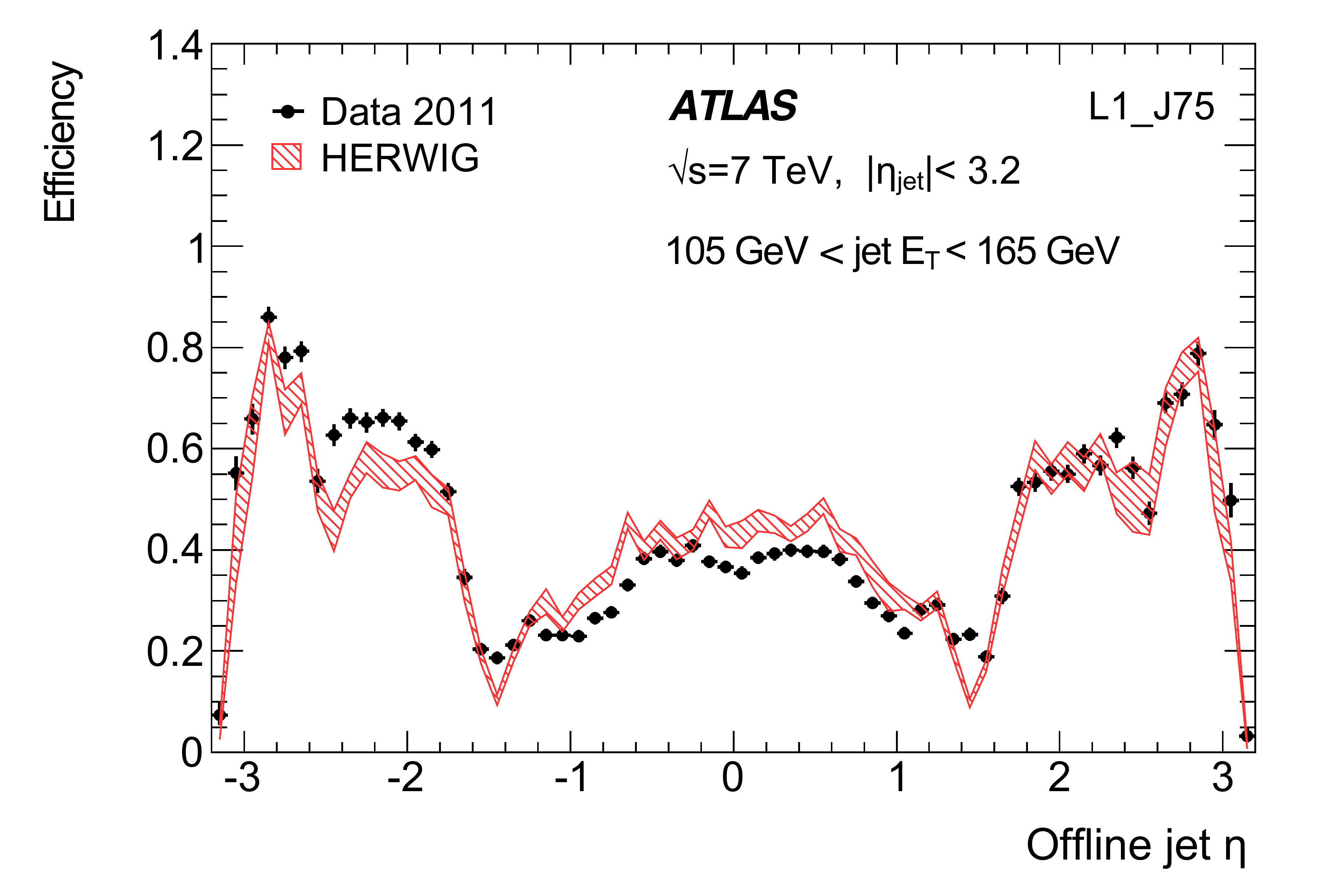}}
  \subfigure[]{
    \includegraphics[width=0.48\textwidth]{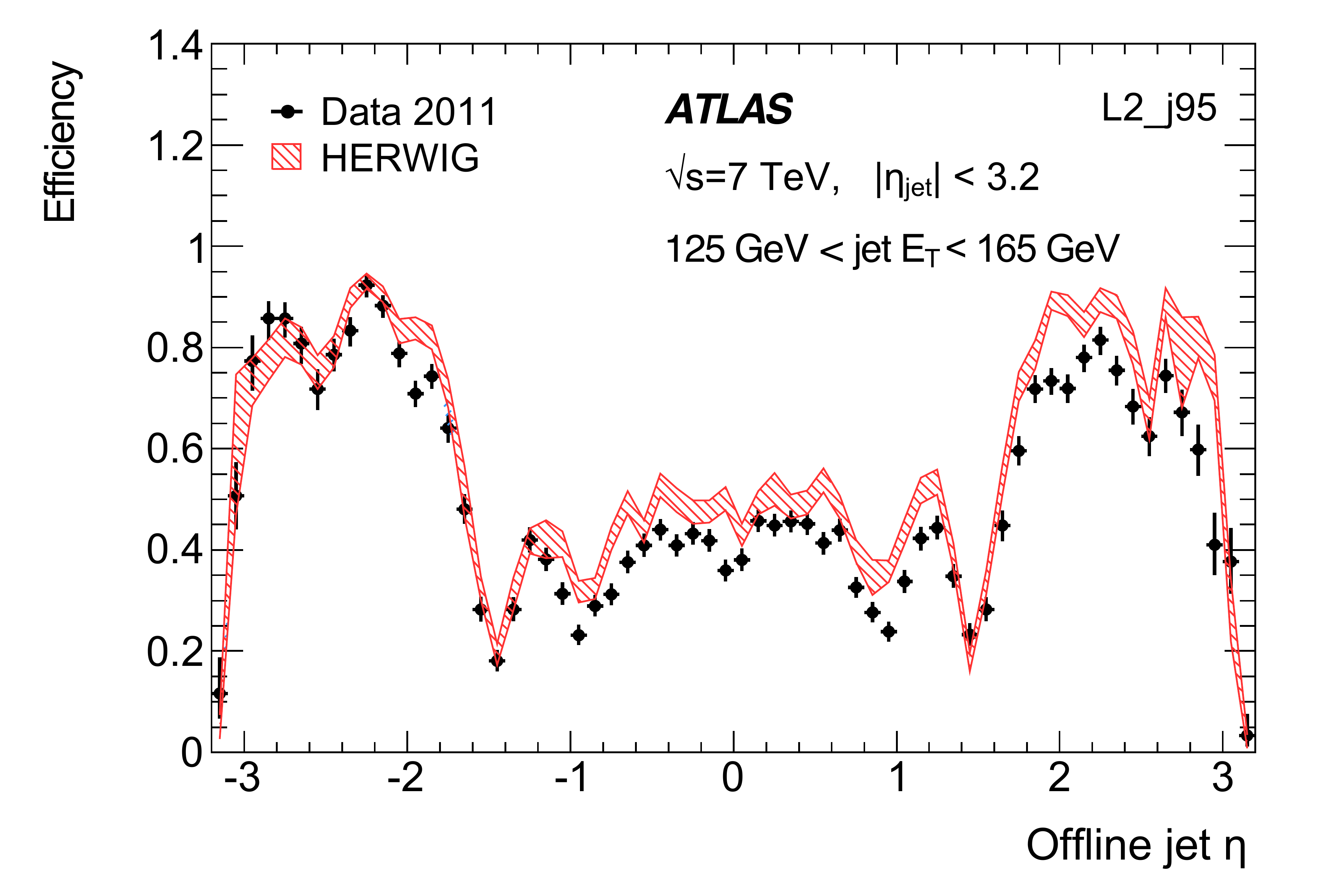}}
  \subfigure[]{
    \includegraphics[width=0.48\textwidth]{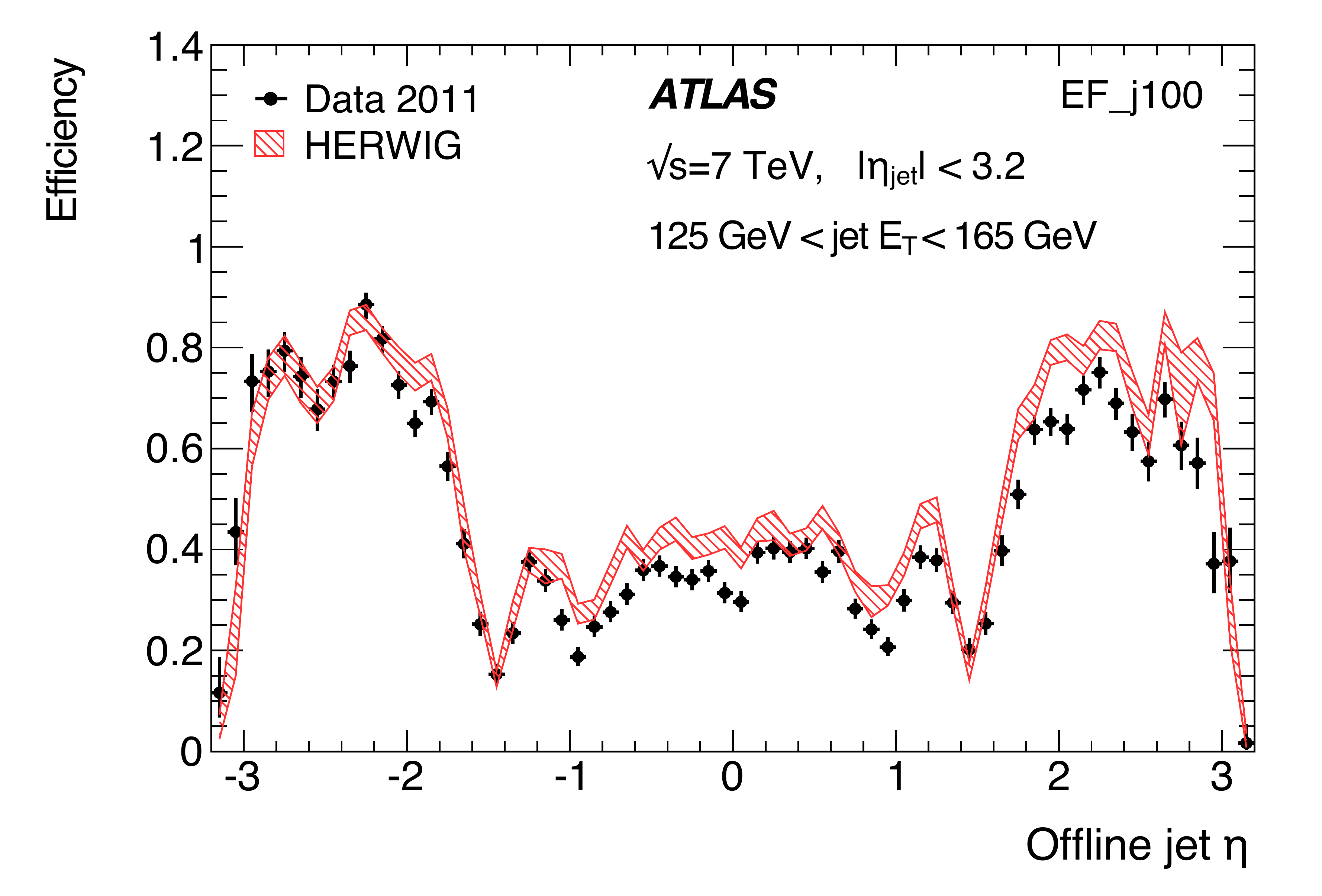}}
  \caption{The single inclusive jet trigger efficiency integrated over
    the \ET\ of the rising edge of the trigger, as a function of
    \eta\ for the triggers: (a)~L1\_J75; (b)~L2\_j95; (c)~EF\_j100.
    The data are shown as the solid points with error bars with
    the simulated sample shown as a shaded band. \statonly }
  \label{fig:eta}
\end{figure}

To quantify the behaviour of the trigger efficiency in the vicinity of
the rising edge as a function of \pseudorapidity, it is informative to
study the efficiency, differential in \eta, but integrated over the
\ET\ interval defined by the 1\% and 99\% efficiency points of the 
sample as a whole.  Figure~\ref{fig:eta} shows this integrated single inclusive
jet trigger efficiency, as a function of \eta\ for the trigger chain
consisting of thresholds of 75, 95 and 100\,\GeV\ at L1, L2 and the EF,
respectively.  A lower efficiency is seen near $|\eta| = 1.5$,
corresponding to the crack region between the barrel and endcap
calorimeters where the measured energy in the calorimeter will be lower.  
These variations are related to the detector geometry
and detector conditions, and are very strongly correlated with the
offsets observed in the previous section, where for instance, the
larger (negative) offset seen in the barrel results in fewer jets
passing the trigger threshold. Related to what was seen in
Section~\ref{sec:hltres}, a small asymmetry is observed between the
positive and negative barrel regions.

\subsubsection{The \multijet\ trigger efficiency}
\label{section:multijet_efficiencies}

A \multijet\ trigger requires that $N$ jets in the event pass certain
\ET\ thresholds. For the triggers considered in this study, all jets
must be reconstructed in the central part of the calorimeter
($|\eta|<2.8$).  When searching for final states with large jet
multiplicities in the high energy environment of the LHC, the
requirement of several jets means that a \multijet\ trigger is more
likely to remain unprescaled than its single jet counterpart.

However, the principal disadvantage of a \multijet\ trigger is an
overall loss in efficiency due to limitations in both transverse
energy and angular resolution at L1 and L2. This loss in efficiency
is compounded by the jet multiplicity requirement in the trigger, but 
is less significant for offline jets when they  are geometrically
isolated. The primary reasons for these 
inefficiencies  in the trigger are the use of the square sliding window 
and reduced granularity at L1, and the limited RoI size used for the 
reconstruction of jets at L2.

\Multijet\ triggers have been used in signal selection and \multijet\ background estimation 
in searches for the Higgs boson, supersymmetry, and other, beyond-the-SM, 
processes~\cite{multijet1, multijet2, multijet3}.
During 2011, \multijet\ triggers requiring between three and six jets were available, with \ET\ thresholds
at the EF ranging from 30 to 100\,\GeV.

For a \multijet\ trigger efficiency, when requiring a signature containing $N$ jets with a single 
common threshold, 
the efficiency will essentially be determined by the efficiency for triggering on the $N$-th leading 
jet in \et. For simplicity, only \multijet\ triggers with a single common threshold are considered 
here. For \multijet\ efficiencies, it is therefore more useful to determine the {\em event level 
efficiency}, determined as a function of the \et\ of this $N$-th jet.

\begin{figure}[thp]
  \begin{center}
    \subfigure[]{
      \includegraphics[width=0.49\textwidth]{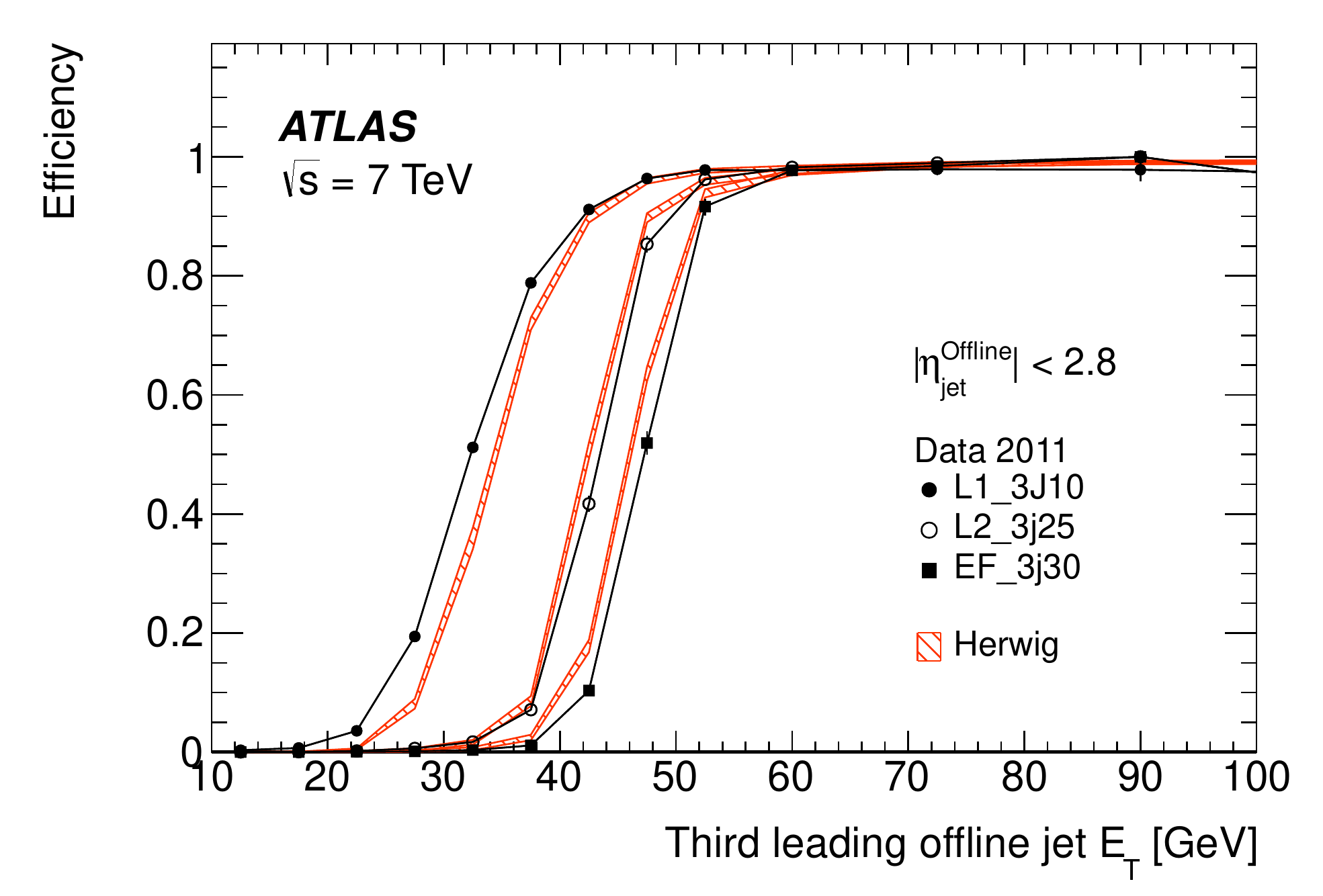}}
    \subfigure[]{
      \includegraphics[width=0.49\textwidth]{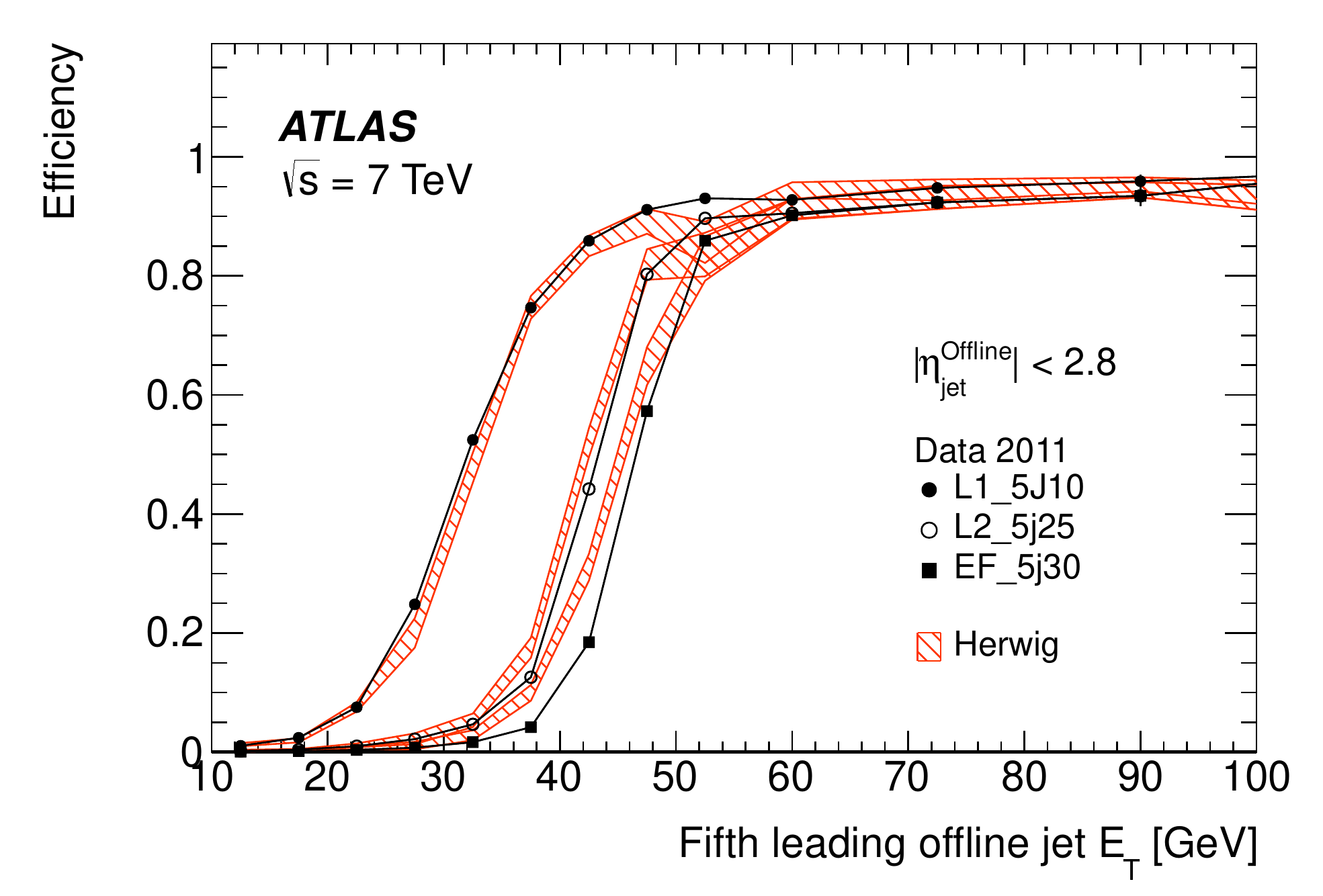}}
    \caption[multijetPlots]{ The efficiency for the three-jet and five-jet
      chains with a 30\,\GeV\ threshold at the EF, as a function of: (a) the
      third jet \et\ for the three jet chains; and (b) as a function of the
      fifth jet \et\ for the five jet trigger chains. Shown are the
      absolute trigger efficiencies: the L2 efficiency also includes that 
      for L1 and the EF efficiency includes that from both L1 and L2.
      \fullerror }
    \label{fig:MJets}
  \end{center}
\end{figure}

The characteristics of \multijet\ triggers are illustrated in Figure
\ref{fig:MJets}, which shows the efficiency for the lowest \ET, 
three jet, and five jet trigger chains.  The reference triggers were chosen
to have a combination of a lower jet multiplicity and a lower
\ET\ requirement, compared to the trigger being studied, so that they
are fully efficient over the rising edge of the trigger being studied.
For the three jet trigger chain, the reference trigger at L1 required
the event to pass either the random seeded, 10, 15, or 20\,\GeV\ EF
triggers, operating beyond their respective plateaux. For the three
jet chain at L2 and EF, the L1 threshold at 10\,\GeV\ was required, with
pass-through at L2 and EF. For the four jet and five jet trigger
chains the requirement of three EF jets above 30\,\GeV\ was used as the
reference trigger.  In contrast to the single inclusive jet trigger
analysis, no jet matching is applied from one level to the next, and
no jet isolation is imposed unless specifically stated.  When the jet
multiplicity requirement is increased from three to five, the plateau
efficiency decreases and the uncertainties on the simulated sample 
increase, due to the smaller Monte Carlo sample size.

In order to allow a very approximate quantitative comparison of the efficiency for a selection of
jet triggers with different multiplicities, a fit to
the efficiency distributions for four \multijet\ trigger chains 
has been performed and the relevant parameters extracted. A
sigmoid function was chosen to parameterise the efficiency,
\begin{equation}
  \mbox{$\epsilon$}(\et) = c_3 + (c_0 - c_3)
  \left[
    1+\exp\left(-\frac{{E_{\mathrm{T}}} - c_1} {c_2}\right)
  \right]^{-1}
  \label{sigmoid}
\end{equation}
where $c_0$ is the plateau efficiency in percent, $c_1$ is the
{\em midpoint} of the rising edge, in \GeV, $c_2$ -- also in \GeV\ -- is related to
the width or {\em sharpness} of the rising edge, and $c_3$ is the
residual efficiency in the region before the trigger begins to
turn on.

\begin{figure}[thp]
  \begin{center}
    \includegraphics[width=0.49\textwidth]{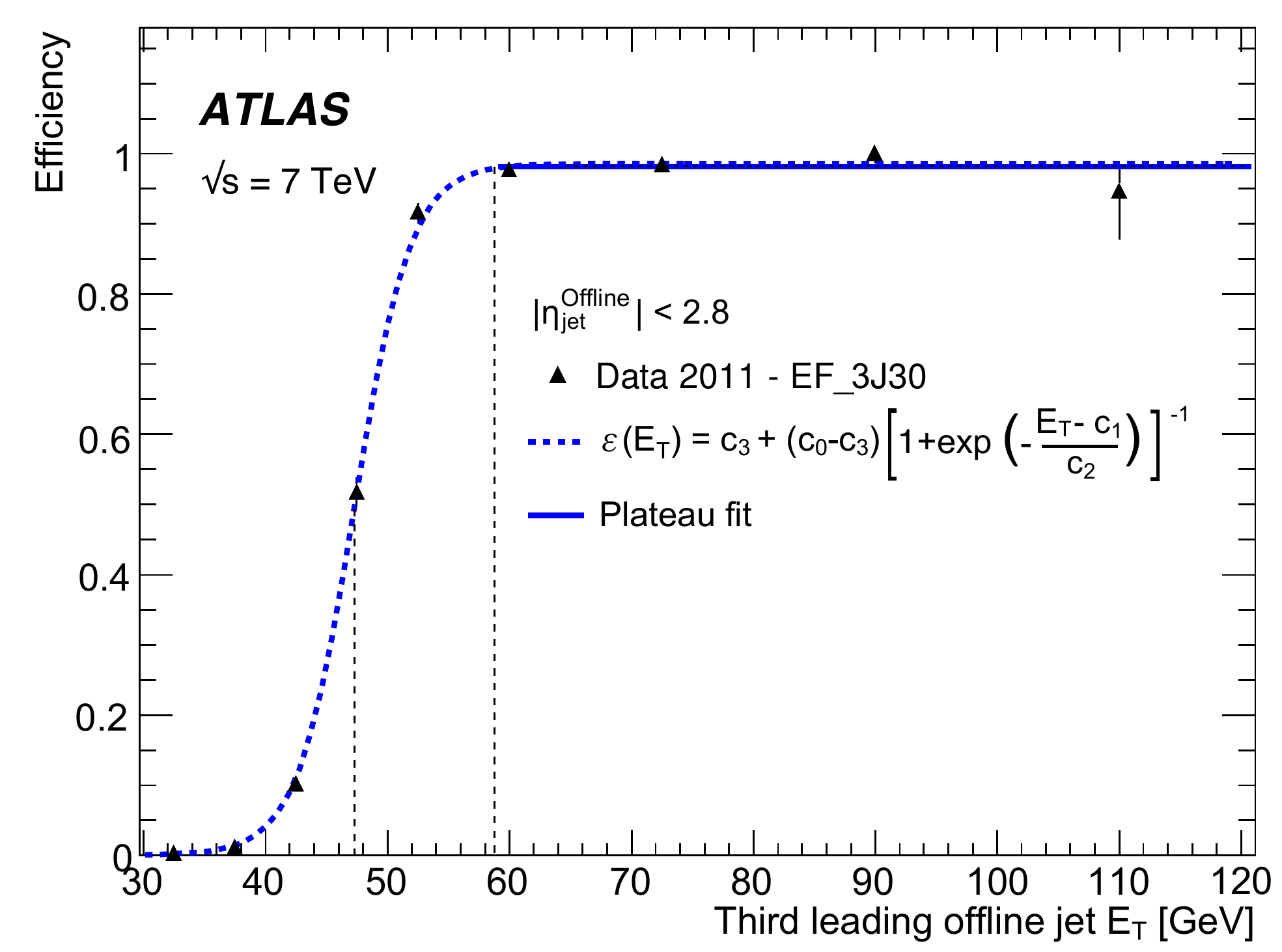} 
    \caption[sigmoidPlot]{The sigmoid fit to the rising edge of the efficiency and the fit purely to 
      the plateau region for the EF\_3j30 trigger, without any jet isolation requirement. 
      The parameter $c_0$ represents the plateau efficiency, $c_1$ represents the midpoint of the 
      rising edge, $c_2$ is related to the sharpness of the rising edge, and $c_3$ is 
      the efficiency prior to the rising edge. The horizontal solid blue line indicates the plateau 
      efficiency.  The vertical dashed lines indicate the rising edge midpoint ($c_1$), and the 
      start of the plateau ($c_1 + 5c_2$).}
    \label{fig:exSig}
  \end{center}
\end{figure}

The plateau efficiency was also determined using the parameters from
the sigmoid fit and, additionally, fitting a constant to the region $\ET> c_1 +
5c_2$, corresponding approximately to the region where the efficiency is 
above 99\% of the ultimate value.  This provides
an alternative determination of the plateau efficiency. Figure
\ref{fig:exSig} shows example fits for the EF\_3j30 chain.

Table \ref{tab:MultiEffs} displays the plateau efficiency and
parameters describing the efficiency at each trigger level, for the
lowest \ET\ single, three, four and five jet trigger chains. This
table highlights the loss in plateau efficiency with increasing jet
multiplicity, the consistency of the rising edge midpoint between
different jet multiplicities, and the general reduction of sharpness 
of the rising edge for higher multiplicities.

\begin{table*}
  \begin{center}
    \caption{
      \label{tab:MultiEffs}
      The plateau efficiency from the linear fit, and the midpoint \ET\ and
      sharpness of the rising edge from the sigmoid fit, for the single, 
      three, four, and five jet trigger chains,
      each with an EF threshold of 30\,\GeV\ and without offline jet
      isolation.  The plateau efficiency decreases with increasing jet
      multiplicity.  }
    \vspace{2mm}
    \begin{tabular}{lS[table-format=4.2]lS[table-format=4.2]lS[table-format=3.2]l}
      \hline 
      Trigger  & \multicolumn{2}{c}{Plateau [\%]} & \multicolumn{2}{c}{Midpoint [\GeV]}  & \multicolumn{2}{c}{Sharpness [\GeV]}   \\
      \hline 
      L1\_J10 & 98.00 & $\pm$ 0.04 & 30.77  & $\pm$ 0.04 & 4.10 & $\pm$ 0.03\\
      L2\_j25 & 99.65 & $\pm$ 0.02 & 43.01  & $\pm$ 0.01 & 1.94 & $\pm$ 0.01\\
      EF\_j30 & 99.75 & $\pm$ 0.02 & 47.09  & $\pm$ 0.02 & 1.94 & $\pm$ 0.01\\
      \\
      L1\_3J10 & 97.3  & $^{+~ 0.3}_{-~ 0.4}$  & 32.0 &$\pm$ 0.1~~ & 2.92 & $\pm$ 0.03\\
      L2\_3j25 & 98.6  & $^{+~ 0.4}_{-~ 0.5}$  & 43.6 &$\pm$ 0.1~~ & 2.78 & $\pm$ 0.06\\
      EF\_3j30 & 98.1  & $^{+~ 0.5}_{-~ 0.6}$  & 47.3 &$\pm$ 0.1~~ & 2.30 & $\pm$ 0.07\\
      \\
      L1\_4J10 & 95.2 & $\pm$ 0.1  & 30.20 &$\pm$ 0.02 & 3.93 & $\pm$ 0.02\\
      L2\_4j25 & 95.0 & $\pm$ 0.1  & 41.98 &$\pm$ 0.02 & 3.06 & $\pm$ 0.02\\
      EF\_4j30 & 94.7 & $\pm$ 0.1  & 46.30 &$\pm$ 0.02 & 2.74 & $\pm$ 0.02\\
      \\
      L1\_5J10 & 93.4  & $\pm$ 0.3   & 31.50  & $\pm$ 0.04  & 3.71  & $\pm$ 0.02\\
      L2\_5j25 & 91.3  & $\pm$ 0.5   & 42.84  & $\pm$ 0.06 & 2.47  & $\pm$ 0.04\\
      EF\_5j30 & 91.1  & $\pm$ 0.5   & 46.56  & $\pm$ 0.07  & 3.17  & $\pm$ 0.04\\
      \hline
    \end{tabular}
  \end{center}
\end{table*}

The plateau efficiency decreases with increasing jet multiplicity 
because of the limitations of accurately reconstructing jets which are 
not well separated and discriminating 
between them at L1 and L2.

\begin{figure}[h] 
  \begin{center}
    \includegraphics[width=0.49\textwidth]{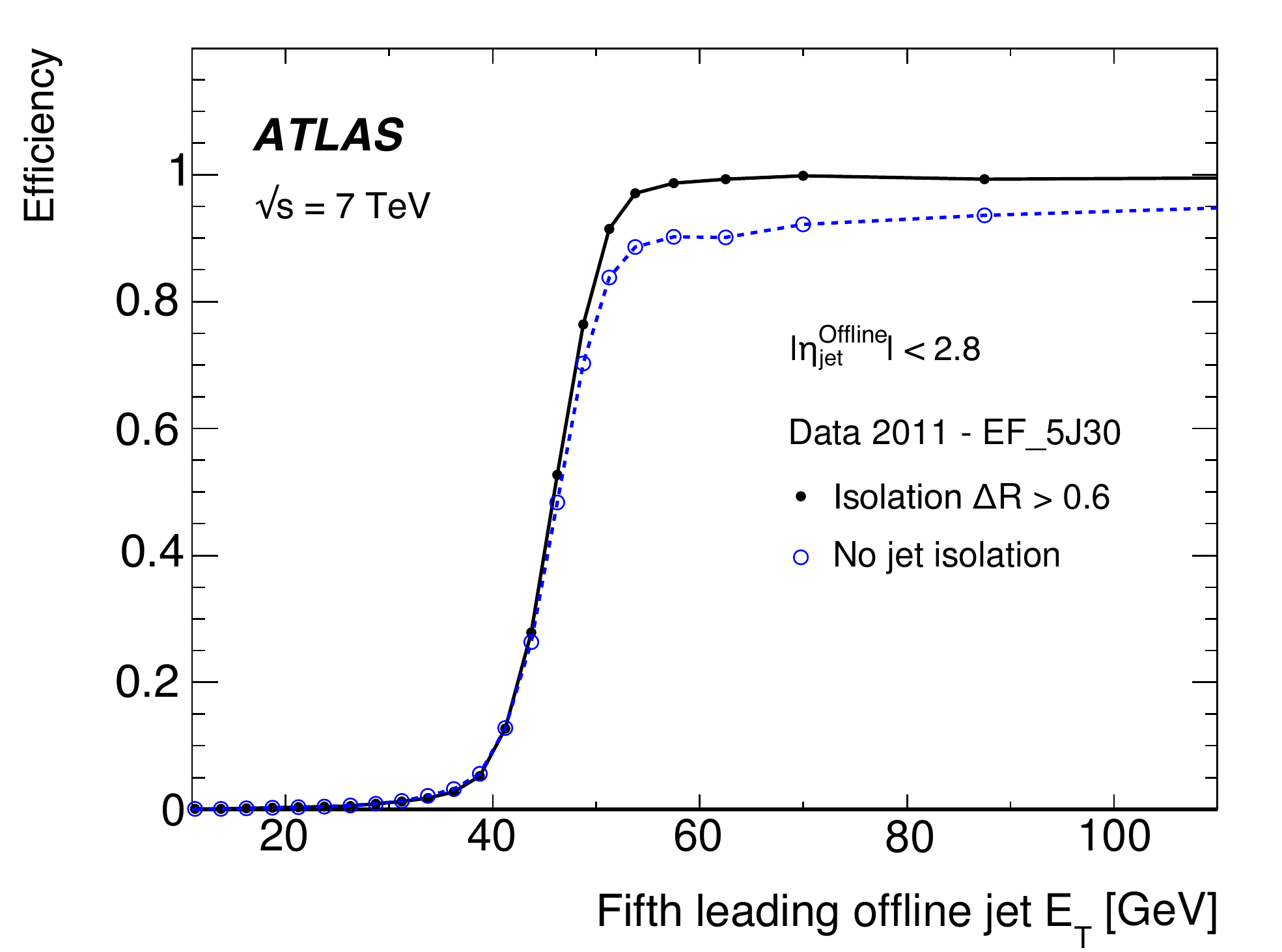} 
    \caption[multijetPlots]{ The efficiency as a function of the fifth
      jet \et\ for the five-jet EF trigger, where no jet isolation is
      required (dashed line) and where jet isolation is required
      (solid line). Shown are the absolute trigger efficiency,
      including both the L1 and L2 efficiencies.  \statonly }
    \label{fig:MJets2}
  \end{center}
\end{figure}

Figure~\ref{fig:MJets2} shows the absolute efficiency including the contributions from L1 and L2, for the five jet trigger
chain as a function of the fifth jet \et.  The solid curve in the
figure shows the same efficiency for events where the
leading offline jets are separated by a distance
$\Delta R > 0.6$ with respect to the corresponding closest jets.  In this case, 
the isolation requirement is applied only
to the four leading jets -- there is no requirement on the isolation
of the fifth leading jet.  The
difference observed with these different isolation requirements
clearly illustrates that this loss in efficiency 
is primarily due to issues in the reconstruction of poorly separated jets 
in the L1 and L2 triggers. 
This effect is shown quantitatively, for the three \multijet\ trigger
chains studied, in Table \ref{tab:MultiEffsIso}.

\begin{table*}[htp]
  \begin{center}
    \caption{ The plateau efficiency from the linear fit, and the
      midpoint \ET\ and sharpness of the
      rising edge from the sigmoid fit, for the three, four and five jet trigger chains,
      with an EF threshold of 30\,\GeV\ and with jet isolation applied
      between the $N$ leading offline jets.  By imposing jet isolation
      the loss in plateau efficiency at the EF is recovered.  }
    \label{tab:MultiEffsIso}  
    \vspace{2mm}
    \begin{tabular}{lS[table-format=4.2]lS[table-format=4.2]lS[table-format=3.2]l}
      \hline 
      Trigger  & \multicolumn{2}{c}{Plateau [\%]} & \multicolumn{2}{c}{Midpoint [\GeV]}  & \multicolumn{2}{c}{Sharpness [\GeV]}   \\
      \hline 
      L1\_3J10 & ~~99.3 & $\pm$ 0.1  &   31.94 & $\pm$ 0.04  &  2.84 & $\pm$ 0.02\\
      L2\_3j25 & ~~99.7 & $\pm$ 0.2  &   43.29 & $\pm$ 0.08  &  2.26 & $\pm$ 0.04\\
      EF\_3j30 & ~~99.5 & $\pm$ 0.2  &   46.96 & $\pm$ 0.09  &  2.09 & $\pm$ 0.05\\
      \\
      L1\_4J10 & ~~99.60 & $\pm$ 0.02 &  30.21 & $\pm$ 0.02  &  3.89 & $\pm$ 0.01\\
      L2\_4j25 & ~~99.64 & $\pm$ 0.03 &  42.15 & $\pm$ 0.01  &  2.47 & $\pm$ 0.01\\
      EF\_4j30 & ~~99.71 & $\pm$ 0.03 &  46.08 & $\pm$ 0.01  &  2.37 & $\pm$ 0.01\\
      \\
      L1\_5J10 & ~~99.4 & $\pm$ 0.1  &   31.32 & $\pm$ 0.02  &  3.61 & $\pm$ 0.01\\
      L2\_5j25 & ~~99.4 & $\pm$ 0.1  &   42.66 & $\pm$ 0.03  &  2.79 & $\pm$ 0.02\\
      EF\_5j30 & ~~99.5 & $\pm$ 0.1  &   45.98 & $\pm$ 0.04  &  2.66 & $\pm$ 0.02\\
      \hline 
    \end{tabular}
  \end{center}
\end{table*}

\section{Jet identification for $pp$ collisions performed by specialised jet triggers}
\label{section:physicsTriggers}
\newcommand{\offlineHT}{$\textrm{H}_{\mathrm T}^{\mathrm {offline}}$}

To further exploit the $pp$ data, jet triggers  designed 
to reconstruct specific physics signatures are used
in the ATLAS trigger.
In 2011 these included \HT\ triggers, cutting on the scalar
transverse energy sum of all jets, and triggers identifying jets with
large radii discussed below.

\subsection{\HT\ triggers}

In many searches for physics beyond-the-Standard Model (BSM), including 
Supersymmetry (SUSY) and other exotic physics signatures, 
the hard process gives rise to a final state containing energetic jets and 
a large missing transverse momentum. The selection adopted to discriminate 
the signal process from the background in such searches typically 
includes requirements on the
\ET\ and the scalar sum of transverse momenta of all selected physics
objects.  Missing transverse momentum triggers~\cite{Casadei:1331180} can be used in
such searches; however, an alternative approach is the use of
\HT\ triggers which  reconstruct the total scalar sum of jet
transverse energy (\HT) in an event at the EF.  The \HT\ triggers are
useful for physics analyses that study, or search for, events with large
overall \ET\ in the final state.  In this case, the requirement of large
\HT\ can help to control the trigger rate without requiring a very
energetic leading jet, although a leading jet with some \ET\ may still 
be required to seed the reconstruction. 
Because the resolution of the missing transverse momentum reconstruction 
in the trigger is poor for small values, using an \HT\ based trigger 
is a realistic alternative to using a missing transverse momentum trigger for final states where 
the  missing transverse momentum is small. 

The \HT\ triggers were introduced to the trigger menu in 2011, the
primary motivation being the selection of events for searches for SUSY
in events with no leptons~\cite{SUSY0l}.
Single, and \multijet\ \HT\ triggers exist, where the single jet \HT\ 
triggers are seeded by a standard single inclusive trigger and the \multijet\ 
\HT\ triggers are seeded by a standard \multijet\ trigger. These seeding triggers 
are required since the calculation of \HT\ without such a seeding trigger
would require full jet finding in all events, which would be computationally 
prohibitive in the trigger. 

To illustrate the \HT\ trigger performance a single \HT\ trigger has
been selected, requiring a leading energetic trigger jet, with $\ET > 100$\,\GeV, 
and total $\HT >400$\,\GeV\ at the EF.
The L1 and L2 stages for the \HT\ triggers are identical to those of the
single and \multijet\ chains discussed in
Section~\ref{section:efficiencies}.  Thus the efficiencies shown in
Figure~\ref{fig:eff} for the L1\_J75 and L2\_j95 triggers are relevant for 
the specific \HT\ trigger discussed here.

The quantity \HT\ in the trigger is calculated from all EF jets  
with an \ET\ above a specified threshold, and within $|\eta|< 3.2$, to 
exclude jets reconstructed in the less well understood forward region.  
There are
thus two key factors that affect the performance of an \HT\ trigger;
the leading jet \ET\ requirement and the jet \ET\ threshold for
summing the \HT.  These factors are investigated and presented below.

\begin{figure}[thp]
  \subfigure[]{
    \includegraphics[width=0.48\textwidth]{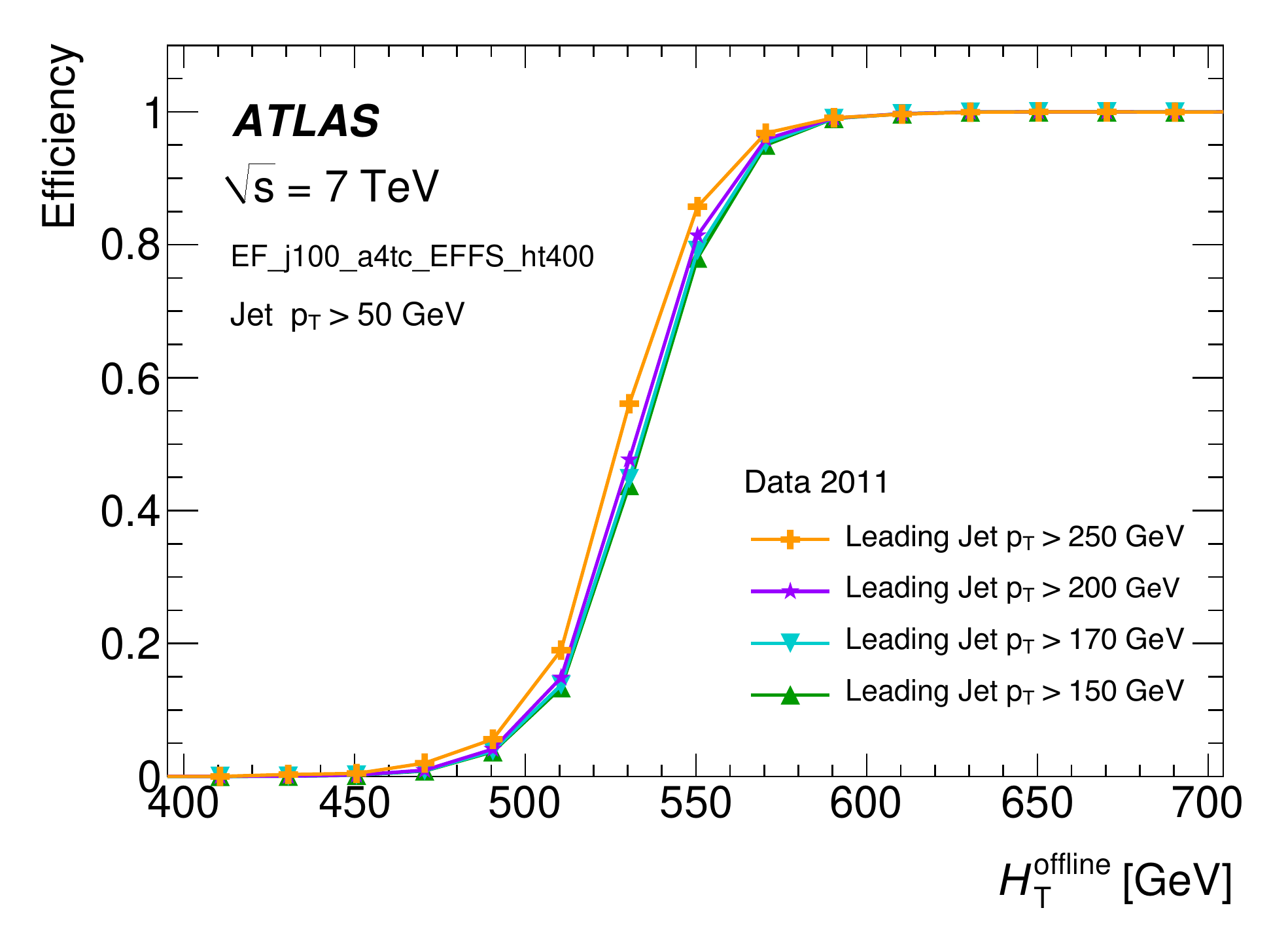}\label{fig:ht_LeadingJetSela}}
  \subfigure[]{
    \includegraphics[width=0.48\textwidth]{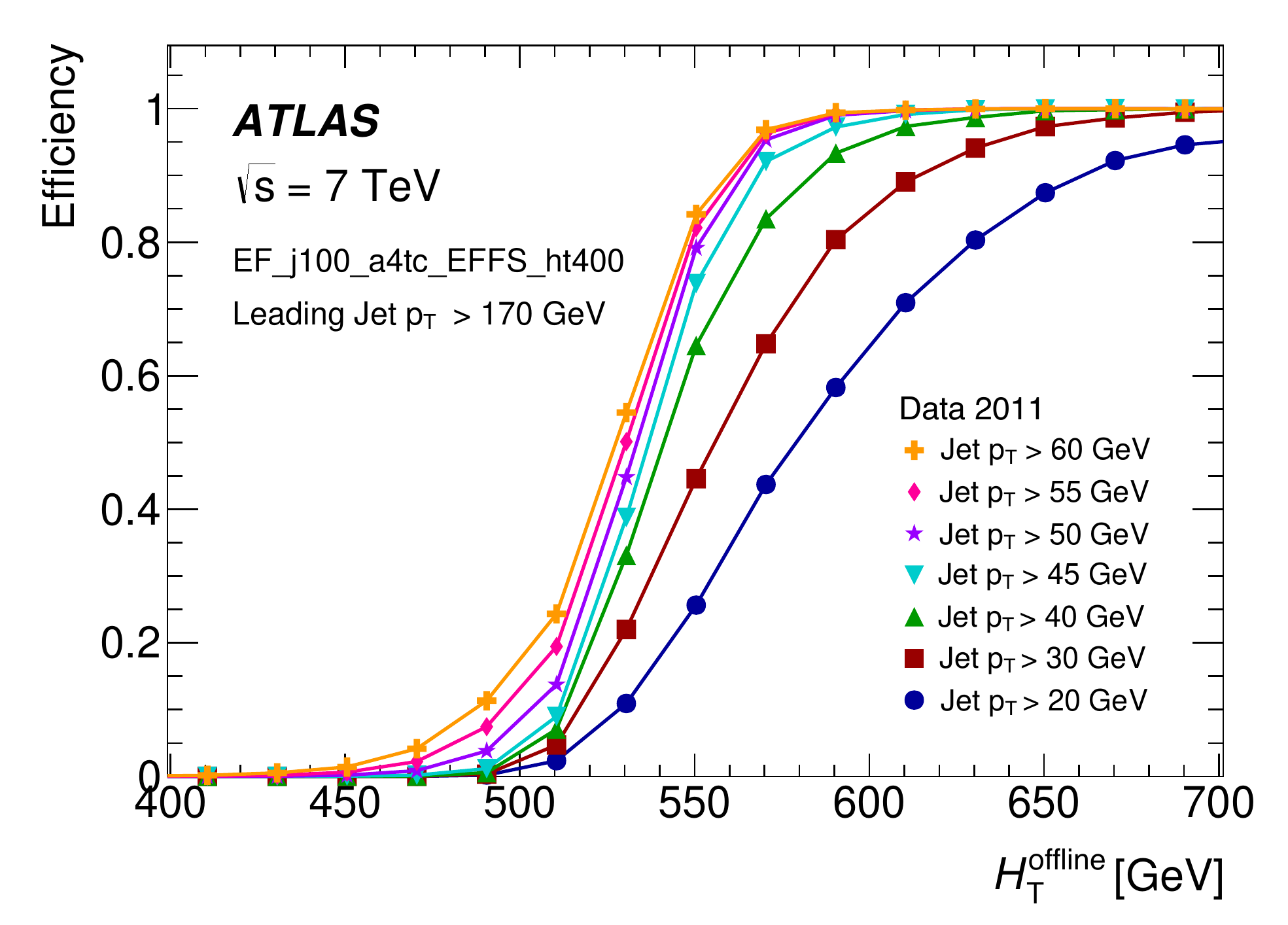}\label{fig:ht_LeadingJetSelb}}
  \caption{The efficiency for the trigger requiring both \mbox{$\HT>400$}\,\GeV\ and a leading jet 
    with \mbox{$\et > 100$}\,\GeV\ at the trigger level shown as a function of offline \HT\ 
    for: (a)~various leading offline jet
    \pt\ selections, where the offline \HT\ is calculated from offline 
    jets with $\pt > 50$\,\GeV; (b)~various offline jet \pt\ selections
    for the definition of offline \HT, where the leading offline jet \pt\ 
    selection is 170\,\GeV.  }
  \label{fig:ht_LeadingJetSel}
\end{figure}

The efficiency with respect to the offline \HT, for the \HT\ trigger 
requiring the leading jet \et\ at the trigger level to be greater than 100\,\GeV, 
and the total trigger \HT\ to be larger than 400\,\GeV, is shown in 
Figure~\ref{fig:ht_LeadingJetSel}
for various choices of the offline \HT\ definition.
Figure~\ref{fig:ht_LeadingJetSela} shows the effect of changing 
the leading offline jet \pt\ requirement in the definition of the offline \HT,
formed in this case from all offline jets with \pt\ greater than 50\,\GeV.  
With the specified trigger thresholds, the efficiency is seen to be relatively 
insensitive to changes in the choice of the leading offline jet \pt\ selection
within the range shown and remains at maximum efficiency
for offline $\HT>600$\,\GeV\ for all illustrated leading offline jet selections.

Figure~\ref{fig:ht_LeadingJetSelb} shows the effect of changing 
the \pt\ threshold for all offline jets used in the calculation of the 
offline \HT.   
In this case the leading offline jet \pt\ threshold used in the offline 
definition is 170\,\GeV. The trigger is seen to  maintain 
full efficiency only for offline $\HT>600$\,\GeV\ for those definitions 
where the selected offline jets are required to have \pt\ greater than 
approximately 50\,\GeV.
Definitions where the offline jet \pt\ selection is reduced to 
40\,\GeV, are seen to incur no corresponding loss in efficiency
only when the offline \HT\ is greater than 650\,\GeV.

In conclusion the performance of the  \HT\ trigger when seeded by a single 
high \ET\ jet is more sensitive to the choice of jet \ET\ threshold for
calculating \HT\ than the choice of leading jet \ET. This is because 
the choice of the leading jet is primarily a selection on the events from 
which the jets used in the \HT\ calculation are taken and does not significantly 
affect the value of \HT\ in that event, whereas the selection of the overall 
jet threshold used in the definition of \HT\ will change the calculated value 
of  \HT. Therefore as long 
as the leading jet threshold is chosen such that the single inclusive trigger 
is maximally efficient for the leading jet threshold used in the offline definition
of \HT\ then the \HT\ trigger will be maximally efficient given a suitable 
choice of offline threshold for the remaining jets.

\subsection{\LargeR\ jet triggers}

Physics analyses studying the properties of, or searching for, new or 
heavy particles decaying into boosted hadronic final states, may
include kinematic regions where the decay products are more collinear.
Such events may not be triggered by standard ($R=0.4$)
\multijet\ triggers, if the jets are too close to one another to
be resolved.  In such situations, \largeR\ jet triggers are useful.
For example, the decay products of a top quark produced with an
\ET\ above 300\,\GeV\ might be contained within a single jet with about
twice the radius of a standard jet. 
One such  ATLAS study involving boosted top quarks
where the hadronic decay products can not be resolved as individual
jets, is the search for new heavy resonances decaying into $t\bar{t}$
pairs~\cite{jettop}.
\LargeR\ jet triggers are useful in such
searches~\cite{jetlepton, jetboosted} and also jet substructure
studies~\cite{jetmass}.

The large-$R$ jet triggers at the EF are essentially the same as the 
standard jet
triggers discussed earlier; they are seeded by L1 and L2 triggers and
use the same reconstruction algorithms as the standard jets but
with a larger jet radius, namely $R=1.0$, in
contrast to the $R=0.4$ of the standard jet triggers.

The EF\_j240\_a10tc trigger, designed to target such final
states as described above, is described in this section.  This was the
lowest \ET\ unprescaled \largeR\ jet trigger for the 2011 running
period.

\begin{figure}[thp]
  \begin{center}
  \includegraphics[width=0.48\textwidth]{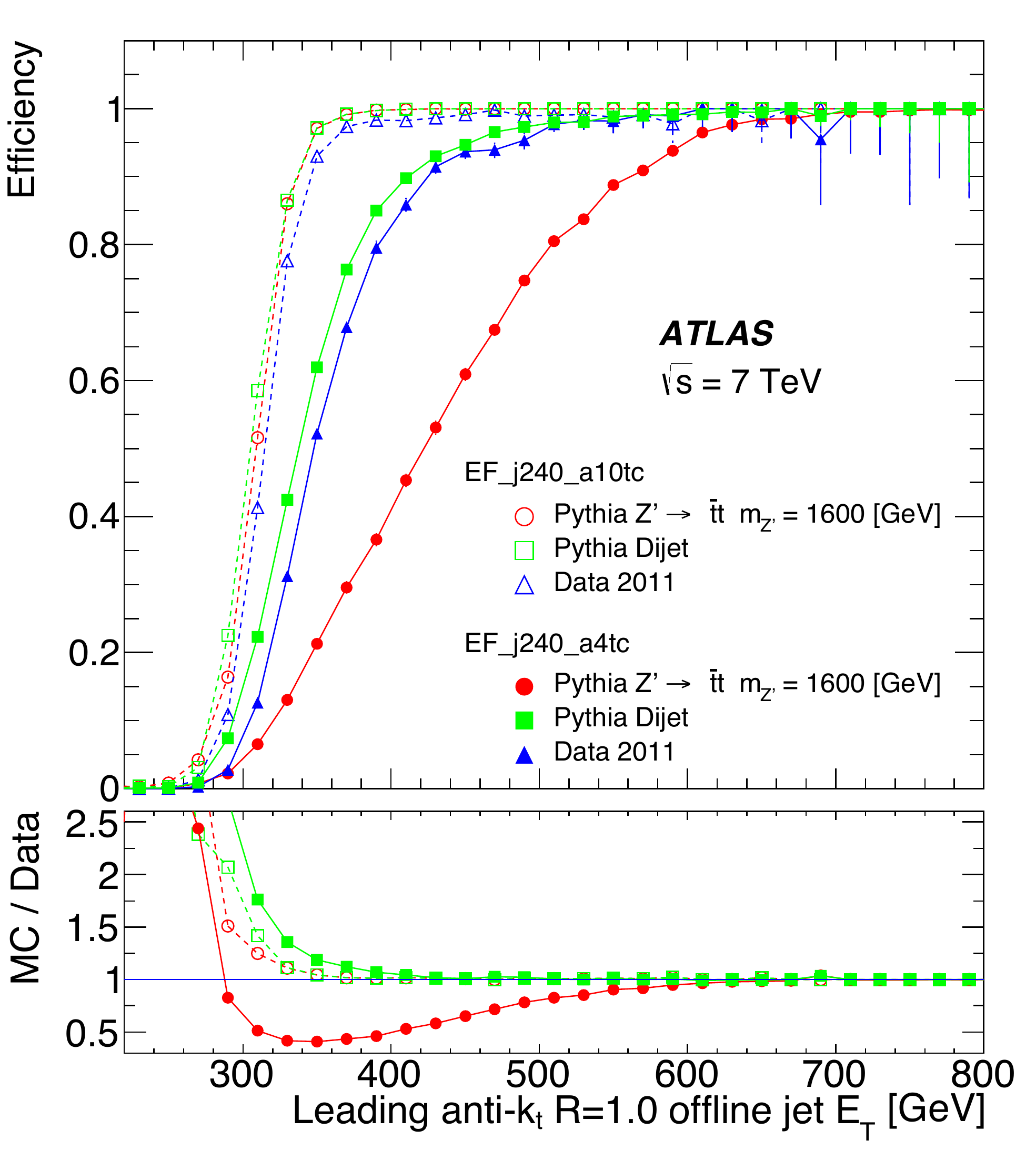}
  \caption{The efficiency of the \largeR\ jet trigger,
    EF\_j240\_a10tc (open markers), compared to the standard
    EF\_j240\_a4tc trigger (filled markers), both with respect to offline $R=1.0$ jets, for both data and simulation.}
  \label{fig:fat_jets_eff}
\end{center}
\end{figure}

Two \pythia~\cite{pythia6} Monte Carlo samples were selected to study such scenarios, both
containing a large number of high \ET\ jets.  The first sample, labelled \pythia\ Dijet, 
contains predominantly light quarks and gluons from the hard process,
and the second sample, labelled \pythia\ $Z^{\prime}\rightarrow t\overline{t}$,
models the production  of a BSM heavy $Z^\prime$ gauge boson, decaying 
to a top anti-top quark pair.  For the specific
\pythia\ Dijet sample chosen, the \ET\ of the leading jet in each event lies in the range 
300 to 600\,\GeV.

In Figure~\ref{fig:fat_jets_eff} the trigger efficiency of the
\largeR\ jet trigger is compared with that of the standard jet trigger
with the same energy threshold.  
The single inclusive jet trigger
efficiency has a sharper rising edge when applied to the \pythia\ Dijet 
sample, in which the jets have most of their energy occupying a narrow cone, 
than when
applied to the 
\pythia\ $Z^\prime\rightarrow t\overline{t}$
sample, in which the jets are the result of the decay of heavy, boosted objects. 
The large-$R$ jet trigger has both a sharper rising edge than the standard 
jet trigger, and similar behaviour for all samples. This suggests that the 
\largeR\ trigger is less sensitive than the standard jet trigger to the 
quark or gluon nature of the jets studied.

\begin{figure}[thp]  
  \subfigure[]{
    \includegraphics[width=0.48\textwidth]{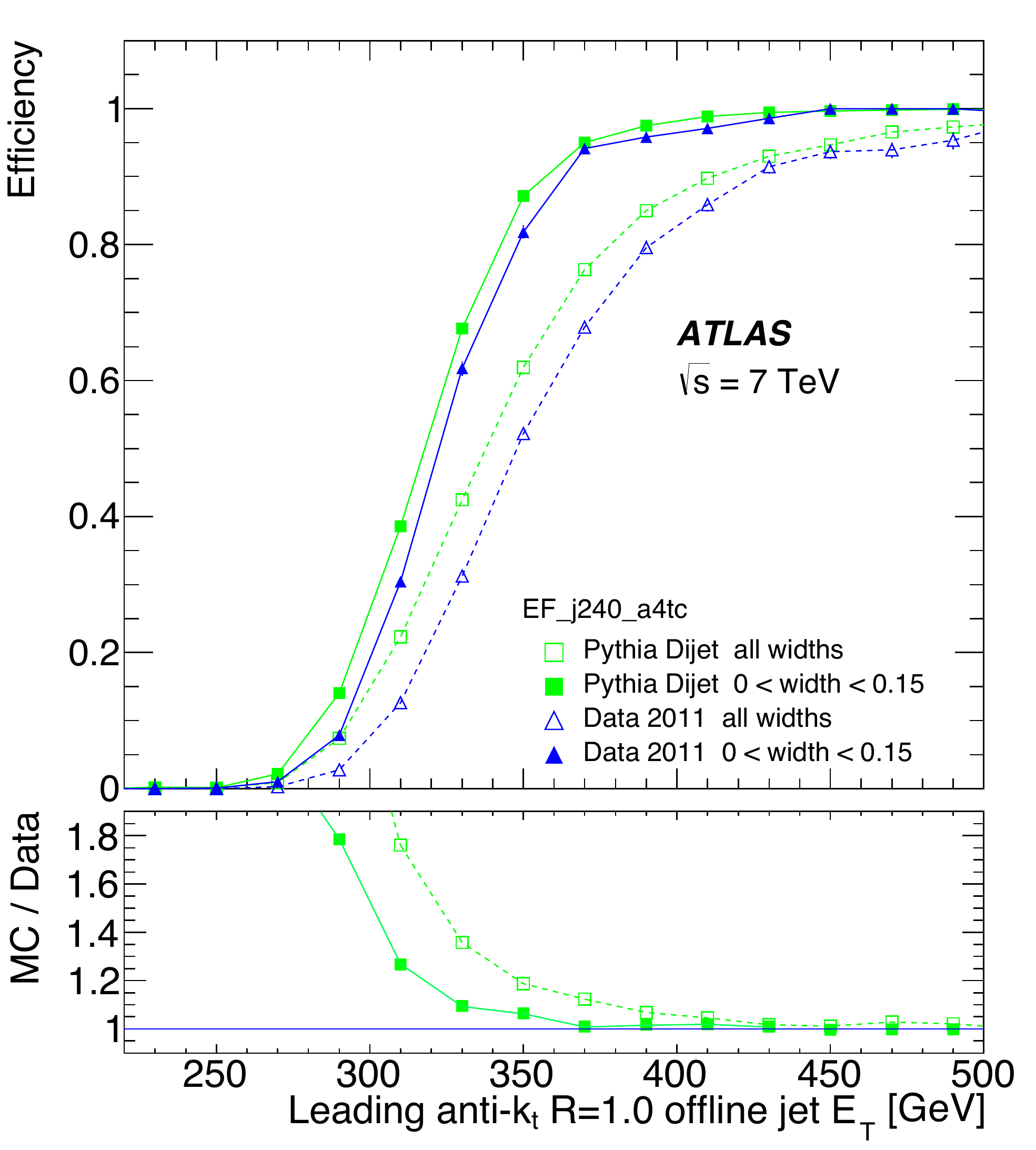}\label{fig:fat_jets_widtha}}
  \subfigure[]{
    \includegraphics[width=0.48\textwidth]{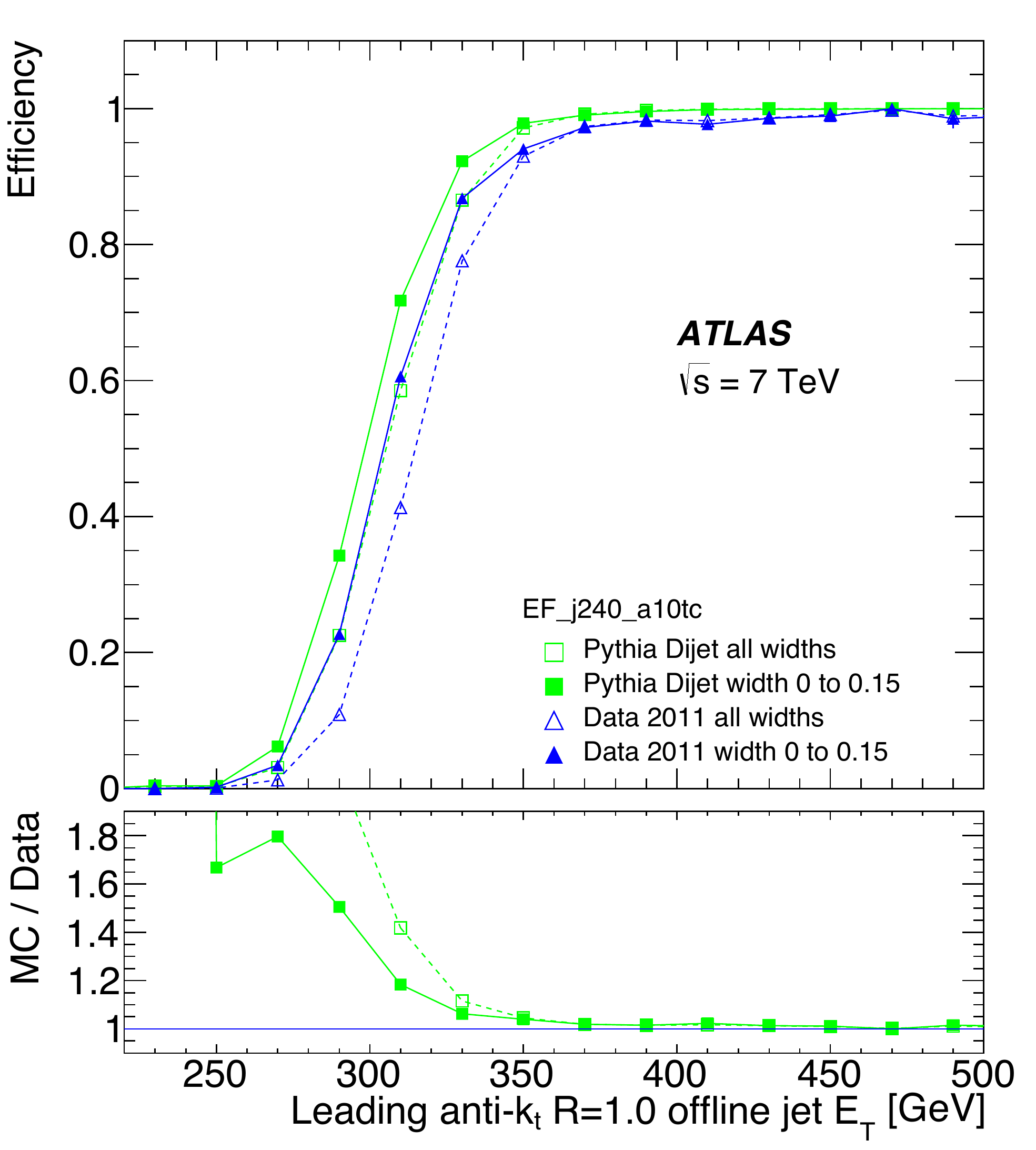}\label{fig:fat_jets_widthb}}
  \caption{The effect of jet width on efficiency, for data and Monte Carlo, 
    with respect to offline $R=1.0$ jets: (a) when the
    standard jet trigger, EF\_j240\_a4tc, is used; and (b) where the \largeR\ jet
    trigger, EF\_j240\_a10tc, is used. 
    Inclusive jet widths (open markers) and widths restricted to 
    $w<0.15$ (filled markers) are shown.} 
  \label{fig:fat_jets_width}
\end{figure}

In Figure~\ref{fig:fat_jets_width} the sensitivity of the jet trigger
to differences between quark-initiated and gluon-initiated jets is
explored further.  Since gluon-initiated jets are typically wider than 
quark-initiated jets, selecting narrower jets with a width~\cite{jetboosted} 
$w<0.15$, would slightly favour quark-initiated jets from the dijet sample. 
A slightly sharper rising edge than seen in the inclusive sample is observed.
The differences between the efficiencies for the samples with different 
widths are more distinct for the standard jet trigger,
seen in Figure~\ref{fig:fat_jets_widtha}, than the \largeR\ jet
trigger, as shown in Figure~\ref{fig:fat_jets_widthb}.  The performance
of the jet trigger in the simulation is seen to be broadly in agreement with
data  for the \largeR\ jet trigger where the larger jet width selection 
exhibits a sharper rising edge. The rising edge is considerably less sharp
for the standard jet trigger and shows significantly more variability with 
the properties of the sample. 
The \largeR\ jet trigger is therefore more robust to changes of the jet
width, which is a measure of jet substructure and radiation profiles.

The single inclusive jet triggers with $R=0.4$ have a sharply rising edge 
for jets with a narrow energy core, but the performance is reduced for jets
with wider, or multi-pronged energy distributions.  Jet triggers with
\largeR\ not only improve the performance for jets with wide energy
distributions but also improve the performance for jets with a narrow
energy core although at the cost of greatly increasing both  
the sensitivity of the trigger to \pileup, and the trigger rate.

\section{Jet identification for heavy ion collisions}
\label{section:HI}

Heavy ion (HI) collisions differ significantly from $pp$ collisions:
the intrinsic geometry of nuclear collisions results in large
variations of both the track multiplicity and the energy density. The 
data collected in 2011 for the ATLAS HI programme included collisions of 
lead nuclei with a nucleon--nucleon centre-of-mass energy of 2.76\,\TeV.  
Dedicated HI triggers
are required for the very different environment of HI collisions.  
Jets produced in HI collisions can be used as direct probes of the resultant
evanescent, hot, dense medium, and as such represent a very important
tool for physics
studies~\cite{HIresults,jetquenching2,jetquenching3}. Studies of such
jets at ATLAS in 2010 led to the first direct observation of a dijet
asymmetry, or {\em jet quenching}, in Pb+Pb
collisions~\cite{Aad:1309851}.

The study of a full range of observables which characterise the hot
and dense medium formed in HI collisions is possible with specific HI
triggers. In addition to global measurements of quantities such as
particle multiplicity and collective flow, electroweak gauge boson
production, heavy-quarkonia suppression, and the modification of jets
passing through the dense medium are accessible with ATLAS
data~\cite{HIresults,jetquenching2,jetquenching3}.

\begin{figure}[tph]
\begin{center}
  \includegraphics[width=0.5\textwidth]{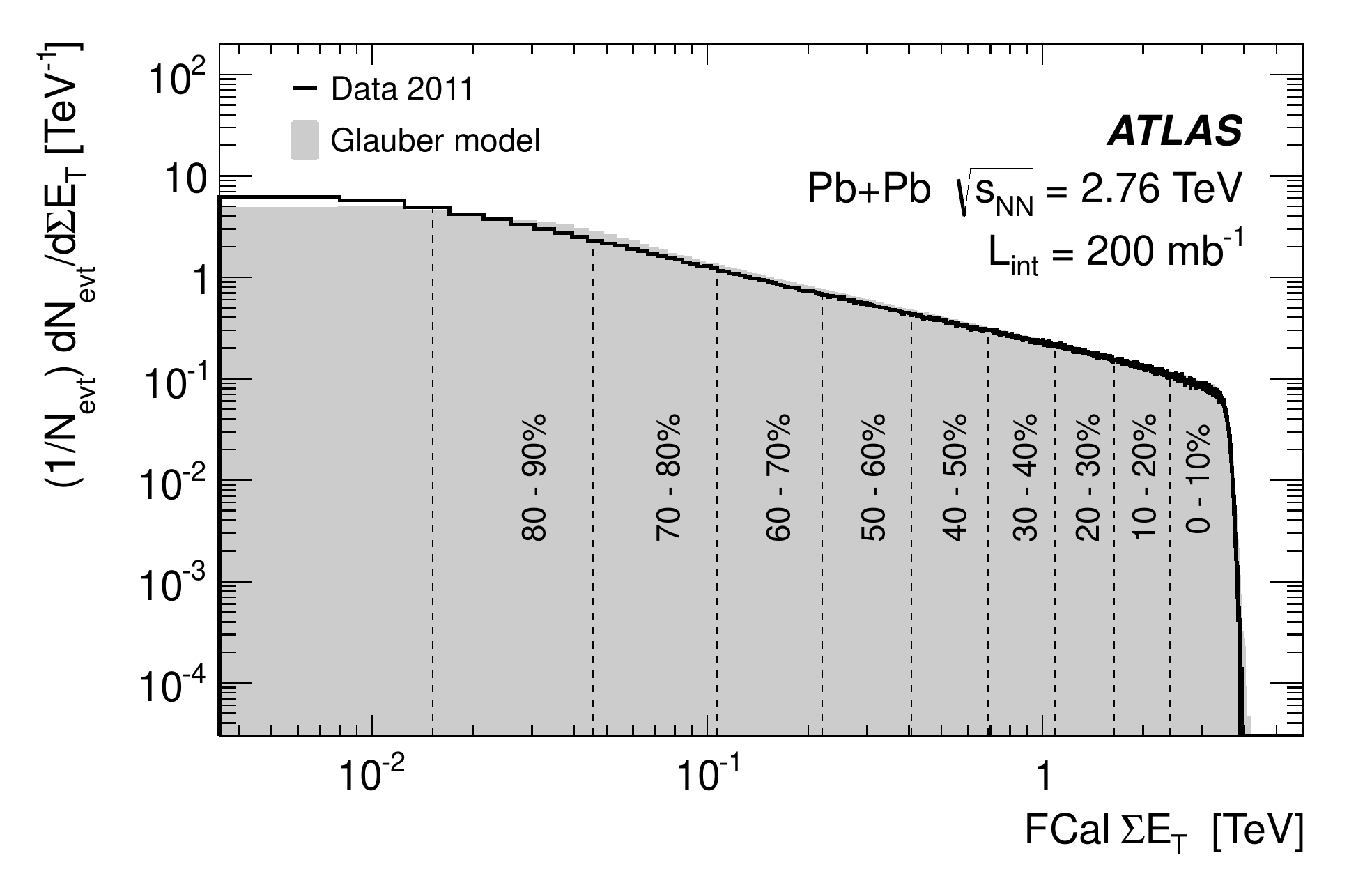}
  \caption{The distribution of the transverse energy deposited in
    forward calorimeter, FCal $\Sigma E_{\mathrm{T}}$, divided into
    10\% centrality bins~\cite{Aad:1379860}.  }
  \label{fig:HI_centrality}
\end{center}
\end{figure}

The dominant issue for jet measurements in the HI environment is the
presence of a large amount of additional energy coming from the
underlying event (UE), additional interactions originating from the
same Pb+Pb collision. The properties of this energy depend on the
impact parameter, or minimum transverse distance between the two
colliding nuclei. A direct measurement of the impact parameter is not
possible and so the {\em centrality} of the collision, defined as the
\ET\ deposited in forward calorimeter, FCal~$\Sigma E_{\mathrm{T}}$,
is used to characterise the UE~\cite{HIcentrality}.  The distribution
of FCal~$\Sigma E_{\mathrm{T}}$ is shown in
Figure~\ref{fig:HI_centrality} and is divided into bins according to
percentiles of the total Pb+Pb cross section.

The 0--10\% centrality bin represents the most central collisions, with the
smallest impact parameter and the largest UE contributions.  The
60--80\% centrality bin represents the most peripheral collisions used 
in this study,
corresponding to the largest impact parameters and the smallest UE
contributions.  Jet reconstruction in HI events typically corrects
for the contributions of the underlying event.

During the first LHC Pb+Pb run, in 2010, all events identified by the
minimum bias (MB) and forward detector
triggers~\cite{atlastrigger,Aad:1125884} were recorded and used for
the HI studies.  However, in 2011 the instantaneous luminosity
substantially increased, such that the peak rate of MB L1 triggers
exceeded \hbox{6 kHz}. Therefore to maintain efficiency for events
containing high \pt\ jets, MB events identified at L1 were passed
through L2 to the EF where specialised HI jet triggers were used to
select events.

The L1 MB triggers based on the total summed transverse energy in the
calorimeter were used as the L1 seeds for HLT triggers which
reconstruct high \pt\ electrons, muons and jets. Events were
transferred directly to the EF if the total transverse energy
deposited in the calorimeters exceeded 10\,\GeV. A detailed description
of the performance of various MB triggers in HI collisions can be
found elsewhere~\cite{Aad:1473425}.

Jets at the EF were reconstructed across the entire calorimeter
(including the forward region) using the \antikt\ algorithm with
radius parameter $R=0.2$ from projective towers of size $\Delta \eta
\times \Delta \phi = 0.1 \times 0.1 $ formed from the summation of
calorimeter cell energies. The small $R=0.2$ radius parameter was
chosen in order to be less sensitive to fluctuations in the underlying event, 
the 
contribution of which is estimated and subtracted event-by-event from each
jet at the calorimeter cell level after the jet finding. The
subtraction is performed separately in each 0.1 $\eta$ region and in
each calorimeter sampling layer ($i$). The background subtracted cell
energies ($E_{{\mathrm T}j}^{\mathrm {sub}}$) are calculated according to:

\begin{equation}
E_{{\mathrm T}j}^{\mathrm {sub}} = E_{{\mathrm T}j} - A_{j}\rho_{i}(\eta_{j})
\end{equation}   
   
\noindent where $E_{\mathrm{T}\it{j}}$ is the measured cell \ET,
$\rho_{\it{i}}$ is the average \ET\ density in a given $\eta$ region
and layer, $A_{\it{j}}$ is the cell area and $j$ runs over all
calorimeter cells.  Cells within jet candidates are excluded from the
estimate of average UE energy density $\rho$ to reduce biases.  Jet
candidates are required to have at least one tower with
$E_{\mathrm{T}}> 3$\,\GeV\ and a ratio of maximum tower \ET\ to average
tower \ET\ greater than four.

\subsection{Performance of the heavy ion triggers}

The performance of the HI jet trigger is evaluated here using the
Pb+Pb collision data recorded near the end of the 2011 data taking period
which  corresponds to an integrated luminosity of
140\,$\mathrm{\mu b}^{-1}$.

\begin{figure}[thp]
\begin{center}
  \includegraphics[width=0.5\textwidth]{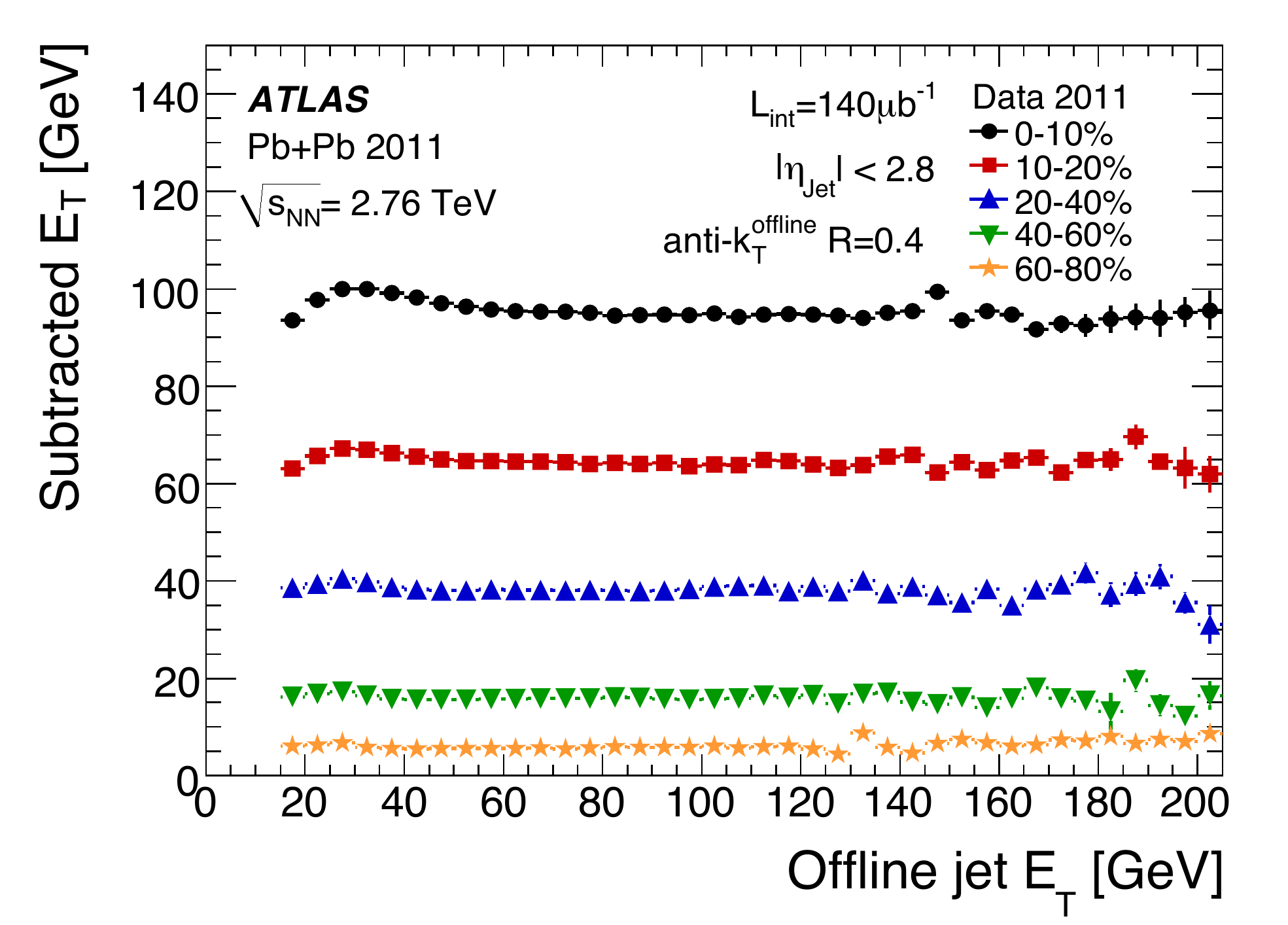}
  \caption{The mean transverse energy subtracted from offline
    jets with $R=0.4$, as a function of jet \ET\ for five centrality 
    bins for the heavy ion data.  }
  \label{fig:HI_centralityb}
\end{center}
\end{figure}

Offline jets are reconstructed from calorimeter towers, using the
\antikt\ algorithm, with $R=0.2$, 0.3, 0.4 and 0.5 in the region
$|\eta|<2.8$.  Unlike the trigger jets, the offline energy is corrected
for the lower hadronic response of the non-compensating ATLAS
calorimeters, using calibration constants obtained from Monte Carlo simulation
using \pythia~\cite{pythia6} embedded in the \hijing~\cite{HIJING} event
generator. In the offline reconstruction for HI events, an event-by-event
correction for elliptic flow, a long-range correlation
originating from the azimuthal momentum anisotropy of particle
emission, and a second iteration of the \antikt\ algorithm are made to
improve the performance of the UE
estimation~\cite{ATLAS:2012gna}. Figure~\ref{fig:HI_centralityb} shows
the mean estimated UE contribution to be subtracted from an $R=0.4$
offline jet as a function of the jet \ET\ for different centrality
bins. A small variation of the estimated UE contribution at the lowest
\ET\ is corrected for in the offline analysis. More details regarding offline jet
reconstruction in HI events can be found
elsewhere~\cite{Aad:12081967v1}.

In HI events it is not uncommon for fluctuations in the UE to create
regions with high \ET\ in the calorimeter that do not originate from
hard-scattering processes but which can nevertheless be reconstructed 
as jets. To
remove these jets it is required that offline calorimeter jets are
matched to a single 
electromagnetic cluster
with $\ET>7$\,\GeV\ or to at least one {\em track jet} - a jet formed using 
tracks from charged particles rather than calorimeter energy deposits.
In this case, track jets are reconstructed using the
\antikt\ algorithm with $R=0.4$ applied to tracks with
$\pt > 4$\,\GeV. This procedure is referred to as the fake-jet rejection
(FJR). Except where noted, the offline jet studies in this section
include FJR.

The efficiency of a trigger is defined as the per jet probability to
satisfy the trigger requirements as a function of offline jet
\ET. Only offline jets matching trigger jets within $\deltaR < 0.4$
contribute to the efficiency. The efficiency of the triggers with 
thresholds at 15, 20 and 25\,\GeV, respectively, are studied.

\begin{figure}[thp]
  \subfigure[]{
    \includegraphics[width=8.cm]{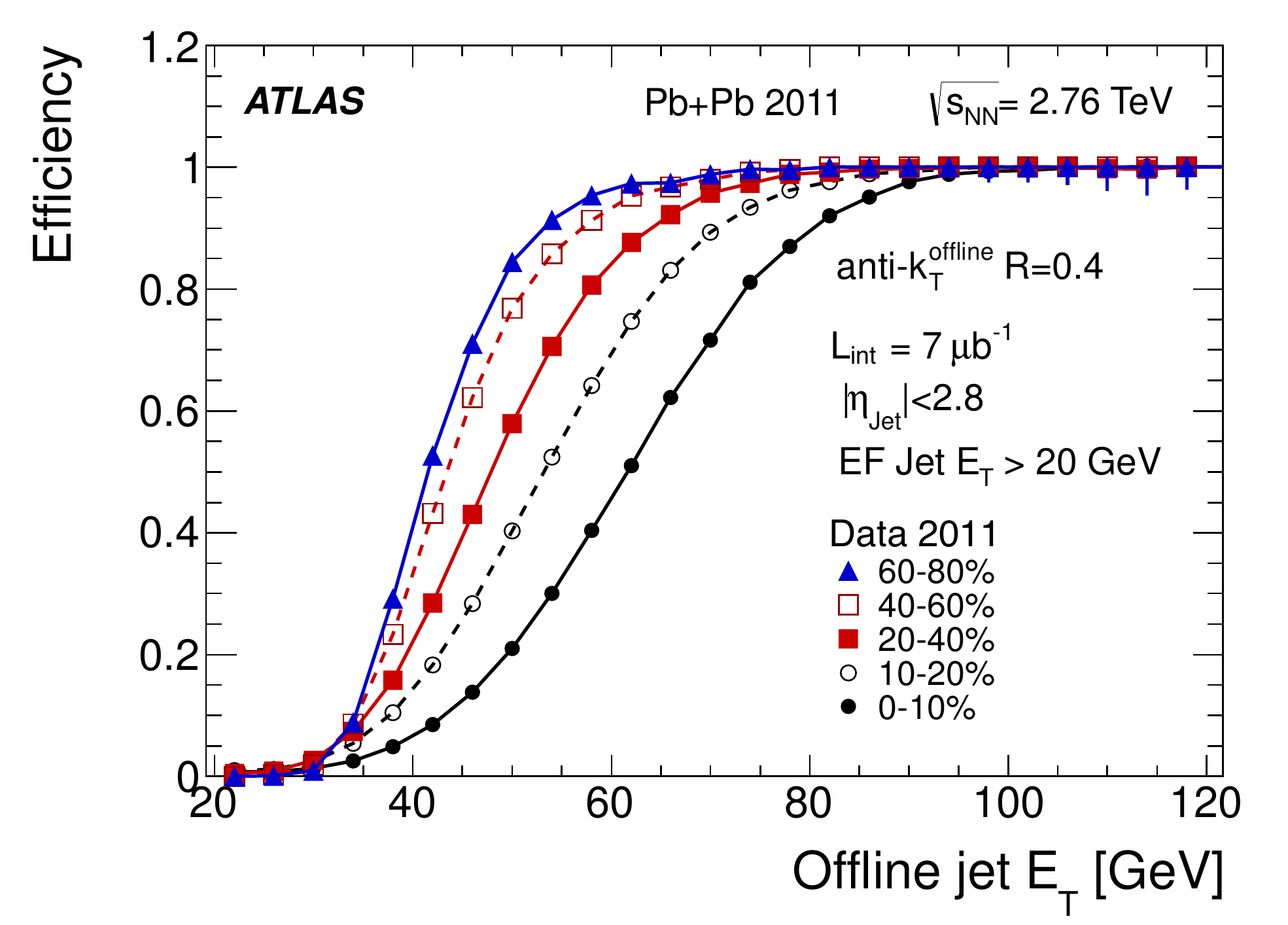}\label{fig:HI_aX_eff_FJRa}}
  \subfigure[]{
    \includegraphics[width=8.cm]{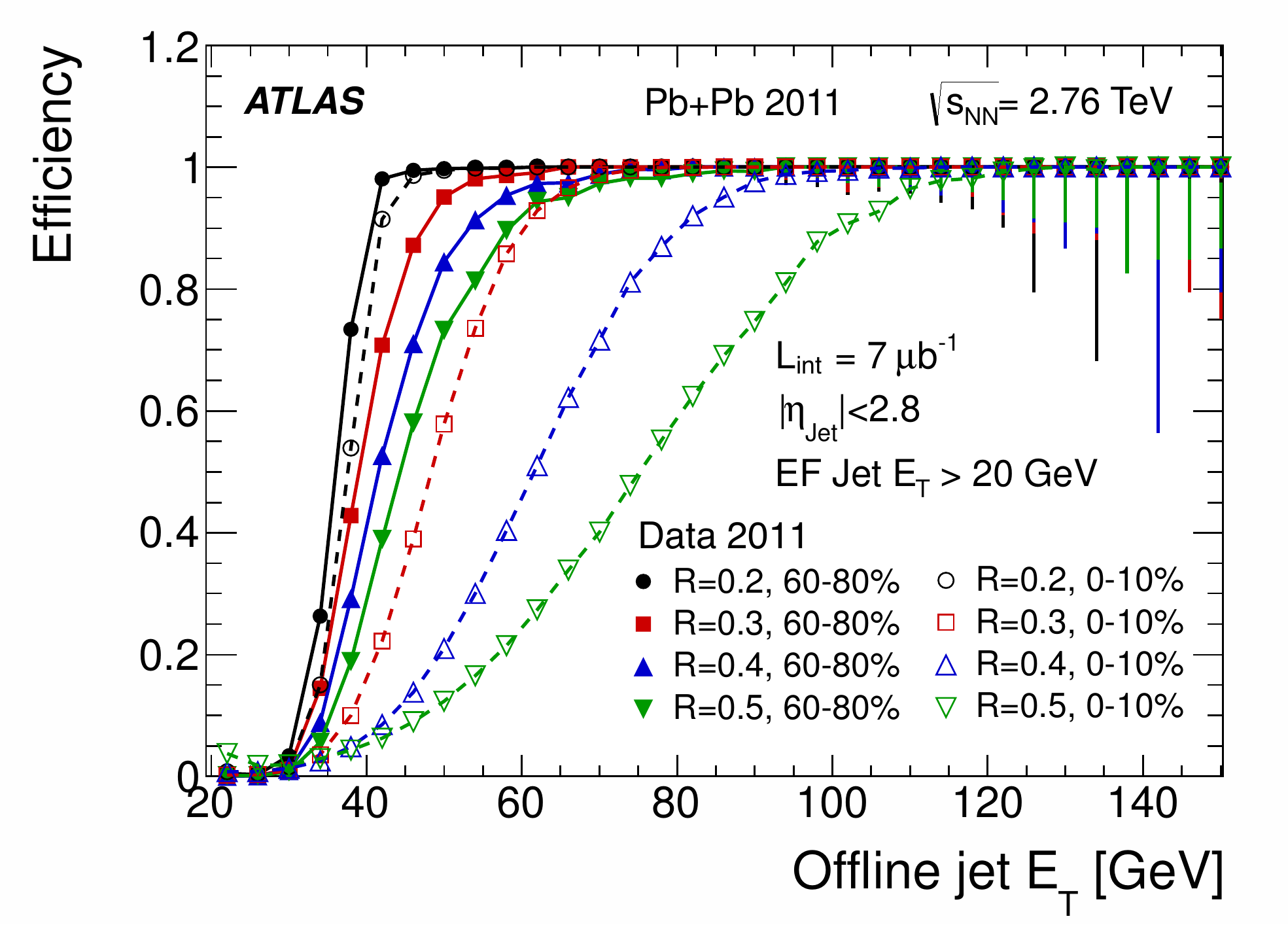}\label{fig:HI_aX_eff_FJRb}}
  \caption{ The trigger efficiency versus jet \et\ for heavy ion events: (a) for \antikt~$R=0.4$ offline jets
    in five centrality bins for a trigger threshold of 20\,\GeV; (b)
    in central and peripheral collisions for
    \antikt~$R=0.2$, 0.3, 0.4 and 0.5 offline jets, also with a trigger threshold of 20\,\GeV.  }
  \label{fig:HI_aX_eff_FJR}
\end{figure}

The performance of the jet reconstruction by the trigger over a range of
centralities and radius parameters typically used in HI analyses is
illustrated in
Figure~\ref{fig:HI_aX_eff_FJR}. Figure~\ref{fig:HI_aX_eff_FJRa}
shows the trigger efficiency for $R=0.4$ offline jets for the jet
trigger with \ET\ threshold of 20\,\GeV.  The efficiency decreases with
increasing centrality: the 95\% efficiency point of the trigger is
reached at 60\,\GeV\ in the most peripheral collisions and at
90\,\GeV\ in the most central collisions. Full efficiency is reached
around 75\,\GeV\ and
100\,\GeV\ respectively. Figure~\ref{fig:HI_aX_eff_FJRb} compares
efficiencies for the four radius parameters in the most central and
in the most peripheral collisions. Here it is observed that the
centrality dependence of the efficiency is more pronounced for larger
radius parameters, as the sharpness of the efficiency curves
degrades from peripheral to central collisions and from smaller to
larger offline jets. This reduction in efficiency for wider jets is
expected due to a degradation of energy and angular resolution.

\begin{figure}[htp]
  \subfigure[]{
    \includegraphics[width=8.cm]{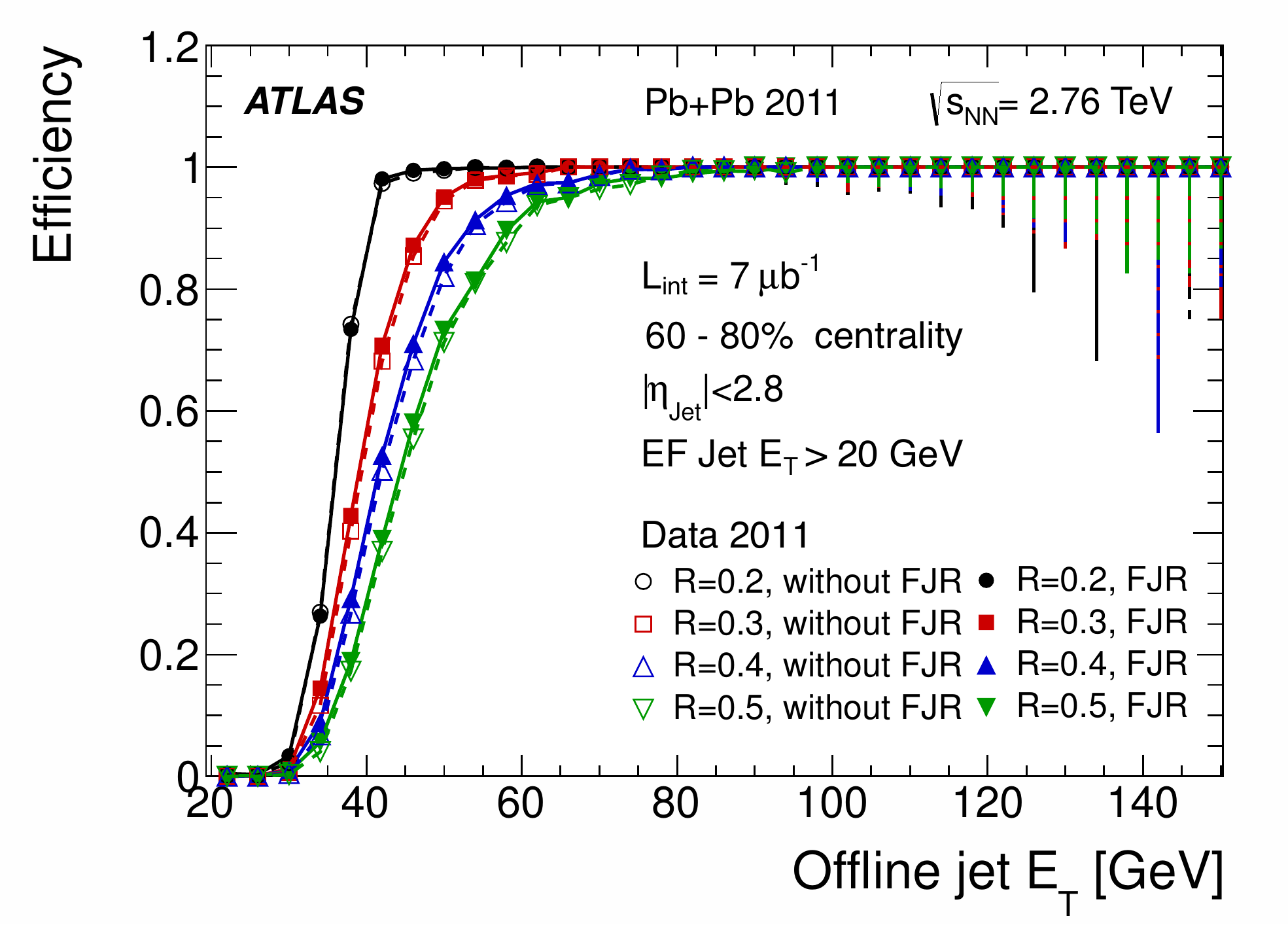}}
  \subfigure[]{
    \includegraphics[width=8.cm]{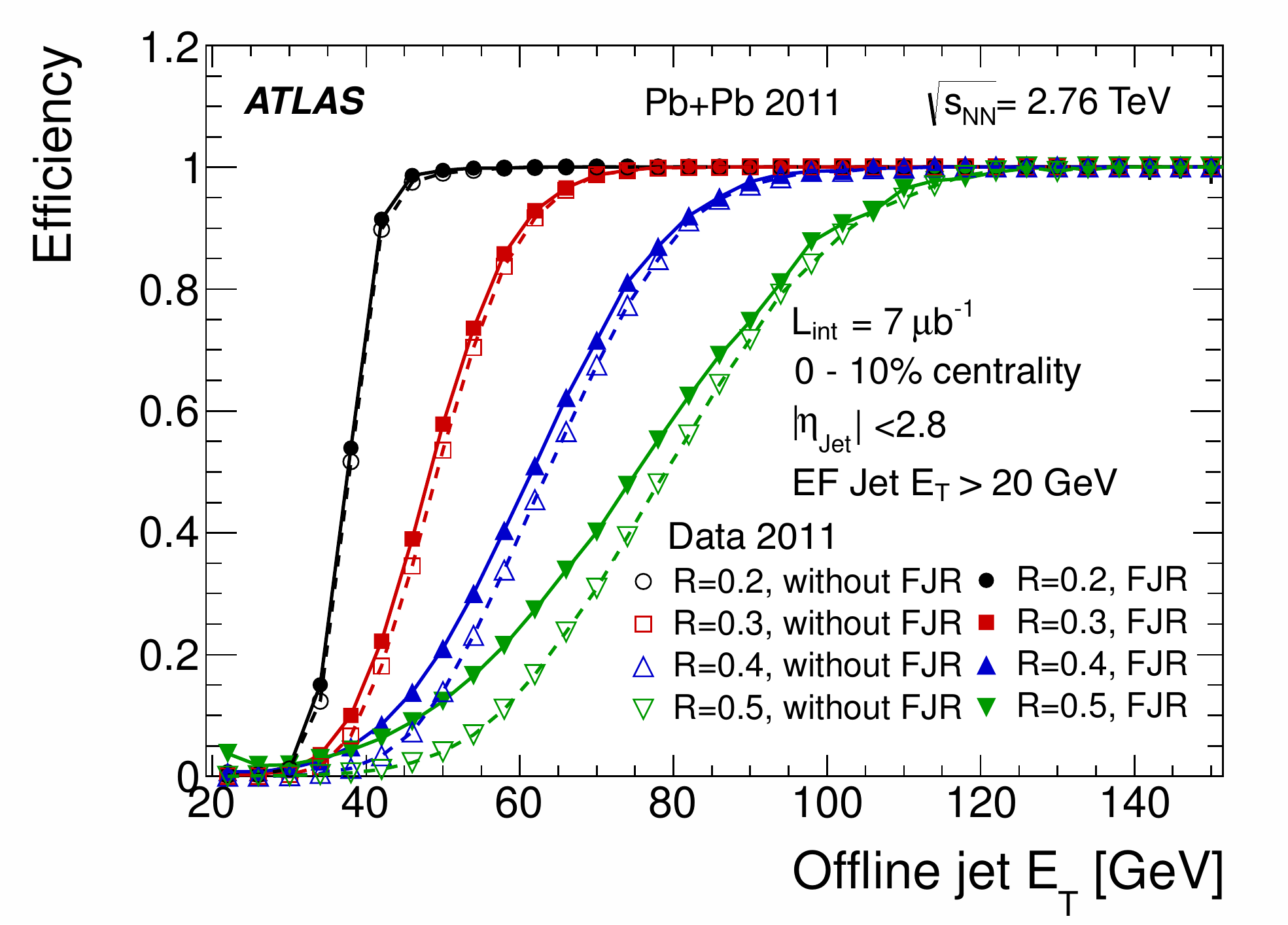}}
  \caption{The trigger efficiency for heavy ion events with a requirement of an 
    EF jet with $\et>20$\,\GeV:
    (a) for peripheral collisions; and (b) for central
    collisions. Shown are results for offline jets reconstructed with the \antikt\ algorithm using $R=0.2$, 0.3, 0.4 and 0.5, 
    both with (closed points), and without (open points) fake-jet rejection.} 
  \label{fig:HI_aX_eff}
\end{figure}

Figure~\ref{fig:HI_aX_eff} illustrates the influence of the FJR on the
efficiency in (a) peripheral and (b) central collisions.  Efficiencies
are shown for offline jets for different radius parameters with and
without FJR being applied. The efficiency is observed to be slightly
lower without FJR.  This difference is more marked for central
collisions, and increases with the increasing size of the offline
jet. This behaviour is caused by two effects: firstly by the increased
UE particle multiplicity in central collisions, leading to
a greater number of jets being reconstructed from underlying-event fluctuations, 
and secondly by the increased sensitivity of the trigger jets with 
larger radius parameter to these UE fluctuations.

\begin{figure*}[htp]
  \subfigure[]{
    \includegraphics[width=0.48\textwidth]{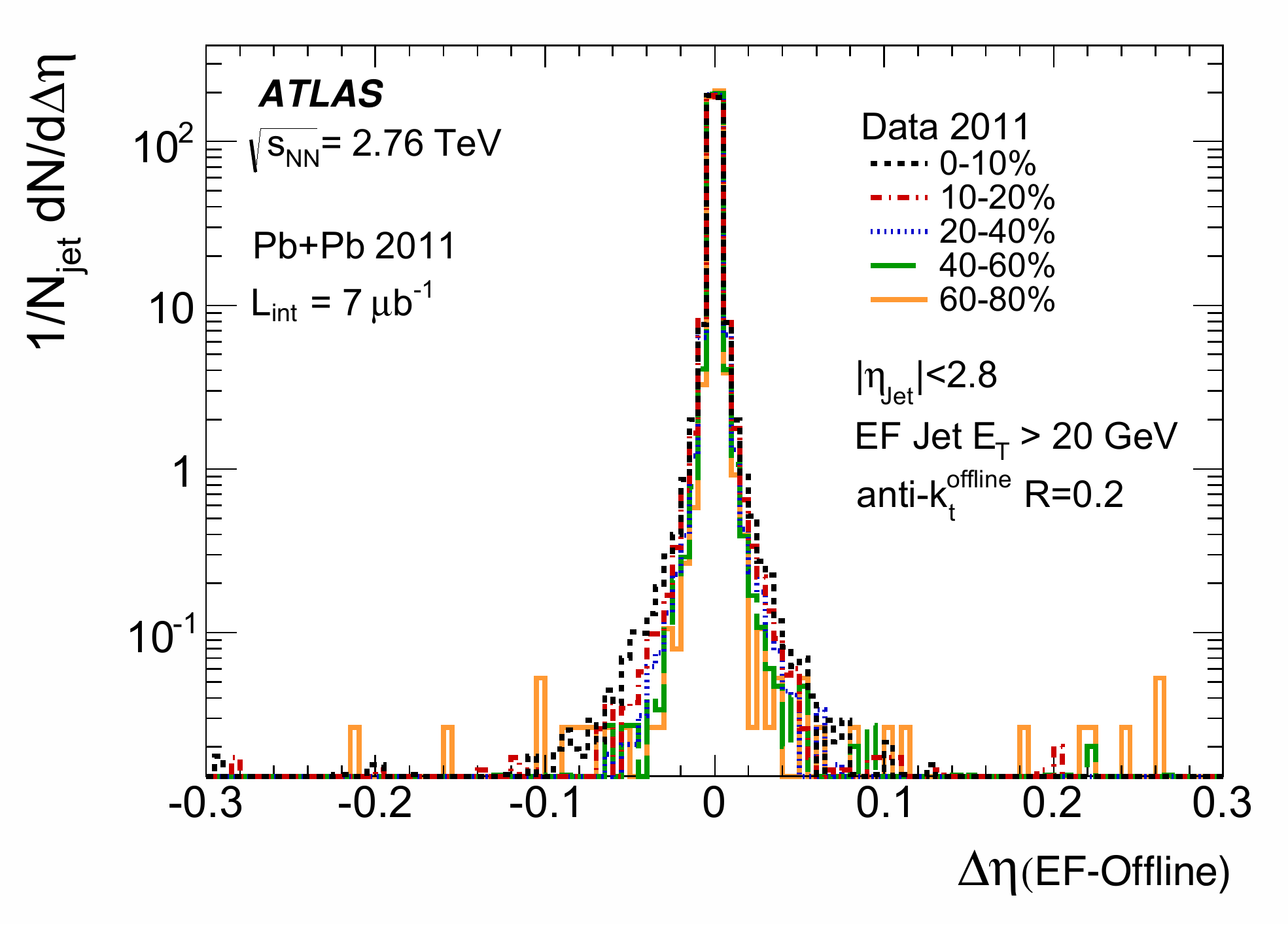}}
  \subfigure[]{
    \includegraphics[width=0.48\textwidth]{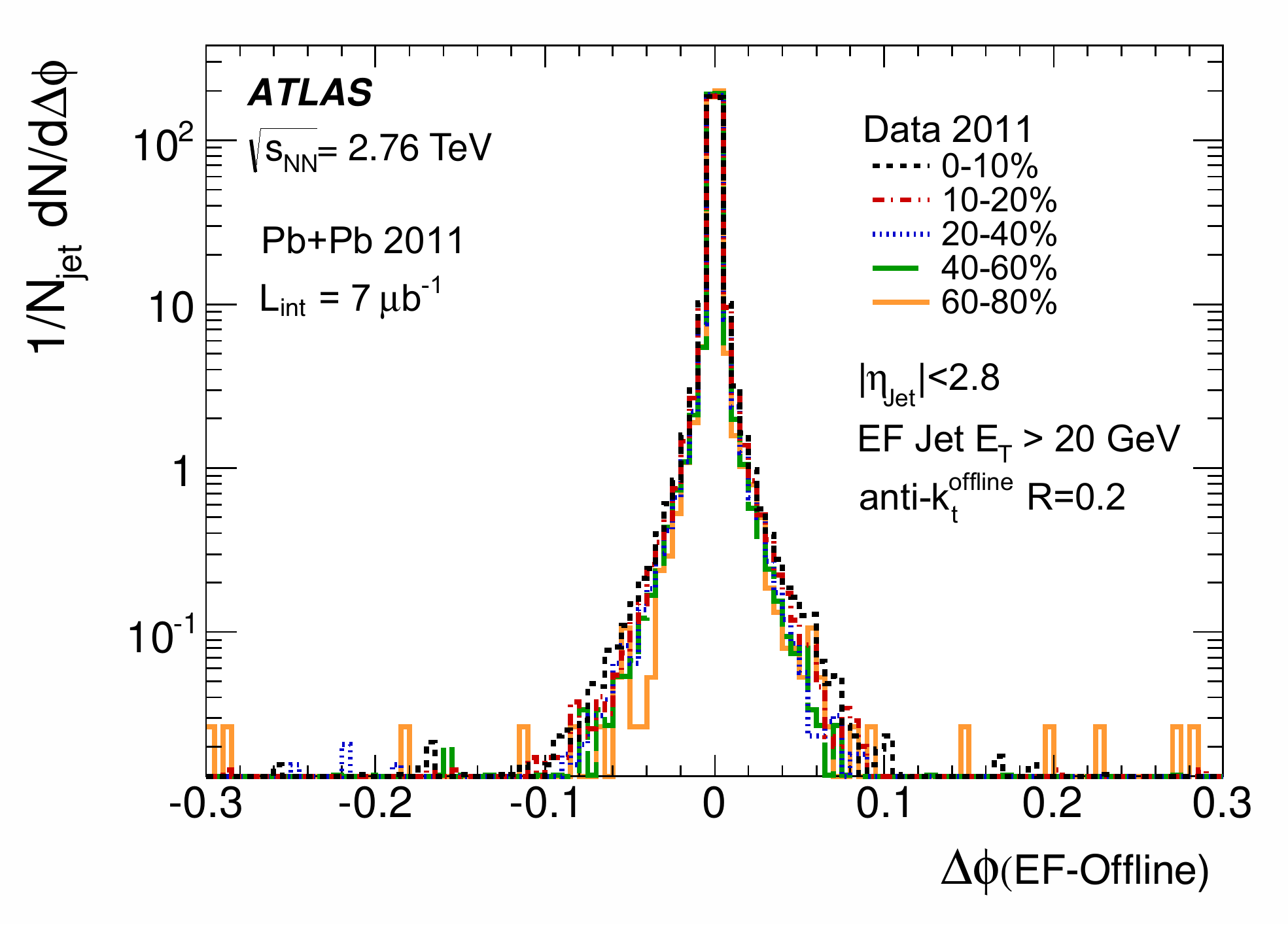}}\\
  \subfigure[]{
    \includegraphics[width=0.48\textwidth]{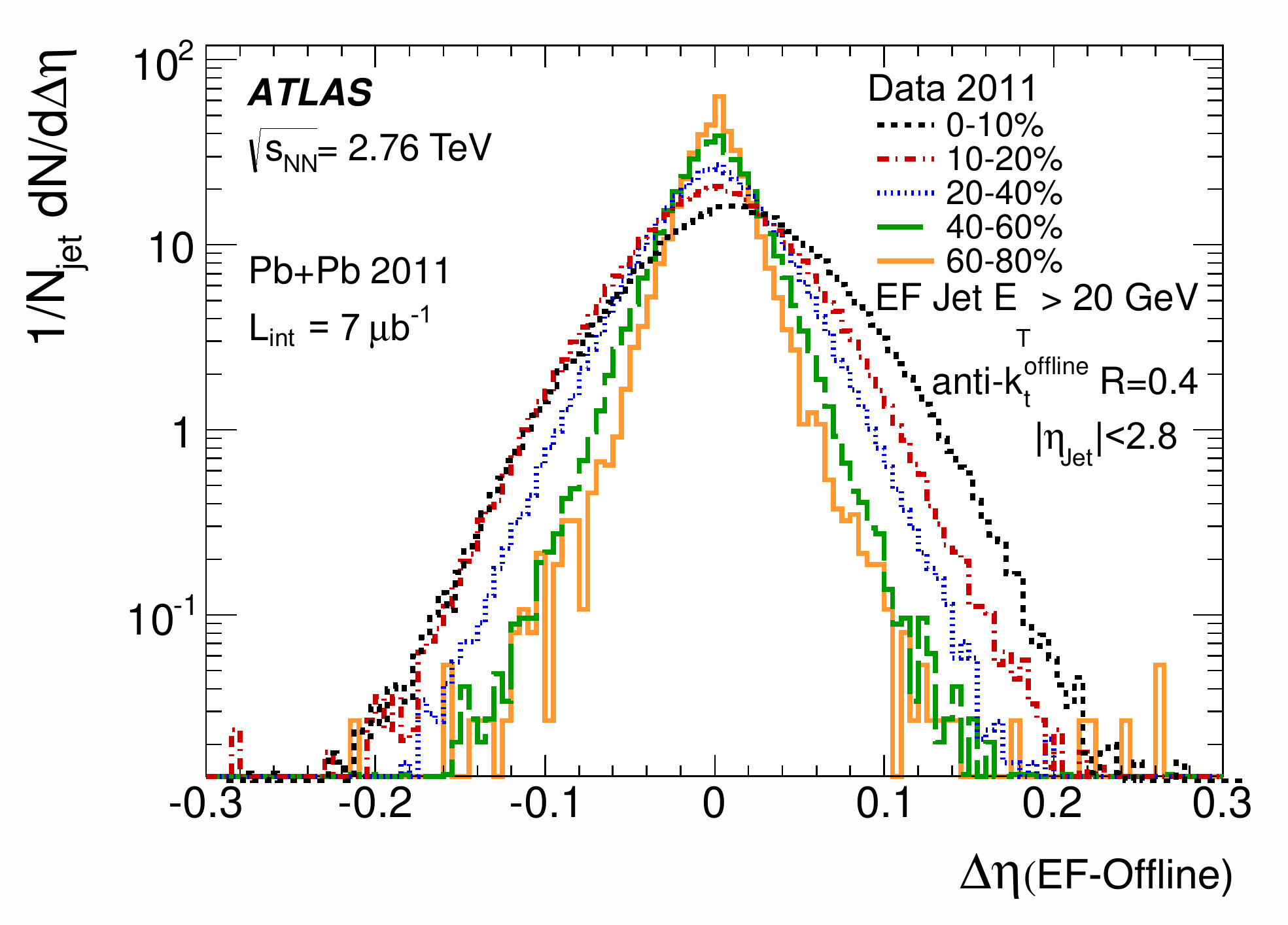}}
  \subfigure[]{
    \includegraphics[width=0.48\textwidth]{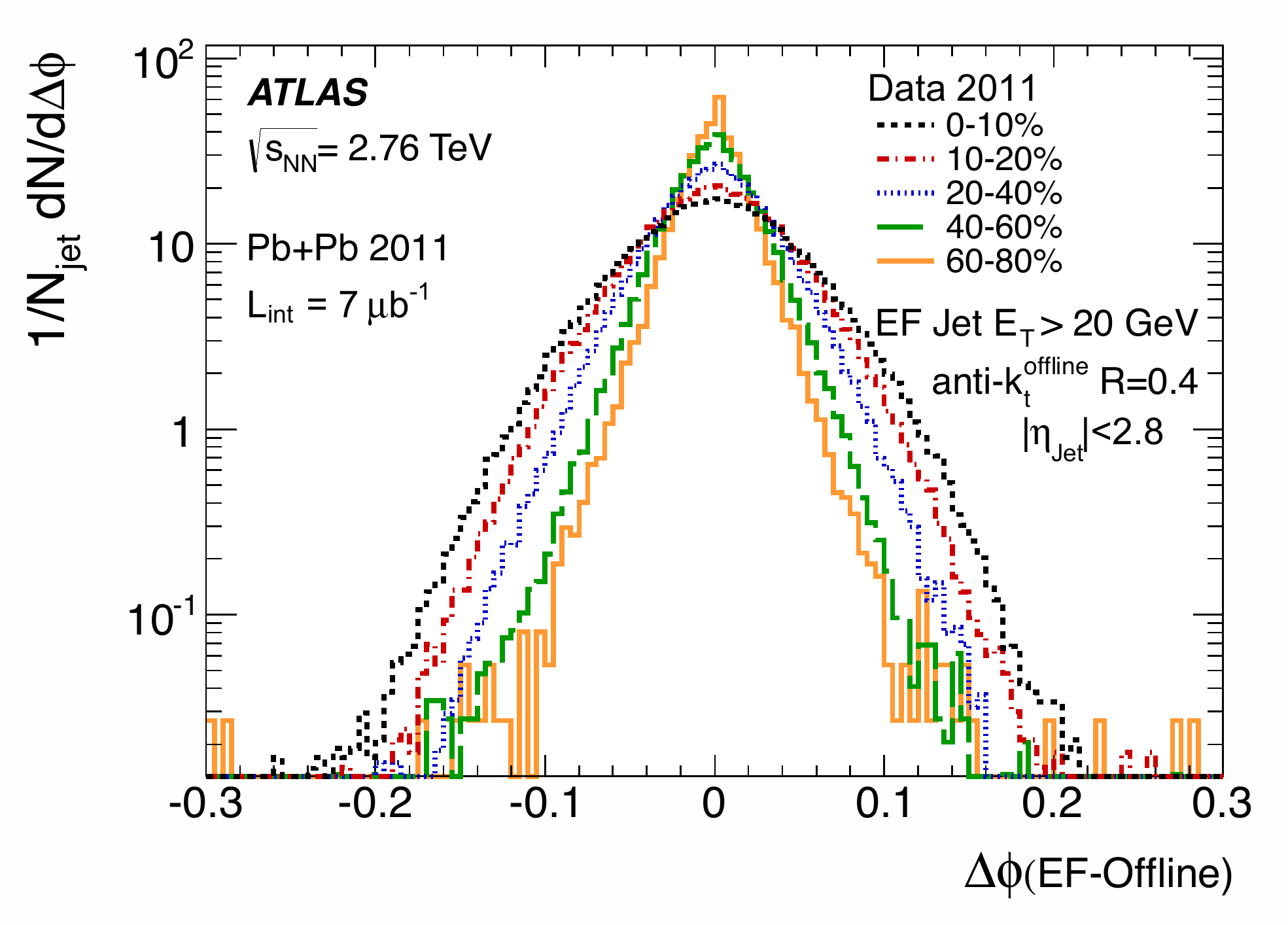}}
  \caption{ 
    The angular resolution for $R=0.2$ jets with $\et>20$\,\GeV\ reconstructed in the EF, with respect
    to $R=0.2$, and $R=0.4$ offline jets, for different centralities in the heavy ion data;
    the residuals in (a) $\eta$ and (b) $\phi$ between trigger jets and offline
    $R=0.2$ jets for different centrality bins;  the residuals in 
    (c) $\eta$ and (d) $\phi$ between trigger jets and offline $R=0.4$
    jets for different centrality bins.  }
  \label{fig:HI_angular_resolutions}
\end{figure*}

The angular resolution of trigger jets with respect to offline jets
with $R=0.2$ and $R=0.4$ is shown in
Figure~\ref{fig:HI_angular_resolutions} for different centrality
intervals. The angular resolution with respect to $R=0.2$ jets shows
very weak centrality dependence. However, the angular resolution with
respect to $R=0.4$ jets degrades with increasing centrality. This is
due to the smearing of the jet direction from the larger underlying-event 
activity.

The heavy ion programme at the LHC will provide crucial information about the
formation of a hot evanescent medium at the highest temperatures and
densities ever created in the laboratory.  The ATLAS jet trigger
algorithm performs well in the HI environment, using the same
\antikt\ algorithm used for $pp$ physics.  A small efficiency
degradation with increasing centrality is observed, which is less
pronounced for smaller radius parameters.  The angular resolution is
good, but shows some centrality dependence for larger radius
parameters.

\section{Summary}

The ATLAS jet trigger has been designed to provide an online
reconstruction of jets matching as closely as possible those from the
offline reconstruction. For this reason, while the RoI approach is mandatory for reasons of
bandwidth limitation at L1 and L2, during Run 1 the jet trigger for the 
EF processed events using the full calorimeter data and using the same 
\antikt\ algorithm as used offline. 

The time required for the complete processing of the full ATLAS jet 
trigger menu per event in the HLT during 2011 had a mean of below 300\,ms, well 
within the required budget for HLT processing.

For the L1 jet trigger, the lowest threshold deployed during 2011
data taking was 10\,\GeV\ at the electromagnetic scale. This trigger
was fully efficient for offline jets above 45\,\GeV.  The lowest threshold
HLT chain which included a L1 jet seed, selected jets reconstructed 
in the HLT with
transverse energy greater than 25\,\GeV\ and 30\,\GeV\ at L2 and the EF, 
respectively. These triggers were fully efficient for offline jets 
above approximately 60\,\GeV.

For unbiased trigger selection of jets with lower \ET, chains 
seeded by a random trigger at L1 with a large prescale and passing 
through L2 -- so not requiring a jet seed at either L1 or L2 -- were 
deployed. After accounting for the large prescale, these randomly 
seeded EF triggers were fully efficient for jets with offline \ET\ 
greater then 25\,\GeV.
 
For offline jets with
$\et>60$\,\GeV\ the jets are reconstructed at the EF in the barrel
region with a resolution in \ET\ with respect to offline jets, of better than 4\% and better than 
2.5\% in the endcaps. The performance in terms of offset and resolution of the 
jet trigger in data is reasonably well modelled by the Monte Carlo 
detector simulation to 
better than 1\%, but slightly worse in the crack regions between the 
barrel and endcap calorimeters. However, the steeply falling jet \pt\ 
spectrum means that small differences in the offset between 
data and the simulation results in differences in the positions 
of the rising edges of the jet trigger when comparing simulation with data. 
Physics analyses typically use data only for which the appropriate jet trigger 
has reached maximal  efficiency in order to ameliorate the effect of these 
differences.

More specialised triggers, intended specifically for searches for signatures of 
new physics, or for measuring the hadronic decay products of highly 
boosted massive objects, were operational in 2011 and are seen to 
perform well. The jet trigger for heavy ion physics was also seen to perform well, 
benefiting significantly from the full scan approach of the Event Filter 
to  reduce the processing time that would have been required by a purely RoI 
based approach in such a 
high occupancy environment.

\section*{Acknowledgements}


We thank CERN for the very successful operation of the LHC, as well as the
support staff from our institutions without whom ATLAS could not be
operated efficiently.

We acknowledge the support of ANPCyT, Argentina; YerPhI, Armenia; ARC, Australia; BMWFW and FWF, Austria; ANAS, Azerbaijan; SSTC, Belarus; CNPq and FAPESP, Brazil; NSERC, NRC and CFI, Canada; CERN; CONICYT, Chile; CAS, MOST and NSFC, China; COLCIENCIAS, Colombia; MSMT CR, MPO CR and VSC CR, Czech Republic; DNRF and DNSRC, Denmark; IN2P3-CNRS, CEA-DSM/IRFU, France; GNSF, Georgia; BMBF, HGF, and MPG, Germany; GSRT, Greece; RGC, Hong Kong SAR, China; ISF, I-CORE and Benoziyo Center, Israel; INFN, Italy; MEXT and JSPS, Japan; CNRST, Morocco; FOM and NWO, Netherlands; RCN, Norway; MNiSW and NCN, Poland; FCT, Portugal; MNE/IFA, Romania; MES of Russia and NRC KI, Russian Federation; JINR; MESTD, Serbia; MSSR, Slovakia; ARRS and MIZ\v{S}, Slovenia; DST/NRF, South Africa; MINECO, Spain; SRC and Wallenberg Foundation, Sweden; SERI, SNSF and Cantons of Bern and Geneva, Switzerland; MOST, Taiwan; TAEK, Turkey; STFC, United Kingdom; DOE and NSF, United States of America. In addition, individual groups and members have received support from BCKDF, the Canada Council, CANARIE, CRC, Compute Canada, FQRNT, and the Ontario Innovation Trust, Canada; EPLANET, ERC, FP7, Horizon 2020 and Marie Sk{\l}odowska-Curie Actions, European Union; Investissements d'Avenir Labex and Idex, ANR, R{\'e}gion Auvergne and Fondation Partager le Savoir, France; DFG and AvH Foundation, Germany; Herakleitos, Thales and Aristeia programmes co-financed by EU-ESF and the Greek NSRF; BSF, GIF and Minerva, Israel; BRF, Norway; Generalitat de Catalunya, Generalitat Valenciana, Spain; the Royal Society and Leverhulme Trust, United Kingdom.

The crucial computing support from all WLCG partners is acknowledged gratefully, in particular from CERN, the ATLAS Tier-1 facilities at TRIUMF (Canada), NDGF (Denmark, Norway, Sweden), CC-IN2P3 (France), KIT/GridKA (Germany), INFN-CNAF (Italy), NL-T1 (Netherlands), PIC (Spain), ASGC (Taiwan), RAL (UK) and BNL (USA), the Tier-2 facilities worldwide and large non-WLCG resource providers. Major contributors of computing resources are listed in Ref.~\cite{ATL-GEN-PUB-2016-002}.

\printbibliography

\newpage 
\begin{flushleft}
{\Large The ATLAS Collaboration}

\bigskip

G.~Aad$^\textrm{\scriptsize 87}$,
B.~Abbott$^\textrm{\scriptsize 114}$,
J.~Abdallah$^\textrm{\scriptsize 65}$,
O.~Abdinov$^\textrm{\scriptsize 12}$,
B.~Abeloos$^\textrm{\scriptsize 118}$,
R.~Aben$^\textrm{\scriptsize 108}$,
M.~Abolins$^\textrm{\scriptsize 92}$,
O.S.~AbouZeid$^\textrm{\scriptsize 138}$,
N.L.~Abraham$^\textrm{\scriptsize 150}$,
H.~Abramowicz$^\textrm{\scriptsize 154}$
% ...
\bigskip
\\
% ...
$^{12}$ Institute of Physics, Azerbaijan Academy of Sciences, Baku, Azerbaijan\\
% ...
$^{65}$ University of Iowa, Iowa City IA, United States of America\\
% ...
$^{87}$ CPPM, Aix-Marseille Universit{\'e} and CNRS/IN2P3, Marseille, France\\
% ...
$^{92}$ Department of Physics and Astronomy, Michigan State University, East Lansing MI, United States of America\\
% ...
$^{108}$ Nikhef National Institute for Subatomic Physics and University of Amsterdam, Amsterdam, Netherlands\\
% ...
$^{114}$ Homer L. Dodge Department of Physics and Astronomy, University of Oklahoma, Norman OK, United States of America\\
% ...
$^{118}$ LAL, Univ. Paris-Sud, CNRS/IN2P3, Universit{\'e} Paris-Saclay, Orsay, France\\
% ...
$^{138}$ Santa Cruz Institute for Particle Physics, University of California Santa Cruz, Santa Cruz CA, United States of America\\
% ...
$^{150}$ Department of Physics and Astronomy, University of Sussex, Brighton, United Kingdom\\
% ...
$^{154}$ Raymond and Beverly Sackler School of Physics and Astronomy, Tel Aviv University, Tel Aviv, Israel\\
% ...
\end{flushleft}


\end{document}